\documentclass[twocolumn,tighten]{aastex62}

\usepackage{color}
\usepackage{amsmath}
\usepackage{latexsym}
\usepackage{amssymb}
\usepackage{bbm}

\providecommand{\keywords}[1]{\textbf{Keywords:} #1}

\newcommand{\bpm}{\begin{pmatrix}}
\newcommand{\epm}{\end{pmatrix}}

\newcount\colveccount
\newcommand*\colvec[1]{
        \global\colveccount#1
        \begin{bmatrix}
        \colvecnext
}
\def\colvecnext#1{
        #1
        \global\advance\colveccount-1
        \ifnum\colveccount>0
                \\
                \expandafter\colvecnext
        \else
                \end{bmatrix}
        \fi
}

\newcommand {\real} {\mathbb{R}}

\newcommand {\cD} {\mathcal{D}}

\newcommand {\cI} {\mathcal{I}}

\newcommand {\cN} {\mathcal{N}}
\newcommand {\cO} {\mathcal{O}}

\newcommand {\cX} {\mathcal{X}}

\newcommand{\BA}{\ensuremath{\mathbf{A}} } %
\newcommand{\BB}{\ensuremath{\mathbf{B}} } %
\newcommand{\BC}{\ensuremath{\mathbf{C}} } %
\newcommand{\BD}{\ensuremath{\mathbf{D}} } %
\newcommand{\BI}{\ensuremath{\mathbf{I}} } %
\newcommand{\BL}{\ensuremath{\mathbf{L}} } %
\newcommand{\BP}{\ensuremath{\mathbf{P}} } %
\newcommand{\BQ}{\ensuremath{\mathbf{Q}} } %
\newcommand{\BS}{\ensuremath{\mathbf{S}} } %
\newcommand{\BU}{\ensuremath{\mathbf{U}} } %
\newcommand{\BV}{\ensuremath{\mathbf{V}} } %
\newcommand{\BY}{\ensuremath{\mathbf{Y}} } %
\DeclareMathOperator*{\E}{\mathbb{E}}

\DeclareMathOperator*{\V}{\mathbb{V}}
\DeclareMathOperator*{\Cov}{\mathbb{C}\text{ov}}

\DeclareMathOperator*{\argmax}{arg\,max}%

\newcommand{\bc}{\ensuremath{\mathbf{c}}} %
\newcommand{\bd}{\ensuremath{\mathbf{d}}} %
\newcommand{\bmm}{\ensuremath{\mathbf{m}}} %
\newcommand{\bvv}{\ensuremath{\mathbf{v}}} %
\newcommand{\by}{\ensuremath{\mathbf{y}}} %
\newcommand{\bzero}{\ensuremath{\mathbf{0}}} %

\newcommand {\bmu} {\mbox{\boldmath $\mu$}}
\newcommand {\BLambda} {\mbox{\boldmath $\Lambda$}}

\newcommand {\bpsi} {\mbox{\boldmath $\psi$}}

\newcommand {\btheta} {\mbox{\boldmath $\theta$}}

\newcommand {\BSigma} {\mbox{\boldmath $\Sigma$}}
\newcommand {\bth} {\mbox{\boldmath $\theta$}}

               {\refstepcounter{theorem}\vspace{2ex}%
                \noindent{\bf Examples \thetheorem} }%
               {\hfill $\diamond$}

               {\refstepcounter{theorem}\vspace{2ex}%
                \noindent{\bf Example \thetheorem} }%
               {\hfill $\diamond$}

\newcounter{problem} 
\renewcommand{\theproblem} {\arabic{problem}}
               {\refstepcounter{problem} \vspace{2ex}%
                \noindent{\bf Problem \theproblem} }%
               { }

               {\noindent{\bf Solution}}%
               {\hfill $\Box$ \\[1ex]}

\newlength{\boxwidth}

\newlength{\fullboxwidth}
\newlength{\fullinboxwidth}
\graphicspath{{figures/}}
\usepackage{graphicx}
\usepackage{cleveref}
\usepackage{algorithm, algorithmic}

\submitjournal{the Astrophysical Journal}

\shorttitle{Cosmic Inference}
\shortauthors{Takhtaganov et al.}

\begin{document}

\title{Cosmic Inference: Constraining Parameters With Observations and
       Highly Limited Number of Simulations}

\correspondingauthor{Zarija Luki\'c}
\email{zarija@lbl.gov}

\author{Timur Takhtaganov}
\affil{
Lawrence Berkeley National Laboratory \\
Berkeley, CA 94720, USA}

\author{Zarija Luki\'c}
\affiliation{
Lawrence Berkeley National Laboratory \\
Berkeley, CA 94720, USA}

\author{Juliane M\"uller}
\affiliation{
Lawrence Berkeley National Laboratory \\
Berkeley, CA 94720, USA}

\author{Dmitriy Morozov}
\affiliation{
Lawrence Berkeley National Laboratory \\
Berkeley, CA 94720, USA}

\begin{abstract}

Cosmological probes pose an inverse problem where the measurement result is
obtained through observations, and the objective is to infer values of
model parameters which characterize the underlying physical system -- our Universe.
Modern cosmological probes increasingly rely on
measurements of the small-scale structure, and the only way to accurately model
physical behavior on those scales, $x \lesssim 65 \, h^{-1}$Mpc, is
via expensive numerical simulations.
In this paper, we provide a detailed description of a novel statistical framework for obtaining
accurate parameter constraints by combining observations with a very limited number of cosmological simulations.
The proposed framework utilizes multi-output Gaussian process emulators that are adaptively constructed using Bayesian optimization methods. 
We compare several approaches for constructing multi-output emulators that enable us to take possible inter-output correlations into account  while maintaining the efficiency needed for inference.
Using Lyman-$\alpha$ forest flux power spectrum, we demonstrate that our adaptive approach requires considerably fewer --- by a factor of a few in Lyman-$\alpha$ $P(k)$ case considered here --- simulations compared to the emulation based on Latin hypercube sampling, and that the
method is more robust in reconstructing parameters and their Bayesian credible intervals.

\end{abstract}


\section{Introduction} \label{sec:intro}

The field of cosmology has rapidly progressed in the last few decades, going from a
largely qualitative picture of the ``Hot Big-Bang'' to the now well-tested 
``Standard Model of Cosmology''. This relatively simple model
describes current observations at a few percent level using only six parameters \citep{Planck2018}.
While this has been a great success --- driven by deluge of observations ---
outstanding questions still remain about the nature of dark matter and dark
energy, primordial fluctuations relating to the
inflation in the early universe, and the mass of neutrino particles.
To make further progress towards answering these questions, new
ground- and space-based observational missions will be carried out probing into highly
non-linear scales of cosmic structure.
Incoming wide-field sky surveys such as the Dark Energy Spectroscopic Instrument \citep{DESI},
the Large Synoptic Survey Telescope \citep{LSST},
the Wide Field Infrared Survey Telescope \citep[WFIRST,][]{WFIRST}, and Euclid \citep{Euclid} will provide precision
measurements of cosmological statistics such as weak lensing shear correlations,
cluster abundance, and the distribution of galaxies, quasars and Lyman $\alpha$
absorption lines.
Inferring values of the physical model's parameters using observations of the mentioned
sky surveys is a problem which belongs to the class of statistical inverse problems.

The application of Markov chain Monte Carlo (MCMC, \cite{Metropolis1953, Gelman2004}) or
similar Bayesian methods requires hundreds of thousands to even millions of forward model
evaluations in order to determine the posterior probabilities of the considered parameters.
When modeling the highly non-linear regime of the structure formation in the universe,
each such evaluation is a high-performance computing simulation costing more than $10^5$ CPU hours, as
even the most elaborate perturbation theory methods break down around scales of $x\lesssim 65 \, h^{-1} {\rm Mpc}$ \citep{Carlson2009}, and thus cannot be reliably used.
Cosmological simulations which from first principles numerically evolve the density field
will therefore be essential for the analysis and scientific inference of the future observational data sets.

While at first it may seem that this cost makes the inference computationally unfeasible,
it is in fact possible to efficiently sample the parameter space with a dramatically reduced
number of simulations, provided that certain smoothness conditions are satisfied.
This is achieved through the development of cosmological \emph{emulators},
that is, computationally cheap \emph{surrogate models} of expensive cosmological simulations.
The pioneering work on these techniques in cosmology was the ``cosmic calibration'' effort \citep{Heitmann2006, SHabib_KHeitmann_DHigdon_CNakhleh_BWilliams_2007a},
resulting in 1\% accurate matter power spectrum emulator \citep{Heitmann2010}.
This was followed by emulating weak lensing observables \citep{Liu2015, Petri2015},
galaxy halo occupation model \citep{Kwan2015},
halo mass function \citep{McClintock2018},
galaxy correlation function \citep{Zhai2018},
1D Lyman $\alpha$ flux power spectrum \citep{Walther2018},
and 21cm power spectrum \citep{Jennings2019}.

In this work we are not concerned with building an emulator for the cosmological simulation models that is accurate over the entire prior range of the input parameter values. Rather we focus on constructing an emulator in an adaptive fashion by preferentially selecting the inputs for the simulation that are more likely to result in the values of the output that are consistent with the observational data. By building up our emulator in this sequential way we strive to avoid performing unnecessary simulations that would be needed to have a globally accurate surrogate.
Similar Bayesian optimization for the construction of an emulator of the 1D Lyman $\alpha$ flux power spectrum has been recently considered in \cite{KKRogers_HBPeiris_APontzen_SBird_LVerde_AFontRibera_2019a}. The authors use a different acquisition function than the one used in this work. Furthermore, they do not consider multi-output emulators as we do but build multiple single-output ones.

The execution of such iterative workflow can be efficiently executed on high-performance computing platforms using the system described in \cite{Lohrmann2017}.
Briefly, as the workflow requires exploration of the parameter space via simulation trials, each of such simulations becomes a job managed by a parallel scheduler. This approach relies on Henson \citep{Morozov2016}, a cooperative multi-tasking system for in situ execution of loosely coupled codes.

Our treatment of multi-output emulation is different from the previous approach of \cite{SHabib_KHeitmann_DHigdon_CNakhleh_BWilliams_2007a} which relied on dimension-reducing techniques. Instead of approximating the power spectrum in the basis obtained from a principal component decomposition of the simulator's covariance structure, we assume a simple separable form for the covariance of the power spectrum as a vector-valued function (similarly to the approach of \cite{SConti_AOHagan_2010a}). This allows us to start with a small number of training inputs for the initial emulator construction and iteratively refine the initial design. Additionally, the separable structure of the covariance function allows us to perform training and prediction with the emulator using Kronecker products of small matrices making it efficient.

In this work, we are using 1D Lyman $\alpha$ (Ly$\alpha$) flux power spectrum as an output quantity of interest.
Following reionization happening around redshift $z\sim8$, the diffuse gas in the intergalatic medium (IGM) is predominantly photoionized,
but the small residual fraction of the neutral hydrogen gives
rise to Ly$\alpha$ absorption observed in spectra of distant
quasar sightlines (for a recent review, see \cite{McQuinn2016}).
This so-called Ly$\alpha$ forest is the premier
probe of the IGM and cosmic structure formation at redshifts $2 \lesssim z \lesssim 6$.
As Ly$\alpha$ absorption at $z \sim 3$ is sensitive to gas at around the cosmic mean, and at redshifts $z \geq 4$ even to the underdense gas in void regions of the universe \citep{Lukic2015}, complex and poorly understood physical processes related to galaxy formation are expected to play only a minor role in determining its structure \citep{Kollmeier2006, Desjacques2006}.
Forward modeling the structure of the IGM for a given cosmological and reionization scenario
is thus a theoretically well-posed problem, albeit requiring expensive cosmological hydrodynamical simulations.
The one-dimensional power spectrum is a summary statistic of the Ly$\alpha$ flux field which measures Fourier-space analogue of 2-point correlations in flux absorption along lines of sight to quasars.
This statistic can be sensibly used to measure cosmological parameters \citep{Seljak2006},
constrain neutrino sector \citep{NPD2015, Rossi2015},
to probe exotic dark matter models \citep{Irsic2017},
or to measure thermal properties of IGM \citep{Walther2018}.
Here, we focus on parameters describing the thermal state of the IGM, similarly as in \cite{Walther2018}.
However, there is nothing specific to the Ly$\alpha$ forest probe, particular data set or simulations in our inference formalism, thus the method we present here can straightforwardly be applied to other cosmological probes as well.

The outline of the paper is as follows.
In Section \ref{sec:physics}, we describe the details of the forward model for the Lyman-$\alpha$ power spectrum.  In addition to hydrodynamical simulations, we also use an approximate model for post-processing the thermal state of the IGM which is described in Section \ref{sec:gimlet-model}.
In Section \ref{sec:inference}, we provide a high-level overview of our main approach to inferring cosmological parameters from measurement data. First, we state the general Bayesian inference problem, and, following \cite{IBilionis_NZabaras_2014a}, show how it can be reformulated using a Gaussian process (GP) emulator as a Bayesian surrogate of the forward model. Next, we provide an outline of the adaptive algorithm developed in \cite{TTakhtaganov_JMueller_2018a} that we use to construct a GP emulator iteratively.
The details of the GP emulator construction and a comparison of the approaches to modeling interactions between emulator outputs are provided in Section \ref{sec:GPs}, as well as in the Appendix \ref{sec:math}. Results of applying this method on \cite{MViel_et_al_2013a} data and inferring thermal parameters of the IGM are given in
Section \ref{sec:inference_numerics}.  Finally, we present our conclusions in Section \ref{sec:conclusions}.
\section{Forward Model} \label{sec:physics}

In this paper, we analyze different ways of inferring the model parameters using flux power  spectrum observations.
To  this end, it is necessary to model the growth of cosmological structure and the thermal evolution of the IGM on scales far smaller (down to
$\mathcal{O}(10 h^{-1}$kpc)) than those described by the linear perturbation theory.
Cosmological hydrodynamical simulations with atomic cooling and UV heating
are the only method capable of modeling this process at
the percent level accuracy \citep[for approximate methods, see][and references therein]{Sorini2016, Lochhaas2016}.
Unfortunately, such simulations are computationally very expensive, $\sim 10^5$
CPU hours or more.
It is therefore desirable to also have a ``reduced'' model, which we can evaluate  for a  large number of points in the chosen parameter space, even if not as accurate as the full simulation model.
In the following we will first review our ``direct'' simulation model, and in Section \ref{sec:gimlet-model} we will present the approximate model based on post-processing the simulation's instantaneous temperature-density relation and the mean flux.

\subsection{Simulations}
\label{sec:simulations}

\begin{figure*}
\gridline{\fig{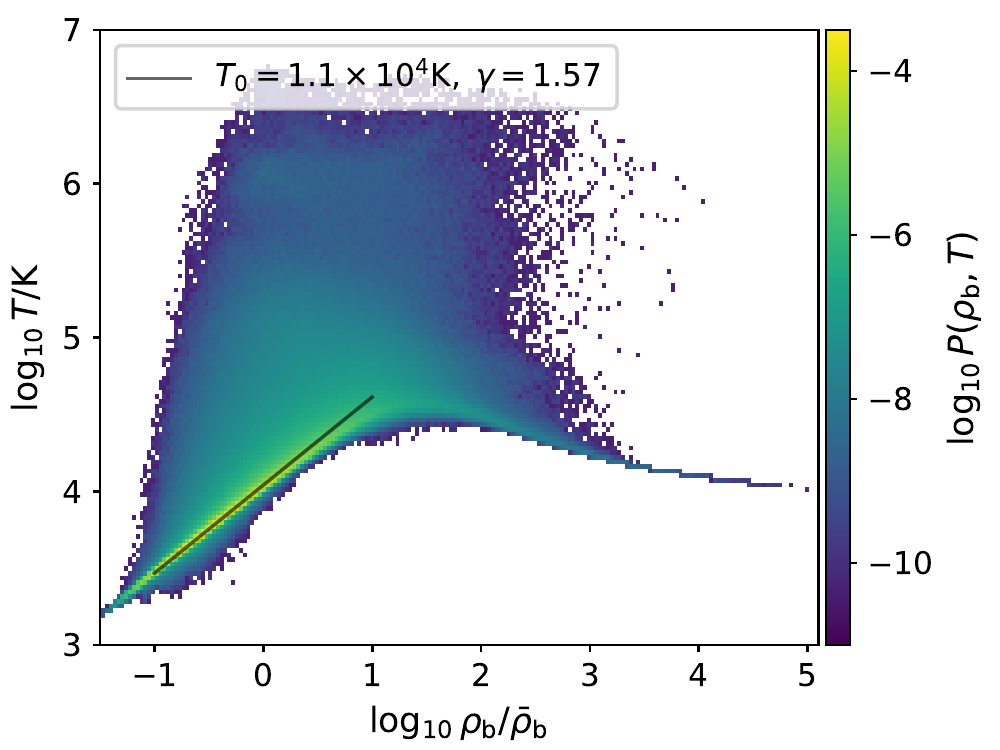}{0.49\textwidth}{(a) Fiducial model}}
\gridline{\fig{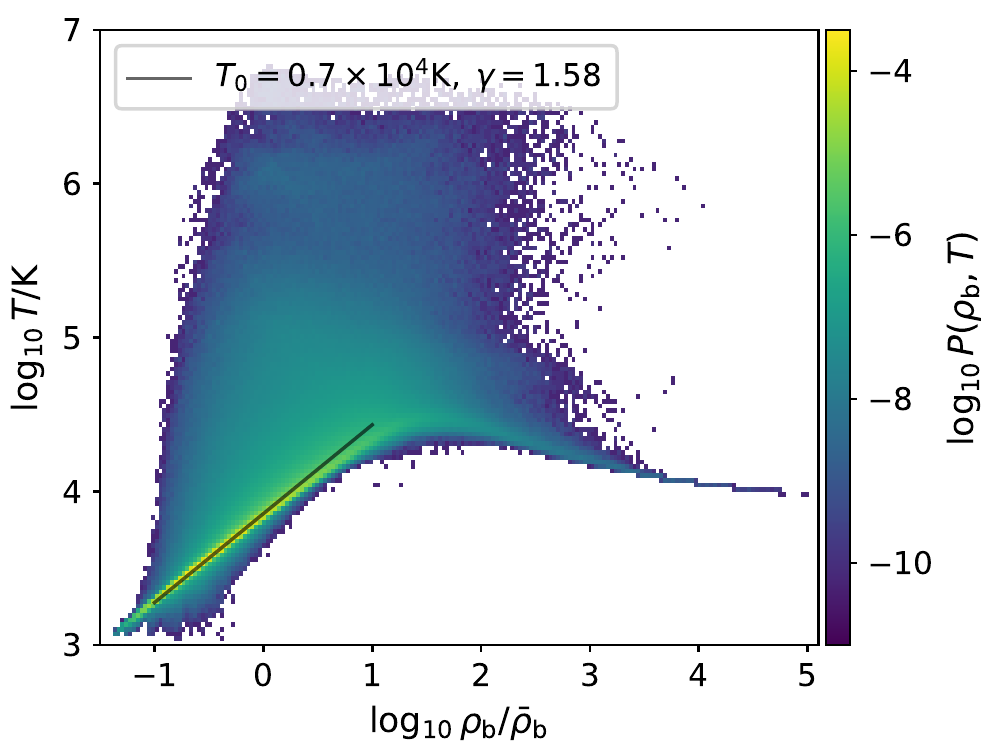}{0.49\textwidth}{(b) Low-$T_0$ model}
          \fig{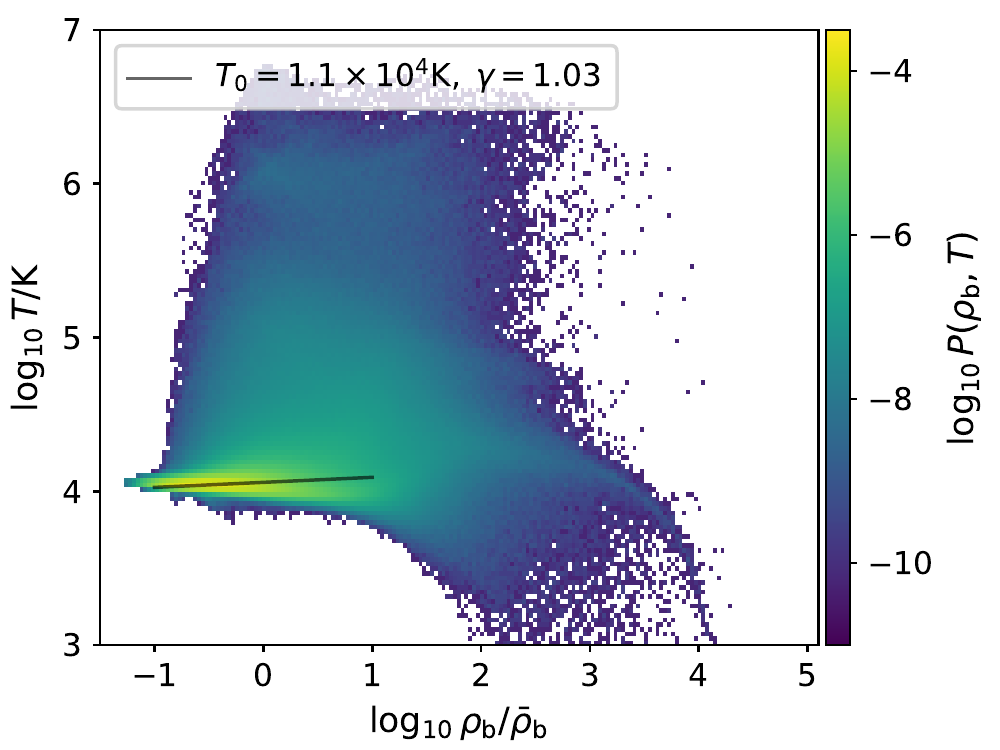}{0.49\textwidth}{(c) Low-$\gamma$ model}
          }
\caption{The density-temperature distribution of gas (volume-weighted
histogram) in three Nyx simulations: (a) "fiducial" one, (b) simulation with lower $T_0$ value
than in the fiducial one, and (3) simulation with lower $\gamma$ value. Other than one
parameter differing, simulations have the same all other parameters, including pressure smoothing scale at this redshift, $\lambda_P$.}
\label{fig:rho-T}
\end{figure*}

The hydrodynamical simulations we use in this paper are part of the
THERMAL  suite of Nyx simulations \citep{Walther2018}
consisting of 75 models, each in $L=20 h^{-1}$Mpc box
with $1024^3$ Eulerian cells and $1024^3$ dark matter particles.
The Nyx code \citep{Almgren2013} follows the evolution of dark
matter modeled as self-gravitating  Lagrangian  particles,  and  baryons  modeled  as  an  ideal gas on a set of rectangular Cartesian grids.
The Eulerian gas dynamics equations are solved using a second-order accurate piecewise
parabolic method (PPM) to accurately capture shocks.
Besides solving for gravity and the Euler equations,
we also include the main physical processes relevant for the Ly$\alpha$
forest.  We consider the chemistry of the gas as having a primordial composition of
hydrogen and helium, include inverse Compton
cooling off the microwave background and keep track of
the net loss of thermal energy resulting from atomic collisional processes \citep{Lukic2015}. All cells are assumed to be optically thin to ionizing radiation, and
radiative feedback is accounted for via a spatially uniform, time-varying 
UV background radiation given to the code as a list of photoionization
and photoheating rates \citep{Haardt2012}.

Thermal histories are generated in a similar way as in \cite{Becker2011} through
rescaling the photoheating by a density dependent factor: $\epsilon = A \Delta^B \epsilon_{\rm hm12}$.  Here, $\Delta = \rho_b/\bar{\rho_b}$ is the baryon overdensity, $\epsilon_{\rm hm12}$ are the heating rates tabulated in \cite{Haardt2012} while $A$ and $B$ are free parameters adjusted to obtain different thermal histories.
Note that while this approach makes it straightforward to change
instantaneous density-temperature
relation in the simulation, changing the pressure smoothing scale is more difficult as
it represents an integral of (an unknown) function of temperature across cosmic time.
We will return to this point later in Section \ref{sec:inference_numerics}.

We choose mock sightlines, or ``skewers'', crossing the domain
parallel to one of the axes of the simulation grid and piercing the cell
centers. Computationally, this is the most efficient approach. This
choice of rays avoids explicit ray-casting and any interpolation of
the cell-centered data, which introduce other numerical and
periodicity  issues.  As a result, from an $N^3$ cell simulation,
we obtain up to $N^2$ mock spectra, each spectrum having $N$ pixels.
We calculate the optical depth, $\tau$, by convolving neutral hydrogen in each pixel
along the skewer with the profile for a resonant line scattering and
assuming Doppler shift for velocity \citep[for details, see][]{Lukic2015}.
We compute this optical depth at a fixed redshift, meaning we do not account
for the speed of light when we cast rays in the simulation;
we use the gas thermodynamical state at a single cosmic time.  The
simulated skewers are therefore not meant to globally mock observed Ly$\alpha$
forest spectra, but they do recover the statistics of the flux in
a narrow redshift window, which is what we need for this work.
We have neglected instrumental noise and metal
contamination in simulated skewers, but this will not be relevant for
the conclusions of this paper.

\subsection{Model parameters}
\label{sec:parameters}

Lyman-$\alpha$ forest simulations include both cosmological parameters as well as astrophysical parameters needed to model the thermal state of the gas, which is significantly affected by hydrogen and helium reionizations.
Our main goal is to test and improve the parameter sampling scheme
and the emulation method used for constraining the parameters; in order to reduce the computational expense, in this work
we will focus our attention on the set of ``standard'' parameters, $\{ T_0, \gamma, \lambda_P, \bar{F} \}$, describing the thermal state of the IGM.
We keep the cosmological parameters fixed and based on the \cite{Planck2014} flat $\Lambda$CDM
model with $h=0.67$, $\Omega_m = 0.32$, $\Omega_b h^2 = 0.022312$,
$n_s = 0.96$, $\sigma_8 = 0.8288$.

The values for thermal parameters $T_0$ and $\gamma$
are obtained from the simulation by
approximating the temperature-density relation as the power law:
\begin{equation}
\label{eq:t-rho}
    T = T_0 \Delta^{\gamma-1} \, ,
\end{equation}
and finding the best fit using a linear least squares method in $\log T$ -- $\log \Delta$ \citep{Lukic2015}.
Therefore, $T_0$ parametrizes the temperature at mean density in the universe, while $\gamma$ is the slope
of temperature-density relation, expected to asymptote $\gamma \approx 1.6$ long after reionization ends.
To determine the pressure smoothing scale, $\lambda_P$, we fit
the cutoff in the power spectrum of the real-space Ly$\alpha$
flux, as described in \cite{Kulkarni2015}.
Real-space Ly$\alpha$ flux is calculated using actual density and
temperature at each cell in the simulation, but omitting all redshift-space
effects such as peculiar velocities and thermal broadening.

\begin{figure*}
\gridline{\fig{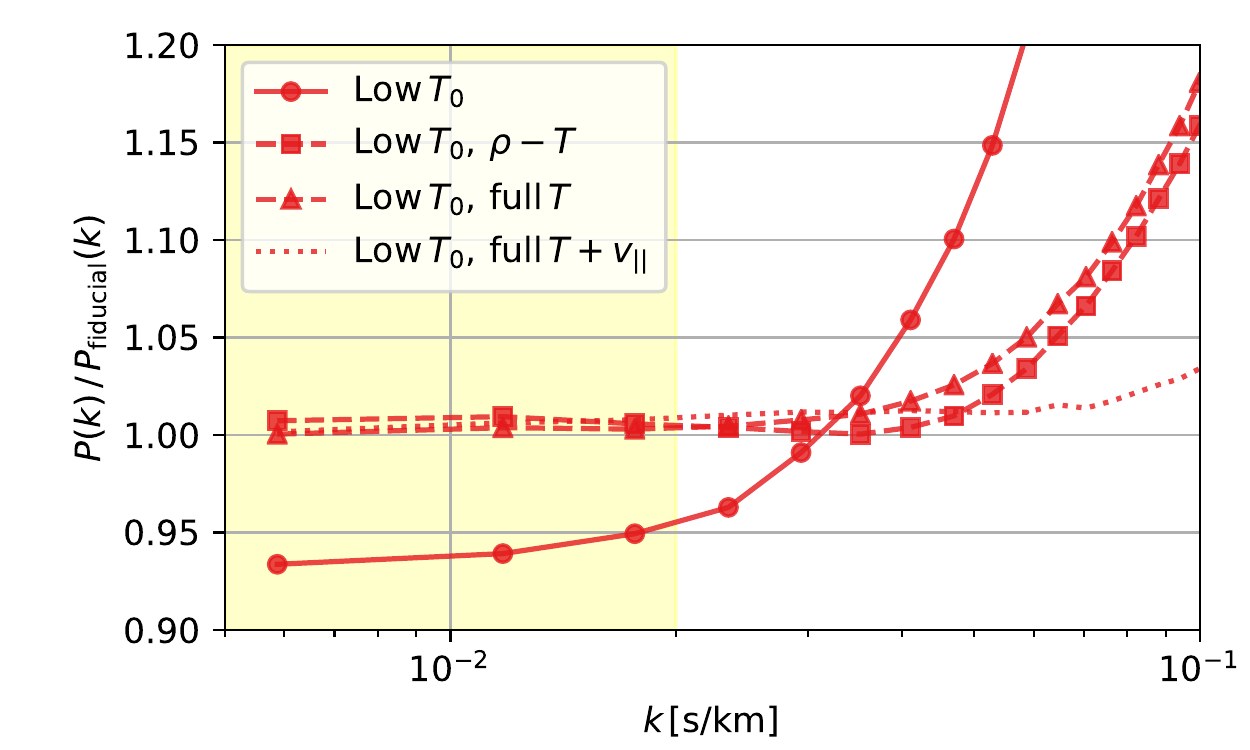}{0.49\textwidth}{}
              \fig{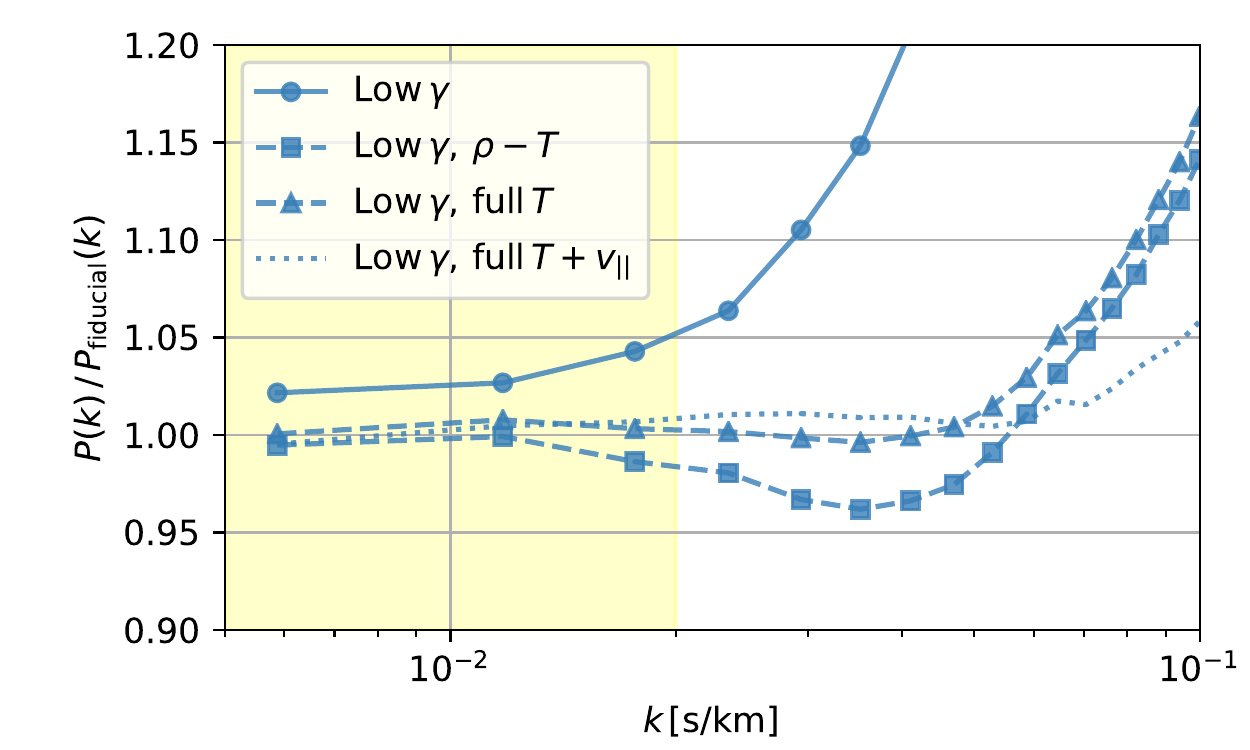}{0.49\textwidth}{}
             }
             \vspace{-0.5cm}
\caption{Power spectra ratios showing the accuracy of different post-processing approaches for rescaling the instantaneous
              temperature in simulations.  Solid lines show ratios of low-$T_0$ (left panel) and low-$\gamma$ (right panel) simulations with respect
              to the fiducial simulation.  Dashed lines lines show same ratios after rescaling simulations to match fiducial one's $T_0$--$\gamma$
              relation assuming all the gas is exactly on the power law (squares) and accounting for the scatter in $T_0$--$\gamma$ (triangles).
              Dotted lines additionally take line of sight velocity from the fiducial simulation demonstrating that most of the remaining rescaling
              error is coming from gas elements moving at different speeds.  Note that over the BOSS/eBOSS/DESI range of $k$ (yellow region),
              this rescaling shows good accuracy.
              }
\label{fig:pk-scaling}
\end{figure*}

In addition to these three parameters describing the thermal state of the gas,
we also need to parameterize the uncertainty in the observed mean flux.
In photo-ionization equilibrium, the mean flux of the Ly$\alpha$
forest is proportional to the fraction of neutral hydrogen, and is
thus degenerate with the amplitude of the assumed UV background.
Therefore, mean flux can be rescaled by finding the constant multiplier of
the optical depth of all spectral pixels in the simulation
so that the mean flux matches the desired value:
$\bar{F} = \langle \exp{(-A \tau_{i,j,k})} \rangle$.
For accuracy considerations of rescaling the mean flux, we refer the reader to \cite{Lukic2015}.

\subsection{Post-processing instantaneous temperature of the IGM}
\label{sec:gimlet-model}

Modification of the mean flux in postprocessing is physically motivated procedure, as explained in the previous section;
however, modifying any other thermal parameter commonly requires running a new simulation.
While modifying $\lambda_P$ (3D pressure smoothing) is inherently difficult due to its
dependence on the whole thermal history,
we can hope to be also able to modify the instantaneous temperature -- the one that determines the
recombination rate and the thermal broadening (1D smoothing).
That way, we can generate different values of \{$T_0$, $\gamma$, $\bar{F}$\}, without the need to re-run expensive simulations.
To test this, we use three simulations which have the same cosmological and numerical parameters, and 
yield the same pressure smoothing parameter $\lambda_p \approx 68 {\rm kpc}$ at redshift $z=3$.
Temperature-density diagrams for these simulations are shown in Figure \ref{fig:rho-T}.
The fiducial simulation has $T_0 = 1.1 \times 10^4$K and $\gamma = 1.57$; the ``low-$T_0$'' simulation differs
from fiducial only in that $T_0 = 7 \times 10^3 K$, while the ``low-$\gamma$'' simulation has $\gamma = 1.03$ and all other
parameters the same as the fiducial model.

In Figure \ref{fig:pk-scaling} we show power spectra ratios of low-$T_0$ and low-$\gamma$
simulations with respect to the fiducial model.  Mean flux is matched in all cases shown.
Solid lines (with circles) are ratios of unscaled simulations, and we can see they are
significantly different over the $k$ range covered by data ($k \lesssim 0.08 \, {\rm km}^{-1} s$).
Two dashed lines with square and triangle points are models where temperature-density 
relation has been rescaled to the fiducial one without and with accounting for the scatter
in the $T-\rho$.  We notice the significant improvement in power spectrum, and we can also
see that scatter in $T-\rho$, as expected from optically thin models, does not play a significant role,
although it help in a case of radical change in $\gamma$ parameter as seen in the right panel of Figure \ref{fig:pk-scaling}.
Finally, dashed line represent the case where we both rescale temperature-density relation,
and use line of sight velocity from the fiducial simulation.  This is not a practical solution, as we wouldn't know these velocities when
rescaling a simulation to a target $T-\rho$ relation, but we want to show that differences in velocity account for most of the remaining error in
this rescaling procedure.

While our approximate, post-processing model does not recover power spectrum at a percent accurate level over the whole range of available data, it is sufficiently accurate for experiments conducted in Section \ref{sec:GPs}.  The essential requirements there are to know the ``true'' answer for a given model, and to be able to evaluate the model a large number of times.
Note also that this rescaling procedure is loosing accuracy at high-$k$ end which is important for thermal constraints and interpreting $P(k)$ from high resolution spectra, but over the $k$ range relevant to BOSS/eBOSS/DESI observations ($k \lesssim 0.02 \, {\rm km}^{-1} s$), the achieved accuracy is $\approx$1\%, which should suffice for many studies.

\section{Adaptive construction of Gaussian process emulators for Bayesian inference} \label{sec:inference}

In this section we outline our main approach to the adaptive construction of the GP emulators. Our approach is designed to solve the specific problem at hand, which is inferring  the parameters of interest that serve as input into the forward model of the power spectrum from the observational data. The main ingredient of our approach is a so-called ``acquisition'' function that guides the selection of the training inputs for the emulator. This acquisition function arises from the form of the likelihood for the measurement data. Thus, before explaining the acquisition process we start by providing a general framework.

We denote the parameters of interest by $\btheta\in\real^p$. We denote the vector of observations by $\bd=(d_1, \dots, d_q)^T$, where each $d_i$ represents a measured value of the Ly$\alpha$ forest flux power spectrum at a certain value of the wavenumber $k$.
The outputs of the forward model of the power spectrum for a given $\btheta$ will be thought of as a $q$-dimensional vector $\BP(\btheta)$ (more on this in Section \ref{sec:GPs}).
We will work under the assumption of the Gaussian measurement noise with zero mean and known covariance $\BSigma_E$. With this specification of the measurement noise we formulate the likelihood function for the observational data which depends on the value of the forward model at $\btheta$:
\[
    L(\btheta | \bd) = \cN_q(\bd - \BP(\btheta) \,|\, \bzero_q, \BSigma_E).
\]
Assuming the Bayesian framework, we model the prior information about the parameters $\btheta$ as a known distribution $p(\btheta)$. Given the prior and the observed measurements $\bd$, the solution of the inverse problem is the posterior density obtained by applying Bayes' rule:
\[
    p(\btheta | \bd) \propto L(\btheta | \bd) p(\btheta).
\]
This posterior density can, in principle, be explored with MCMC methods. However, since evaluating the likelihood function $L(\btheta | \bd)$ requires evaluating the forward model $\BP(\btheta)$, the direct application of MCMC methods for the current application (or any expensive-to-evaluate forward model) is infeasible. This difficulty can be circumvented by using a surrogate model, such as a Gaussian process emulator, in place of the forward model. The Bayesian nature of the Gaussian processes makes them particularly suitable for our framework. Next, we provide a formal review of GP emulators leaving the details out until Section \ref{sec:GPs} and outline our adaptive approach to sequentially adding training inputs for the emulator.

Suppose that we have collected a set of evaluations of the forward model $\BP(\btheta)$ at $n$ input points:
\[
    \cD = \{ \btheta^{(j)}, \BP(\btheta^{(j)}) \}_{j=1}^n.
\]
The information in $\cD$ can be used to obtain a surrogate model specified by a random variable $\BP^{GP}$ with a predictive distribution conditioned on the input $\btheta$ and data $\cD$:
\[
    p\big(\BP^{GP} \,|\, \btheta, \cD\big) = \int p\big(\BP^{GP} \,|\, \btheta, \cD, \bpsi\big) p(\bpsi \,|\, \cD) d\bpsi.
\]
Here $\bpsi$ denotes the \textit{hyperparameters} of the predictive model, $p\big(\BP^{GP} \,|\, \btheta, \cD, \bpsi\big)$ is the predictive distribution of the assumed model given the hyperparameters, and $p(\bpsi | \cD)$ is the posterior distribution of $\bpsi$ given data $\cD$. For the details on obtaining $p(\bpsi | \cD)$, see Appendix \ref{sec:math}.

The solution of the inverse problem with a limited number of forward model evaluations can now be formulated using the likelihood of the observational data evaluated using the surrogate model. This \textit{$\cD$-restricted likelihood} (similarly to \cite{IBilionis_NZabaras_2014a}) is defined as follows:
\[
    L(\btheta \,|\, \bd, \cD) = \int L\big(\btheta \,|\, \bd, \BP^{GP}\big) p\big(\BP^{GP} \,|\, \btheta, \cD\big) d\BP^{GP},
\]
where $L\big(\btheta \,|\, \bd, \BP^{GP}\big) = \cN_q\big(\bd - \BP^{GP}(\btheta) \,|\, \bzero_q, \BSigma_E\big)$.
Once we have an approximation of the distribution $p(\bpsi | \cD)$, we can integrate the product of the two Gaussians and obtain an approximate formula for the likelihood $L(\btheta \,|\, \bd, \cD)$, see \cite{TTakhtaganov_JMueller_2018a} and Appendix \ref{sec:math} for details.

\begin{algorithm}[H]
\caption{Adaptive construction of GP emulators}
\begin{algorithmic}[1]\label{algo:adaptiveGP}
    \REQUIRE Initial design $\big\{\btheta^{(j)}_{}\big\}_{j=1}^{n_{}}$, threshold value $\epsilon_{thresh}$, search space $\cX_{\theta}$, maximum allowed number of forward model evaluations $s_{max}$.
    \ENSURE Adaptive design $\cD$.
    
    \STATE Evaluate $\BP(\btheta)$ for $\btheta\in\big\{\btheta^{(j)}_{}\big\}_{j=1}^{n_{}}$ to obtain $\cD = \big\{\btheta^{(j)}_{},\BP\big(\btheta^{(j)}_{}\big)\big\}_{j=1}^{n_{}}$.
    \FOR{$s$ from $1$ to $s_{max}$}
        \STATE train the GP model using current design;
        \STATE update the current best fit value $g_{min}$;
        \STATE maximize the expected improvement in fit function $\cI(\btheta)$ over the search space $\cX_\theta$, and let $\btheta^{s}=\argmax\limits_{\ensuremath{\boldsymbol{\theta}}\in\cX_\theta} \cI(\btheta)$;
        \IF{$\cI\big(\btheta^{s}\big) < \epsilon_{thresh}\cdot g_{min}$} 
            \STATE \textbf{break}
        \ENDIF
        \STATE evaluate $\BP$ at $\btheta^{s}$ and augment the training set: $\cD = \cD\cup\big\{\btheta^{s}$, $\BP\big(\btheta^{s}\big)\big\}$;
    \ENDFOR
    \RETURN $\cD$.
\end{algorithmic}
\end{algorithm}

Evaluations of the approximate likelihood $L(\btheta | \bd, \cD)$ involve computing the following misfit function between the observational data and the predictions of the GP emulator
\begin{align}\label{eq:misfit_GP}
\begin{split}
    g(\btheta; \cD, \bpsi) = &(\bd - \bmm(\btheta; \cD, \bpsi))^T(\BSigma_E + \BSigma_{GP}(\btheta; \cD, \bpsi))^{-1} \\
    &\times (\bd - \bmm(\btheta, \cD, \bpsi))
\end{split}
\end{align}
for the hyperparameter samples from distribution $p(\bpsi | \cD)$. In \eqref{eq:misfit_GP}, we denote by $\bmm(\btheta; \cD, \bpsi)$ the mean vector of the GP emulator evaluated at $\btheta$, and by $\BSigma_{GP}(\btheta; \cD, \bpsi)$  its predictive covariance. This misfit function captures the discrepancy between the observed values of the power spectrum and the predicted values in the norm weighted by the measurement error and the uncertainty of the emulator. Note that for the inputs $\btheta$ in the training set $\cD$, the mean of the GP emulator $\bmm(\btheta; \cD, \bpsi)$ coincides with the values of the power spectrum $\BP(\btheta)$, and the covariance $\BSigma_{GP}(\btheta; \cD, \bpsi)$ vanishes, and thus the exact value of the misfit (and hence the likelihood) is known. 

We use the misfit function \eqref{eq:misfit_GP} to inform our choice of the candidate inputs to add to the dataset $\cD$ in order to improve the GP surrogate. The ``improvement'' we are looking for is to make the approximate likelihood $L(\btheta | \bd, \cD)$ more accurately resemble the ``true'' likelihood $L(\btheta | \bd)$. The overall accuracy of the GP emulator over the support of the prior is unimportant. 

Our adaptive approach to extending the training set $\cD$ is based on an  acquisition function commonly used in  Bayesian optimization---the so-called ``expected improvement'' (EI) criterion \citep{DRJones_MSchonlau_WJWelch_1998a}. In our version of the EI criterion, we look for an input $\btheta$ that provides the largest \textit{expected improvement in fit} with expectation taken with respect to the posterior distribution of the hyperparameters $p(\bpsi | \cD)$:
\begin{equation}\label{eq:exp_improvement_in_fit}
    \cI(\btheta) \equiv \int [g_{min} - g(\btheta; \cD, \bpsi)]^+ p(\bpsi | \cD) d\bpsi.
\end{equation}
Here $g_{min}$ denotes the smallest misfit to the measurement data for the points in the current training set $\cD$, and $[\,\cdot\,]^+$ takes the positive part of the difference: $[\,\cdot\,]\equiv \max\{ \,\cdot\,, 0\}$.
This formulation allows us to balance the exploration and the exploitation of the GP emulator in an iterative search for a new training input to maximize the \textit{expected improvement in fit} function $\cI(\btheta)$, i.e.,  we search for the input $\btheta$ that provides the largest improvement in fit to the measurement data under the current GP model, conditioned on the misfit being smaller than the current best value for the points in the training set. 
The outline of the algorithm is given in Algorithm \ref{algo:adaptiveGP}. For more details see \cite{TTakhtaganov_JMueller_2018a}.

For our numerical experiments, we take the search space $\cX_\theta$ to be the support of the prior $p(\btheta)$, and we set the threshold value $\epsilon_{thresh}$ to be $1\%$. We solve the auxiliary optimization problem in Step 5 by using multi-start gradient-based optimization, see \cite{TTakhtaganov_JMueller_2018a} for details.
We set the allowed number of simulations $s_{max}$ to a large number so that the effective termination condition is the one on line 6. In practice, $s_{max}$ is dictated by the simulation budget. 
In the next section, we discuss the details of the construction of the GP emulators for modeling the Ly$\alpha$ forest power spectrum.
\section{Gaussian Process  Emulators for the Ly$\alpha$ flux power spectrum}\label{sec:GPs}

We model the power spectrum $P(k,\btheta)$ as a multi-output Gaussian process with outputs corresponding to the fixed values of the wavenumber $k$. Furthermore, we assume a separable structure of the kernel function, meaning that it can be formulated as a product of a kernel function for the input space $\btheta$ alone, and a kernel function that encodes the interactions between the outputs $k$ \cite[Section~4]{MAlvarez_LRosasco_NDLawrence_2012a}. In the following subsections we discuss the details of the construction of GP emulators and contrast our approach to those used in related works.

\subsection{Gaussian process emulators}

We will treat $P(k,\btheta)$ as a function from $\cX_k\times\cX_\theta$ to $\real^q$, where $\cX_k\subset\real$, $\cX_\theta\subset\real^p$, and $q$ is the number of values of the wavenumber $k$ for which we have observations. For a given vector of input parameters $\btheta$, we treat the output of the simulation code as a vector $\BP(\btheta)\in\real^q$ of the values of the power spectrum at fixed values of $k$. 

Similarly to \cite{SConti_AOHagan_2010a}, we model $\BP(\cdot)$ as a $q$-dimensional separable Gaussian process:
\[
   \BP^{GP}(\cdot) | \bpsi, \BSigma_k \sim \cN_q(\bmu(\cdot), c(\cdot,\cdot\,; \,\bpsi)\BSigma_k),
\]
conditional on hyperparameters $\bpsi$ of the correlation function $c:\cX_\theta\times\cX_\theta\times \cX_\psi\rightarrow\real$, and symmetric positive-definite matrix $\BSigma_k\in\real^{q\times q}$. This means that for any two inputs $\btheta^{(1)}$ and $\btheta^{(2)}$, we have $\E\big[\BP^{GP}(\btheta^{(i)}) \,|\, \bpsi, \BSigma_k\big] = \bmu(\btheta^{(i)})$, $i=1,2$, and $\Cov\big[\BP^{GP}(\btheta^{(1)}), \BP^{GP}(\btheta^{(2)}) \,|\, \bpsi, \BSigma_k\big] = c(\btheta^{(1)}, \btheta^{(2)};\bpsi)\BSigma_k$. As indicated by several studies, e.g., \cite{HChen_JLLoeppky_JSacks_WJWelch_2016a}, the introduction of the regression term $\bmu(\cdot)$ does not generally affect the performance of the predictive model, and, in some cases, might have an adverse effect. In our case, adequate results were obtained by simply setting $\bmu(\btheta)\equiv 0$. Furthermore, we set the covariance function $c(\cdot, \cdot\,;\,\bpsi)$ to be squared-exponential with $p+1$ hyperparameters $\bpsi=(\sigma_c, \ell_1, \dots, \ell_p)^T$:
\[
    c\big(\btheta^{(1)}, \btheta^{(2)}; \bpsi\big) = \sigma_c^2 \exp\bigg(-\sum_{i=1}^p \frac{\big(\theta_i^{(1)} - \theta_i^{(2)}\big)^2}{2\ell_i^2}\bigg).
\]
Note that the choice of the covariance function here is purely empirical and does not affect the forthcoming methodology.  

Our treatment of the inter-output covariance matrix $\BSigma_k$ differs from that in \cite{SConti_AOHagan_2010a} and \cite{IBilionis_NZabars_BAKonomi_GLin_2013a}. There, the authors assume a weak non-informative prior on the matrix $\BSigma_k$ and integrate it out of the predictive posterior distribution. Instead, we study four different approaches for treating interactions between outputs,  summarized   in the following Section \ref{sec:multi_GPs}.

\subsection{Approaches for dealing with multi-output models}\label{sec:multi_GPs}

\begin{enumerate}
    \item First, we test a naive approach that emulates each output separately with a single-output GP. We refer to this approach as (MS) as it corresponds to the MS (many single-output) emulator in \cite{SConti_AOHagan_2010a}. This approach has an increased computational cost (which could be alleviated by training in parallel) compared to training only one GP as the following two approaches do. Also, this approach ignores any dependencies between the outputs. 
    \item In our second approach (IND), we treat the outputs as independent given the hyperparameters of the covariance function $c(\cdot, \cdot \,;\,\bpsi)$. This approach has been considered in \cite{IBilionis_NZabaras_2014a} and \cite{TTakhtaganov_JMueller_2018a}. It leads to a simple and efficient implementation of a multi-output emulator with a diagonal $\BSigma_k$, see Appendix \ref{sec:math} for details. 
    \item Our third approach (COR)  assumes that correlations between different outputs are non-zero but are still independent of the parameter $\btheta$. In this case, we fix the correlation matrix a priori and use it to obtain $\BSigma_k$ by rescaling by the training variances. This approach requires specifying the inter-output correlation matrix. In terms of computational efficiency, this approach still allows us to use the Kronecker product structure of the training covariances (see Appendix \ref{sec:math}) and is as efficient as using the diagonal covariance in approach IND.
    \item Our final approach (INP) is related to Approach COR, but it is  computationally more  demanding. Here, we treat $k$ as another input dimension into the GP model with associated covariance kernel being again a squared-exponential. The main computational cost is associated with inverting the training covariance matrices, which become $q$ times larger due to the addition of another input dimension. Conceptually, however, this approach is similar to Approach COR with $\BSigma_k$ having entries specified by the squared-exponential kernel for a fixed (learned) value of the length-scale hyperparameter $\ell_{p+1}$ associated with the $k$ input dimension. The difference to Approach COR is that the inter-output correlation matrix now depends on the training data. As our experiments demonstrate, this additional flexibility provides no discernible advantage. This approach is referred to as TI (time-input) emulator in \cite{SConti_AOHagan_2010a} where an extra dimension is time rather than $k$.
\end{enumerate}

All of our approaches utilize matrix-valued kernels that fall into the category of \textit{separable} kernels, see \cite[Section~4]{MAlvarez_LRosasco_NDLawrence_2012a} for an overview. Specifically, our Approaches IND, COR, and INP are examples of the so-called \textit{intrinsic coregionalization model} or ICM \cite[Section~4.2]{MAlvarez_LRosasco_NDLawrence_2012a}. The ICM approach allows for an efficient implementation of GP-based regression and inference that exploits the properties of the Kronecker product of the covariance matrix, see Appendix \ref{sec:math} for details. For the adaptive algorithm, efficiency is important for solving the auxiliary optimization problem for the expected improvement in fit function $\cI(\btheta)$. Solving this optimization problem requires multiple restarts as the $\cI(\btheta)$ function is highly multi-modal, leading to a large number of evaluations of the GP emulator predictions and their gradients. For the GP-based inference, having an efficient emulator allows us to carry out MCMC sampling of the posterior with minimum computational effort. As the following numerical study of the considered approaches suggests, in our application it is reasonable to expect that the outputs corresponding to different values of the wavenumber $k$ have similar properties with respect to the parameters $\btheta$, therefore, our choice of the separable form of the kernel is justified.

\subsection{Numerical study of multi-output approaches}\label{sec:multi_GP_numerics}
In this section we compare the predictive performance of the different multi-output Gaussian process emulators introduced in Section \ref{sec:multi_GPs} using an approximate model of the power spectrum $P(k,\btheta)$ described in Section \ref{sec:gimlet-model}. To obtain a better picture of the dependence of the results on the choice of design inputs, we build emulators using $10$ to $30$ inputs arranged in a Latin Hypercube Design (LHD) where the minimum distance between the points has been maximized---the so-called maximin LHD (see, e.g., \cite{MEJohnson_MLeslie_DYlvisaker_1990a}). For each design we perform multiple experiments. For each experiment we generate a large test set consisting of $500$ input-output pairs for computing various measures of predictive accuracy. In order to avoid an unnecessary cost associated even with the post-processing procedure of Section \ref{sec:gimlet-model}, we pre-compute the power spectrum on a dense grid in $\cX_\theta$ using our automatized Henson system \citep{Morozov2016, Lohrmann2017} and interpolate the outputs using tri-linear interpolation to obtain the continuous approximation of the power spectrum in $\cX_\theta$.  We have confirmed that the $P(k,\btheta)$ error associated with this interpolation is negligible.
Specifically, we take a grid of $10^3$ input parameters $\btheta=(F, T_0, \gamma)$ covering the box 
\[
\cX_\theta = [0.2, 0.5]\times[3\times 10^3K, 3\times 10^4K]\times[1.0, 2.0] \ .
\]
We restrict our attention to the redshift of $z=4.2$ and $q=8$ outputs corresponding to the following $k$ values $\{3.26\times 10^{-3}, 6.51\times 10^{-3}, 9.77\times 10^{-3}, 1.63\times 10^{-2}, 2.28\times 10^{-2}, 3.26\times 10^{-2}, 5.21\times 10^{-2}, 8.14\times 10^{-2}\} {\rm km^{-1} s}$, which cover the range of values as the \cite{MViel_et_al_2013a} measurements that we use later. In the following we simply number these $k$ outputs from 1 to 8.

\begin{figure}[t]
    \centering
    \begin{minipage}[t]{0.45\textwidth}
        \centering
        \includegraphics[width=\textwidth, trim={0 0 4em 0},clip]{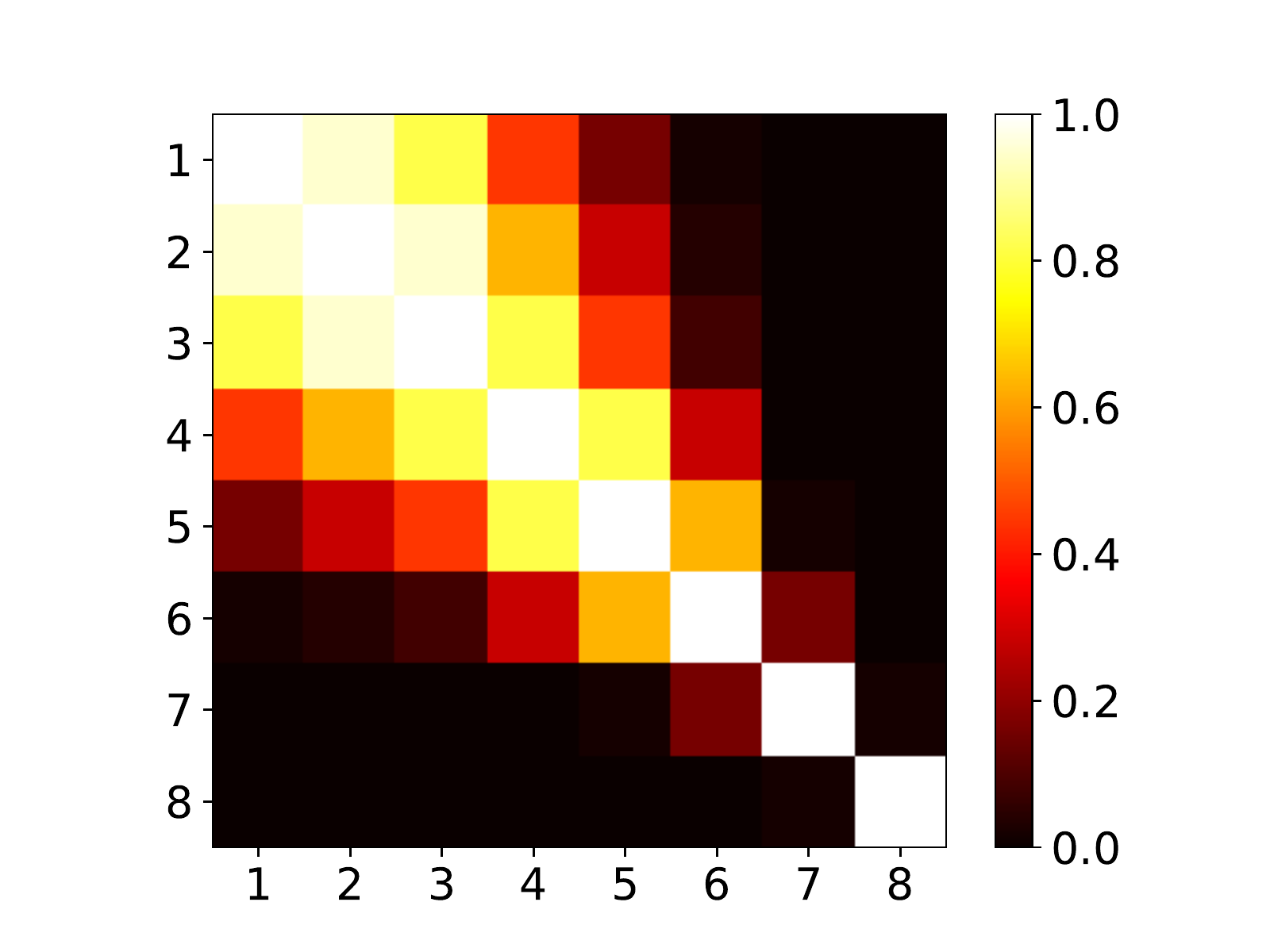}
	\vspace{-1.\baselineskip}
 \caption{\small{Correlation matrix between outputs ($k$-wavemodes) 1 through 8 for the COR emulator.}}
 \end{minipage}%
    \label{fig:cor_matrix}
\end{figure}

We construct the four multi-output GP emulators (MS, IND, COR, and INP) using 
fixed  LHDs with 10, 20, and 30 points in $\cX_\theta$.
In each case, the training output values are normalized (see Appendix \ref{sec:math}).
We fit the hyperparameters of the covariance function using the maximum likelihood approach (MLE in Appendix \ref{sec:math}). Recall that the COR emulator requires a fixed output correlation matrix $\BSigma_k$. We obtain an estimate of this matrix by first using the INP emulator built with LHD 20 design. Upon training of the INP emulator we obtain an estimate of the length scale for its $k$ (output) variable. By plugging-in this estimate into the one-dimensional squared-exponential kernel and evaluating the kernel for the values of $k$ that we consider, we get a desired estimate of $\BSigma_k$. This estimate remains fixed for all experiments with the COR emulator, see Figure \ref{fig:cor_matrix}.

\begin{figure*}[t]
    \centering
    \begin{minipage}[t]{0.36\textwidth}
        \centering
        \includegraphics[width=\textwidth, trim={0 0 4em 0},clip]{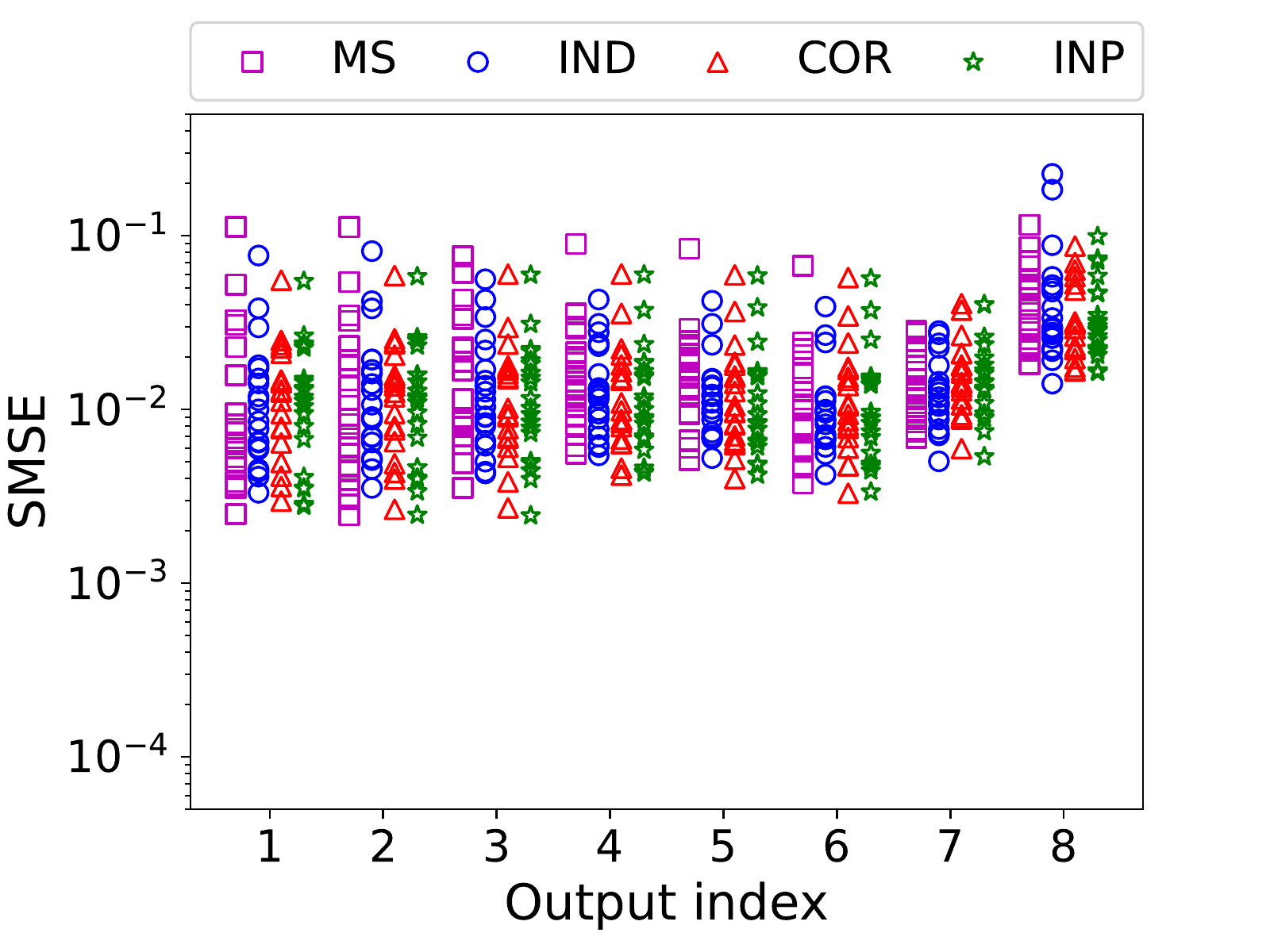}
	\vspace{-0.5\baselineskip}
	{\small{LHD 10}}
    \end{minipage}%
    \begin{minipage}[t]{0.304\textwidth}
        \centering
        \includegraphics[width=\textwidth, trim={7em 0 4em 0},clip]{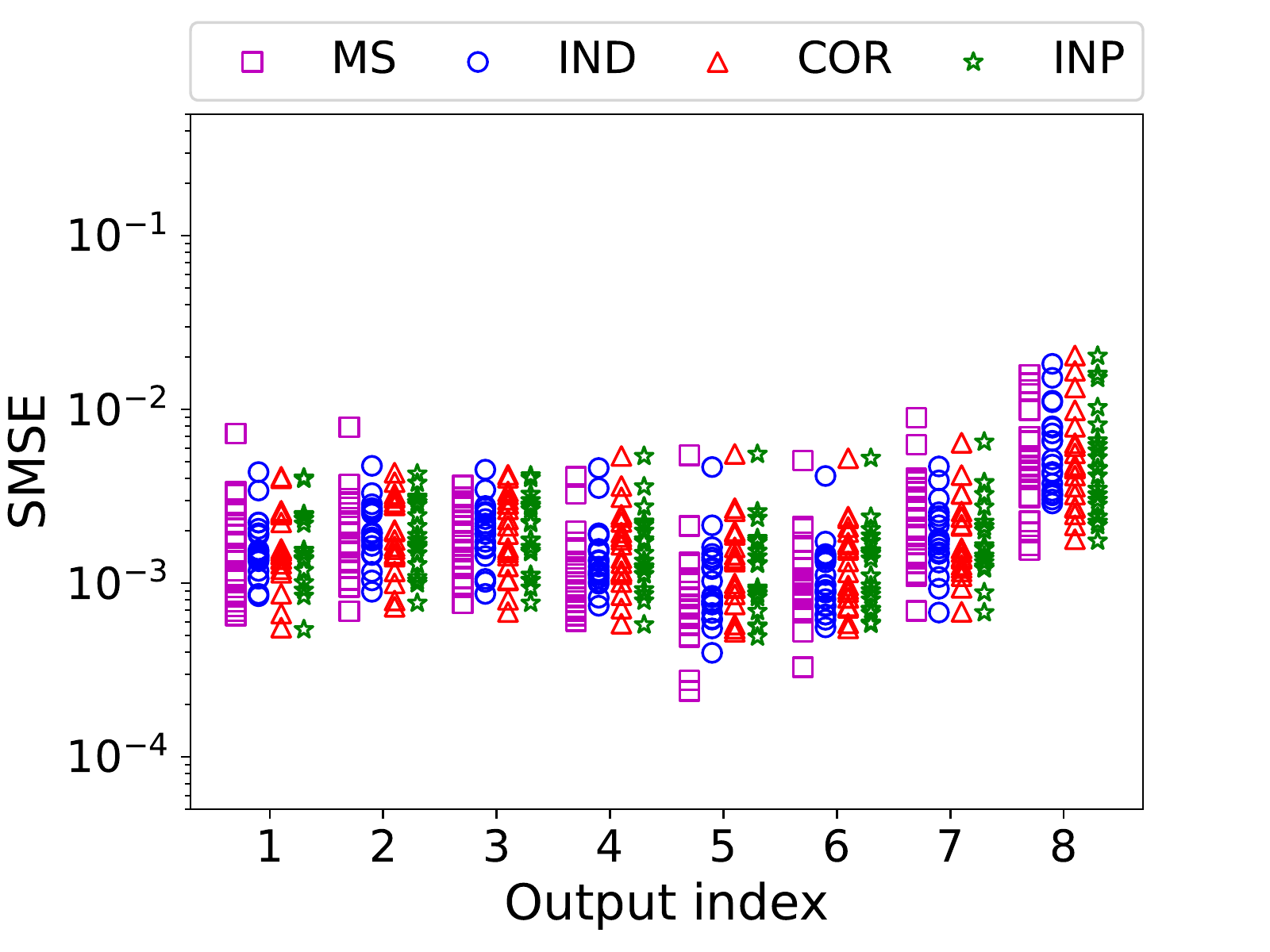}
	\vspace{-1.\baselineskip}
	{\small{LHD 20}}
        \end{minipage}%
    \begin{minipage}[t]{0.304\textwidth}
        \centering
        \includegraphics[width=\textwidth, trim={7em 0 4em 0},clip]{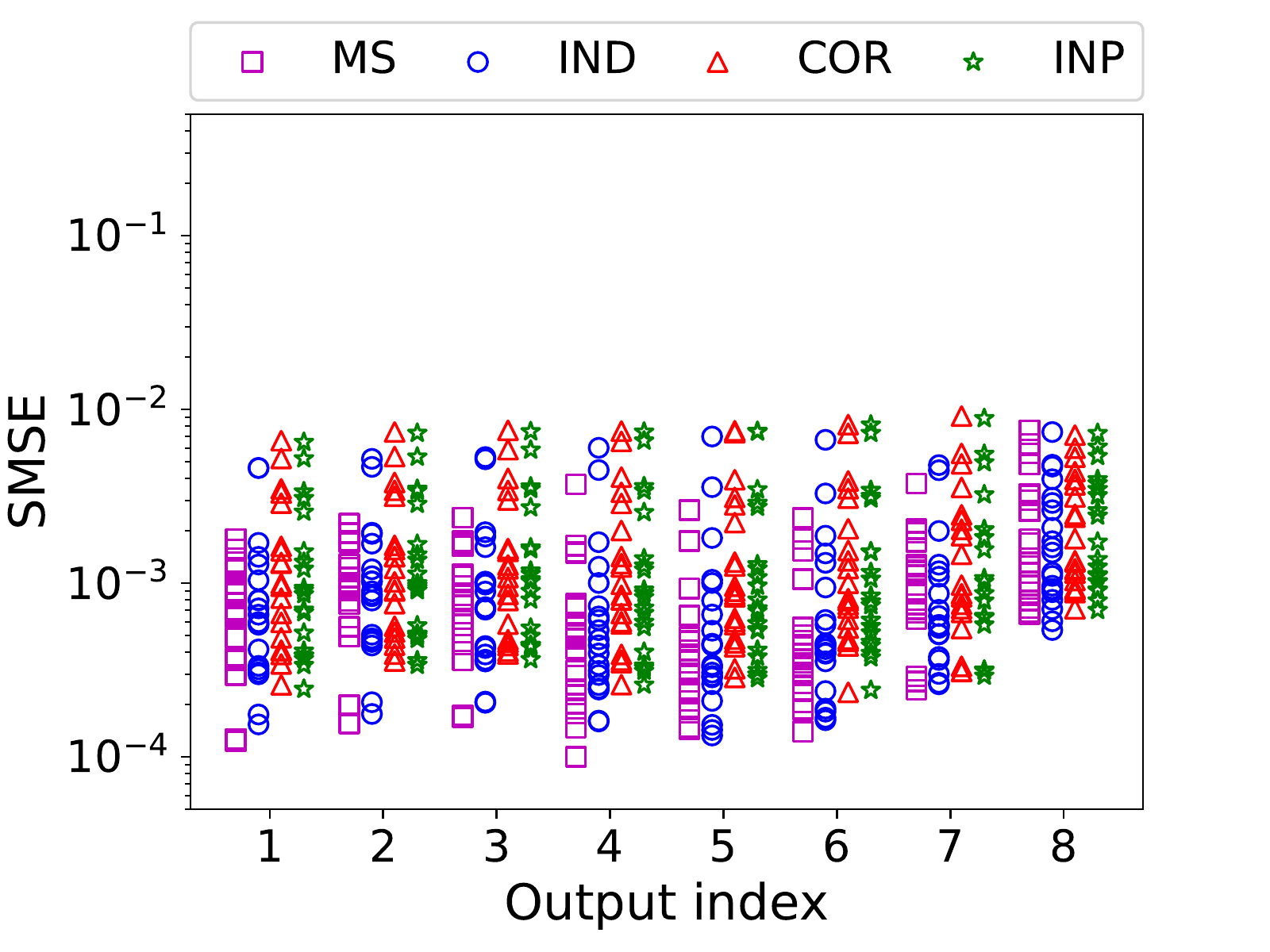}
	\vspace{-1.\baselineskip}
	 {\small{LHD 30}}
    \end{minipage}%
\vspace{1.\baselineskip}
\caption{Standardized mean-squared errors (SMSE) for the four multi-output emulators (MS, IND, COR, INP) for three different LHDs (LHD 10, 20, and 30). Test errors are computed with 500 points. Each experiment is repeated 20 times.}
\label{fig:smse_mult_out}
\end{figure*}

To test the predictive performance of the emulators we use a test set consisting of $N=500$ points in a randomized LHD. We use  the following measures of predictive accuracy.
\begin{enumerate}
    \item Standardized mean squared error (SMSE): this is the mean-squared prediction error scaled by the variance of the test data for each output $j=1,\dots,q$:
    \[
        \text{SMSE}_j =  \frac{\sum\limits_{i=1}^N \big(m_j\big(\bth^{(i)}, \cD \big) - P_j\big(\btheta^{(i)}\big)\big)^2}{\sum\limits_{i=1}^N \big(\overline{P}_j - P_j\big(\btheta^{(i)}\big)\big)^2},
    \]
    where $m_j(\btheta, \cD)$ is the $j$-th component of the vector of predictive means $\bmm(\btheta, \cD)$ of the GP model, and $\overline{P}_j$ is the mean of the test values $P_j(\btheta^{(i)})$, $i=1,\dots,N$.
    
    \item Credible interval percentage (CIP), also known as coverage probability: the percentage of the $100\alpha\%$ credible intervals that contain the true test value. For an emulator that provides adequate estimates of the uncertainty about its predictions, this value should be close to $\alpha$. We plot the CIP against $\alpha\in[0,1]$ and look for deviations from the straight line. This statistic can only be plotted for each output separately.
    
    \item Squared Mahalanobis distance (SMD) between the predicted and the test outputs at a test point $i$: 
    \begin{align*}
        \text{SMD}_i = &\big(\BP(\btheta^{(i)}) - \bmm(\btheta^{(i)}, \cD)\big)\BSigma_{GP}(\btheta^{(i)}, \cD)^{-1} \\
        &\times\big(\BP(\btheta^{(i)}) - \bmm(\btheta^{(i)}, \cD)\big),
    \end{align*}
    where $\bmm(\btheta, \cD)$ and $\BSigma_{GP}(\btheta, \cD)$ are the predictive mean and the predictive covariance of the multi-output GP emulator, respectively. According to the multivariate normal theory, this distance  should be distributed as $\chi^2_{q}$ for all  test points.
    A discrepancy between the distribution of distances for the emulator and a reference distribution indicates a misspecification of the covariance structure between the outputs.
\end{enumerate}

\begin{figure*}[t]
    \centering
    \begin{minipage}[t]{0.36\textwidth}
        \centering
        \includegraphics[width=\textwidth, trim={0 5em 4em 0},clip]{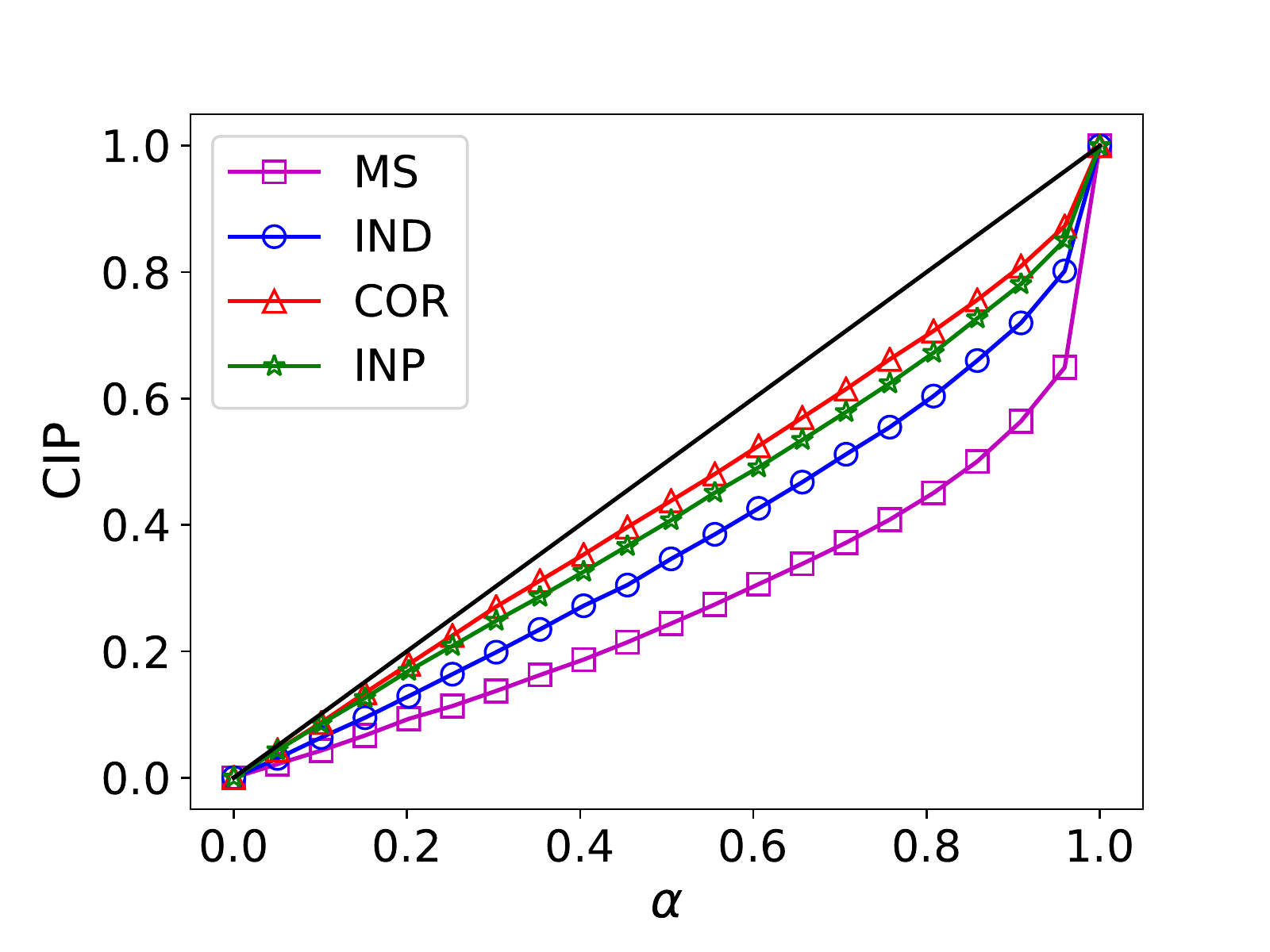}
	\vspace{-1.\baselineskip}
    \end{minipage}%
    \begin{minipage}[t]{0.304\textwidth}
        \centering
        \includegraphics[width=\textwidth, trim={7em 5em 4em 0},clip]{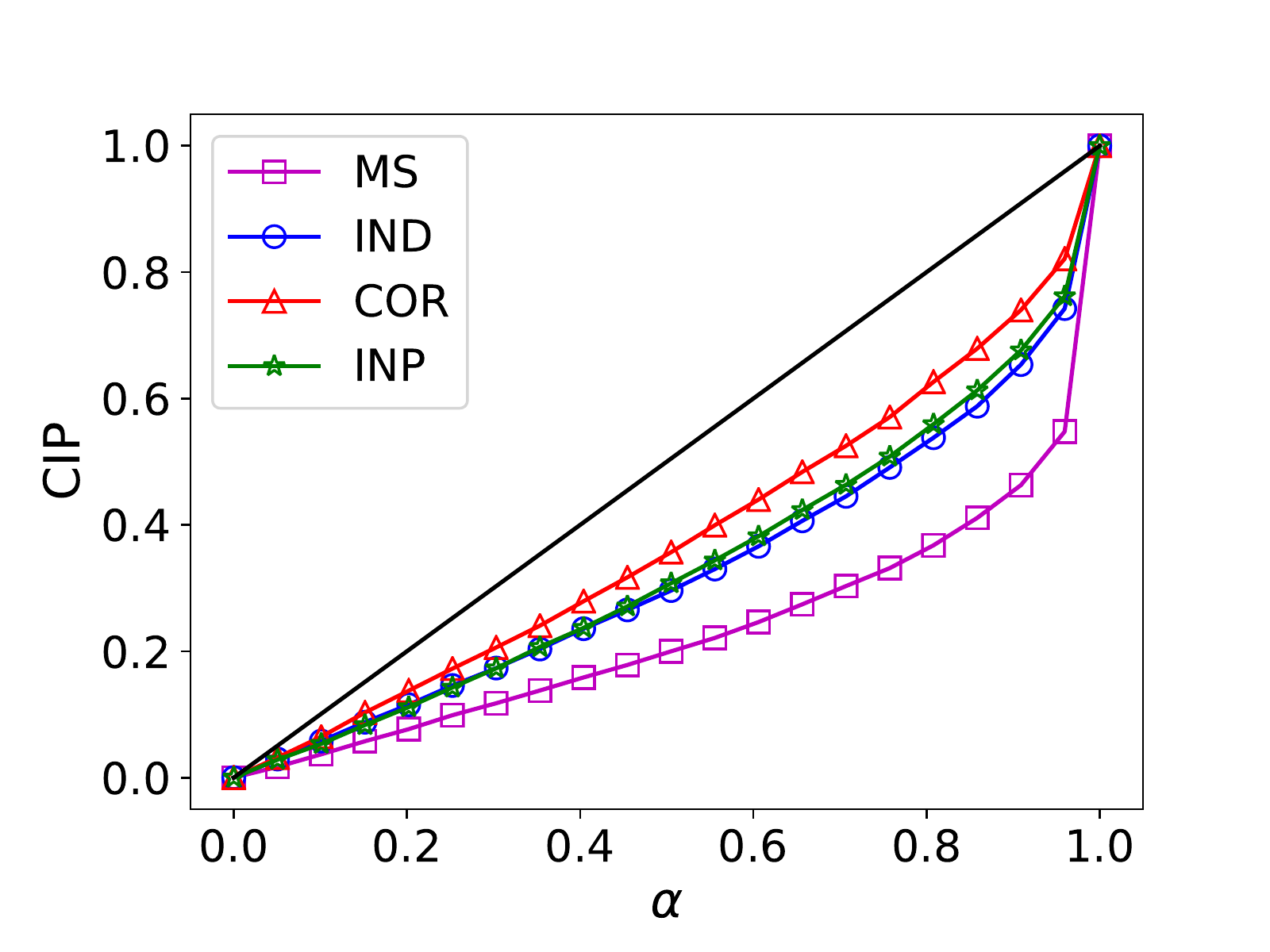}
	\vspace{-1.\baselineskip}
    \end{minipage}%
    \begin{minipage}[t]{0.304\textwidth}
        \centering
        \includegraphics[width=\textwidth, trim={7em 5em 4em 0},clip]{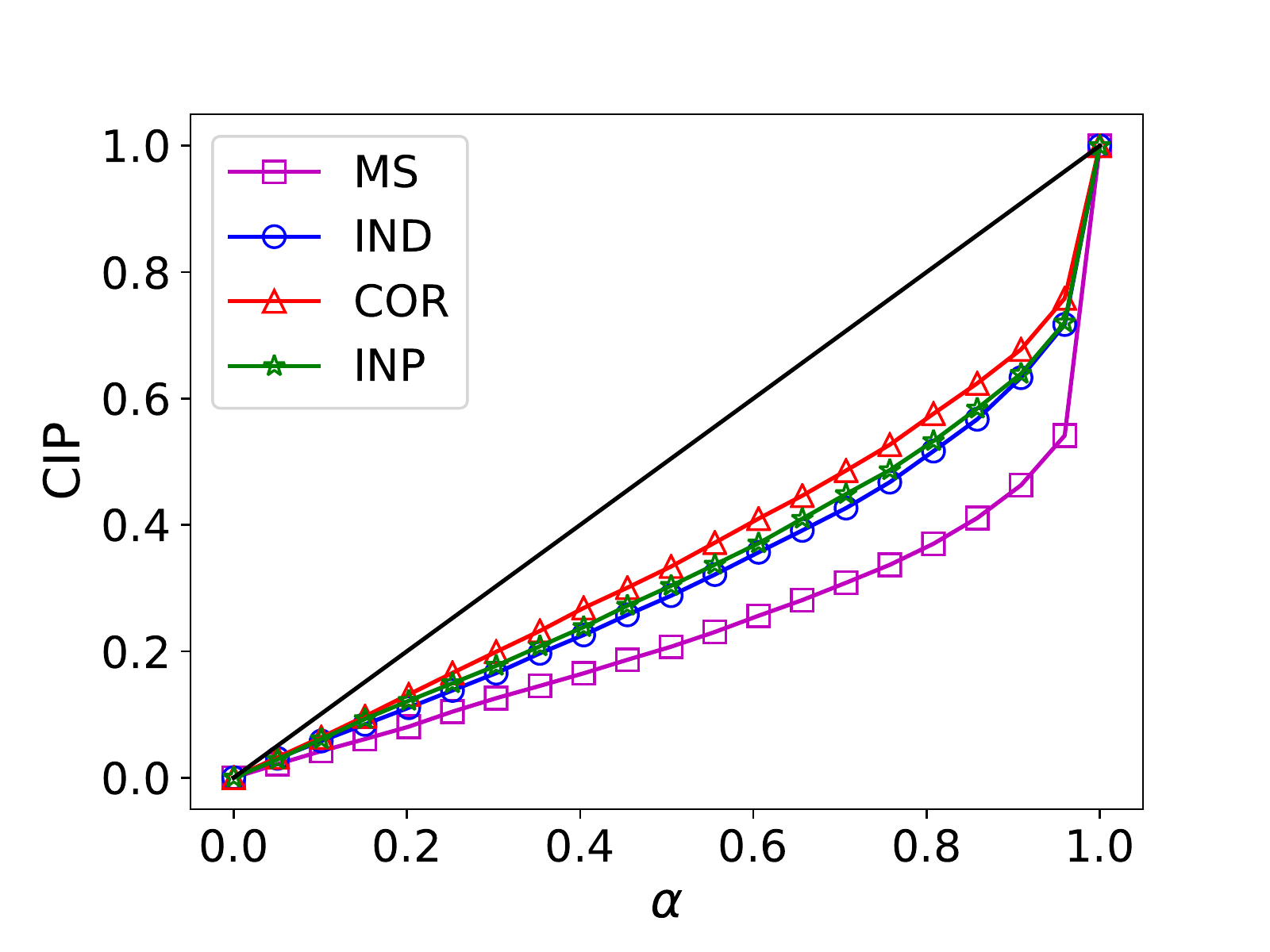}
	\vspace{-1.\baselineskip}
    \end{minipage}%
    \\ \vspace{-0.1em}
    \begin{minipage}[t]{0.36\textwidth}
        \centering
        \includegraphics[width=\textwidth, trim={0 5em 4em 3em},clip]{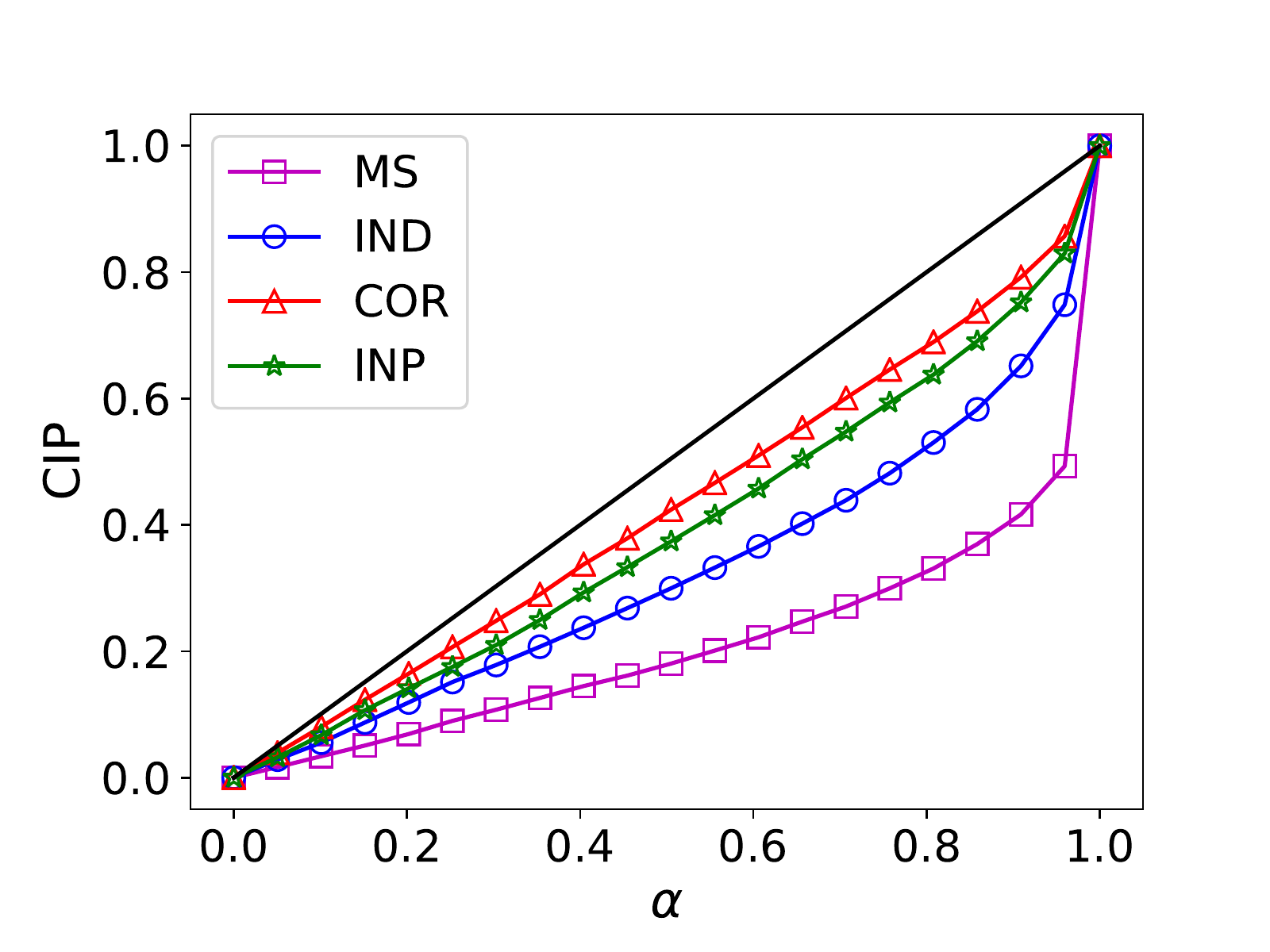}
	\vspace{-1.\baselineskip}
    \end{minipage}%
    \begin{minipage}[t]{0.304\textwidth}
        \centering
        \includegraphics[width=\textwidth, trim={7em 5em 4em 3em},clip]{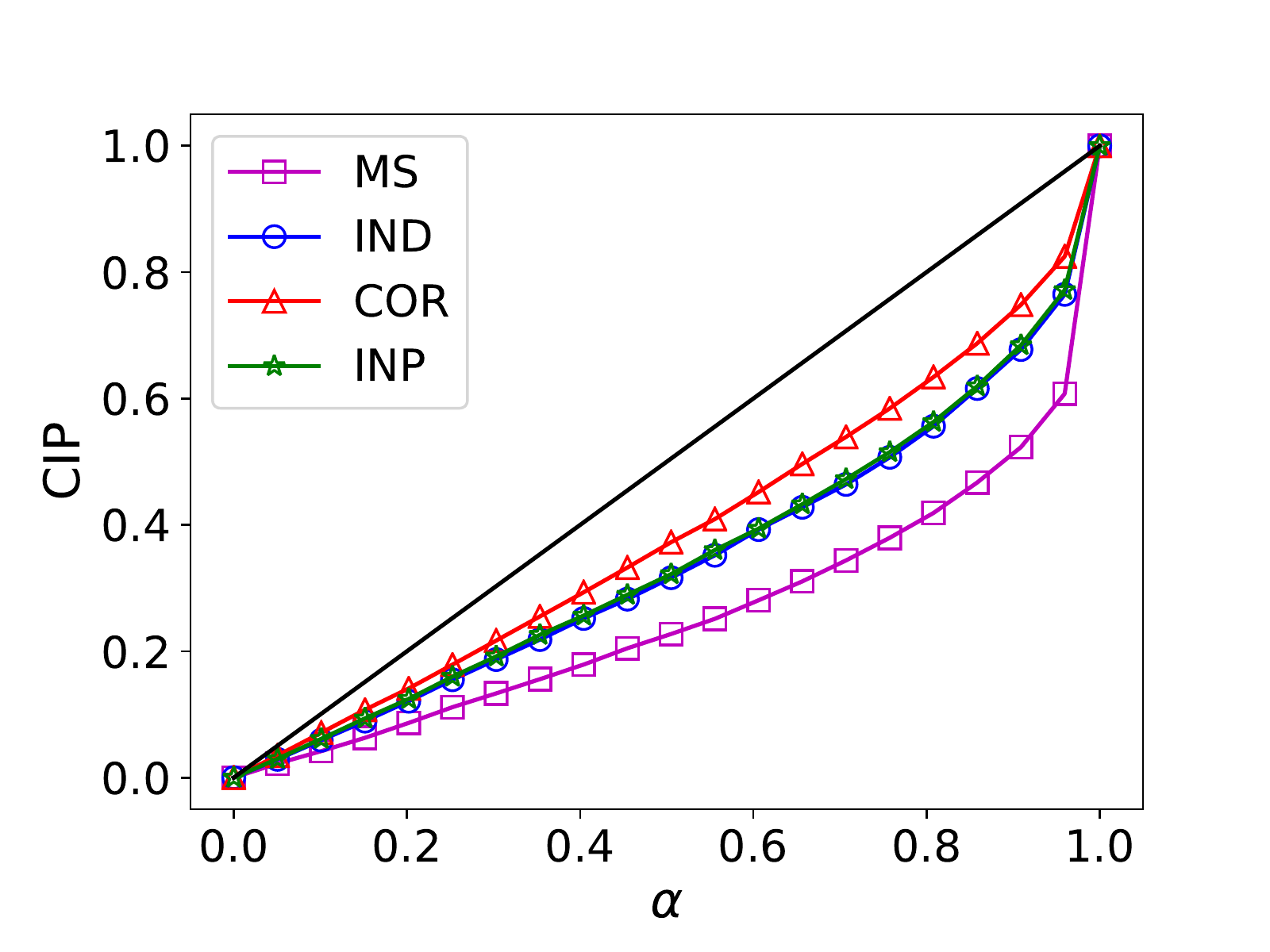}
	\vspace{-1.\baselineskip}
    \end{minipage}%
    \begin{minipage}[t]{0.304\textwidth}
        \centering
        \includegraphics[width=\textwidth, trim={7em 5em 4em 3em},clip]{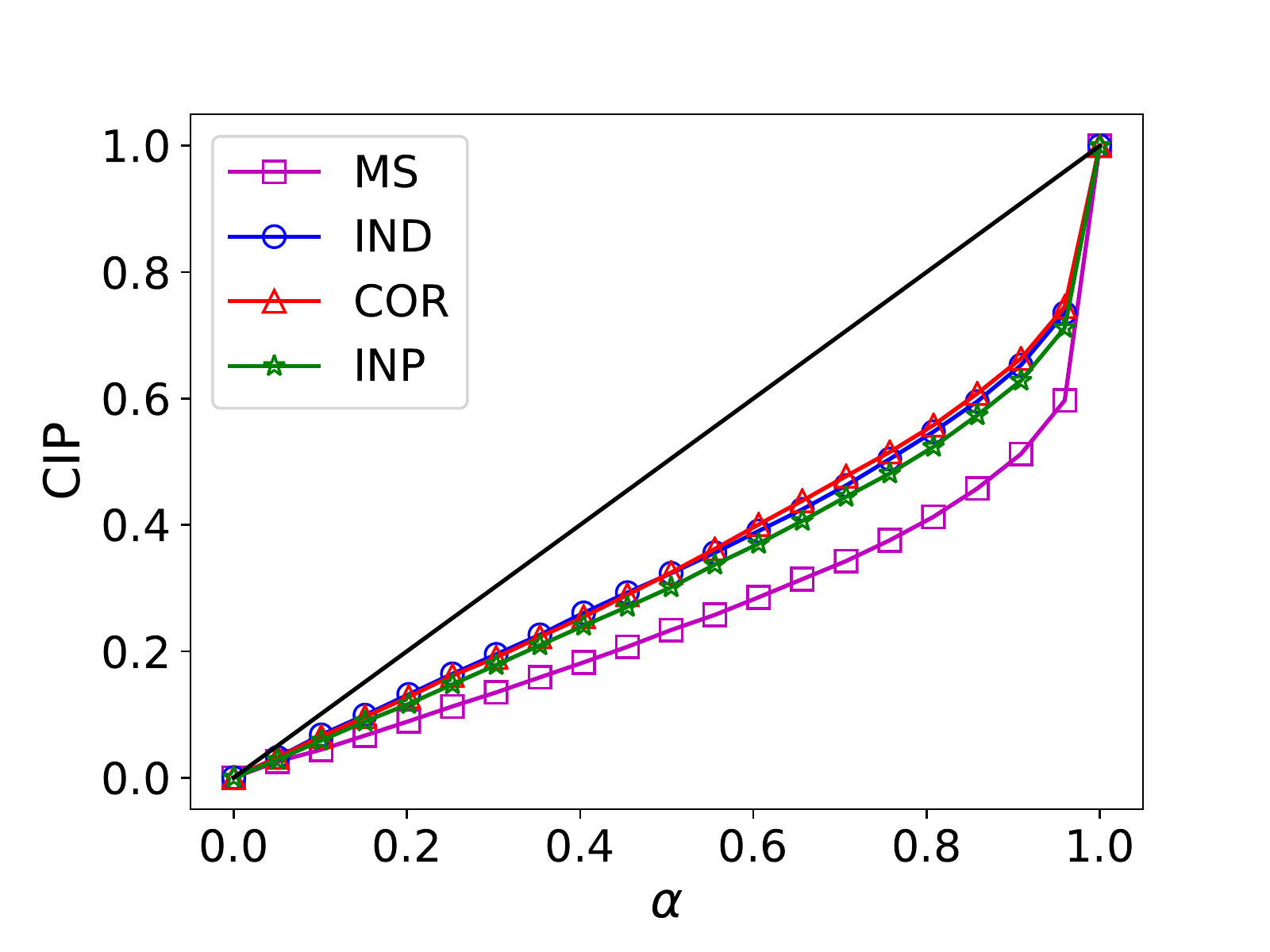}
	\vspace{-1.\baselineskip}
    \end{minipage}%
    \\ \vspace{-0.1em}
    \begin{minipage}[t]{0.36\textwidth}
        \centering
        \includegraphics[width=\textwidth, trim={0 0 4em 3em},clip]{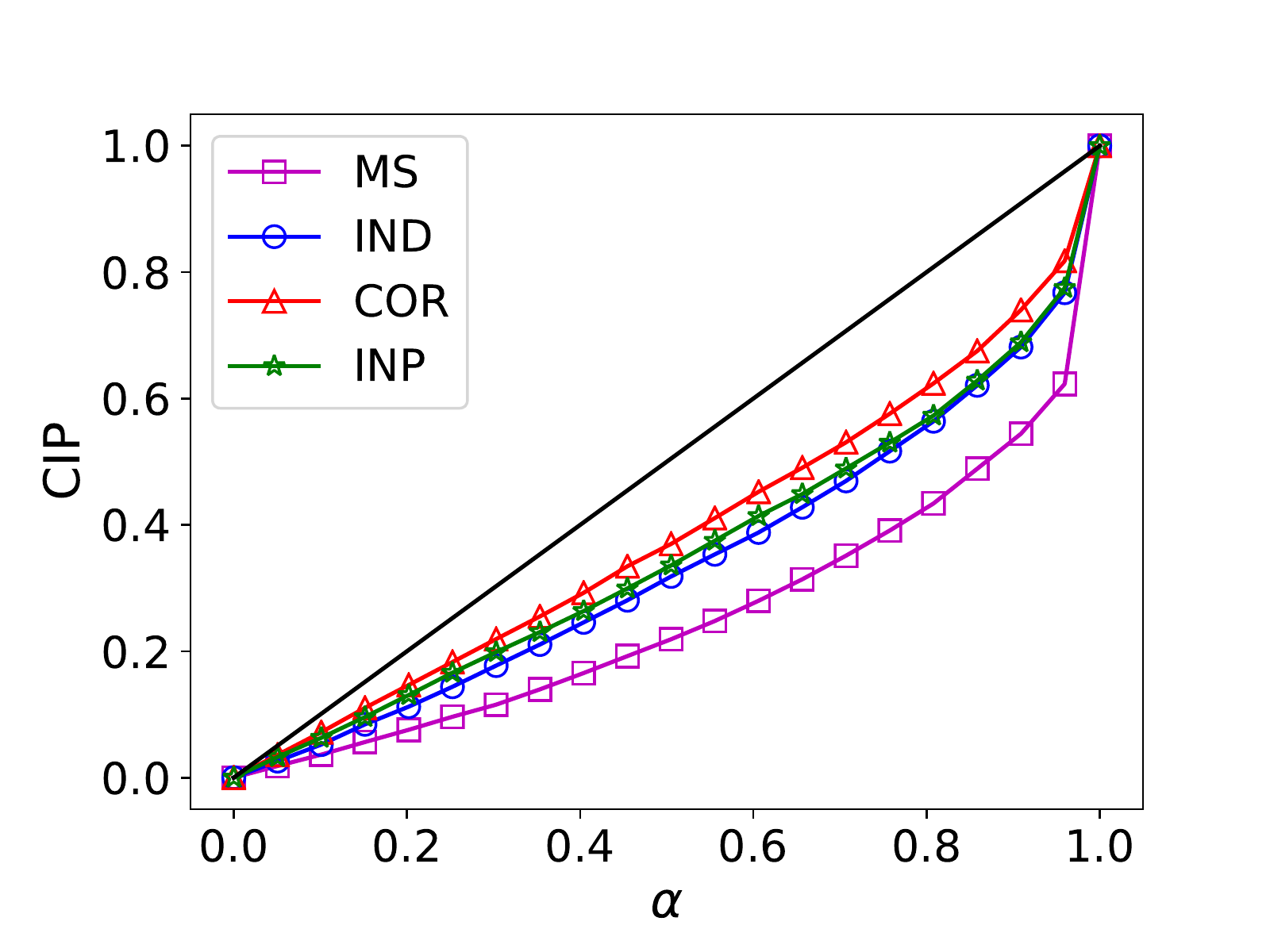}
	\vspace{-1.\baselineskip}
        {\small{LHD 10}}
    \end{minipage}%
    \begin{minipage}[t]{0.304\textwidth}
        \centering
        \includegraphics[width=\textwidth, trim={7em 0 4em 3em},clip]{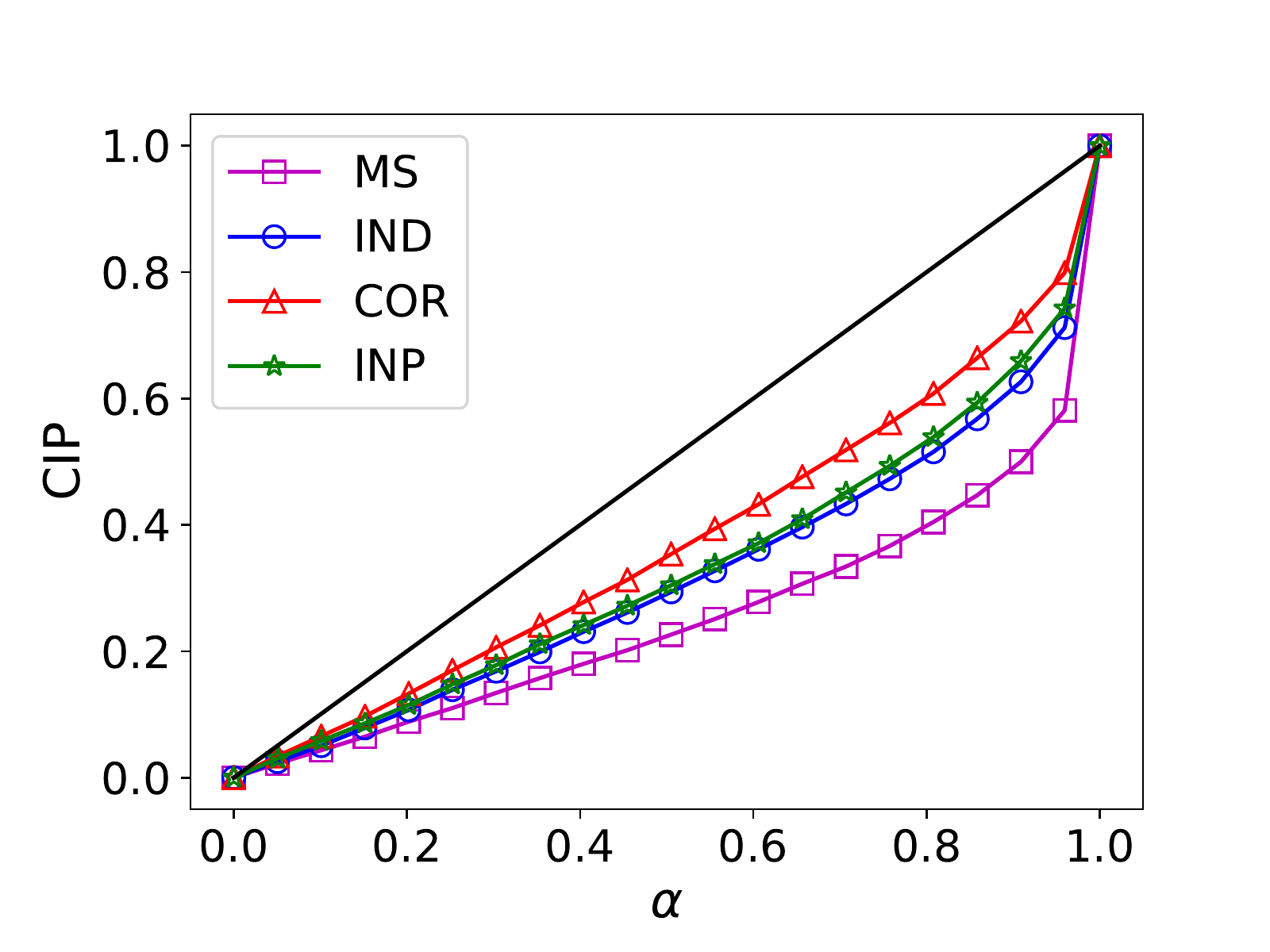}
	\vspace{-1.\baselineskip}
        {\small{LHD 20}}
    \end{minipage}%
    \begin{minipage}[t]{0.304\textwidth}
        \centering
        \includegraphics[width=\textwidth, trim={7em 0 4em 3em},clip]{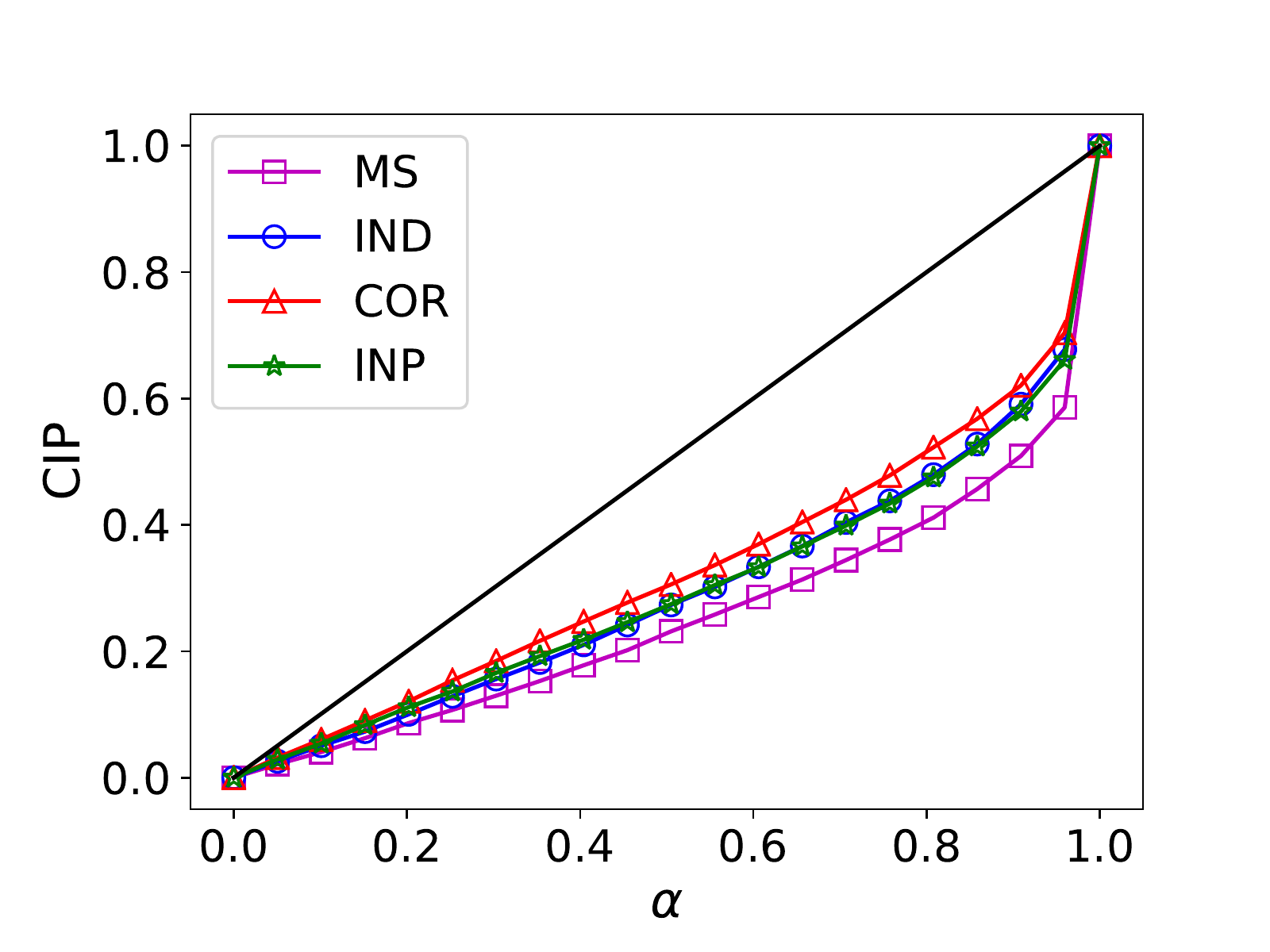}
	\vspace{-1.\baselineskip}
        {\small{LHD 30}}
    \end{minipage}%
\vspace{1.\baselineskip}
\caption{Percentage of credible intervals (CIP) containing the true test values for the four multi-output emulators (MS, IND, COR, INP) for three different LHDs based on 500 test points and averaged over 20 experiments. Top row - output 1, middle row - output 4, bottom row - output 7. The closer the colored graphs are to the black line,  the better.}
\label{fig:cip_mult_out}
\end{figure*}

Figure \ref{fig:smse_mult_out} reports the SMSE estimates for the three LHDs. In each case we repeat the experiment $20$ times, meaning that we generate $20$ training and $20$ test sets for each of the three designs and use them to train and test each of the four emulators. We observe a noticeable spread of the estimates between the experiments with the same number of design points regardless of the chosen emulator or output index. Increasing the size of the training design generally improves accuracy but does not necessarily combat the variation in results for different experiments. All four emulators demonstrate similar output-marginal accuracy, and the difference in errors for different outputs is low, with the exception of output 8, which has consistently higher relative errors for all four emulators. The COR and the INP emulators appear to have a similar spread of the error values indicating that fixing the output covariance $\BSigma_k$ a priori rather than allowing it to depend on training data has little effect on accuracy. The MS and the IND emulators achieve smaller errors, but also have relatively larger spreads  between experiments, which  suggests higher dependence on the design than in the cases of COR and INP emulators.

While there is little that separates different multi-output approaches in terms of per-output SMSE, the results for CIP and SMD provide a more complete picture.

Figure \ref{fig:cip_mult_out} reports the CIP estimates for the three LHDs and for selected output indices (1, 4, and 7, in the top, middle, and bottom rows, respectively). Here we report the  CIP values averaged over the $20$ experiments. In general, we observe that all four emulators underestimate the uncertainty in their predictions, i.e., they are over-confident in their predictions. This trend becomes more pronounced with growing training design size. We view this as an artifact of the MLE approach to GP training, see discussion in \cite[Section~3]{TTakhtaganov_JMueller_2018a}. Among the emulators, the MS emulator exhibits the worst performance  and is most over-confident in its predictions in all cases. The other emulators compensate for the shortcomings of the MLE approach by accounting for correlations between the outputs. The IND emulator does it in a least effective way, and hence performs worse than the COR and the INP emulators. The COR emulator on average comes closest to providing accurate uncertainty estimates.

\begin{figure*}[t]
    \centering
    \begin{minipage}[t]{0.36\textwidth}
        \centering
        \includegraphics[width=\textwidth, trim={0 0 4em 0},clip]{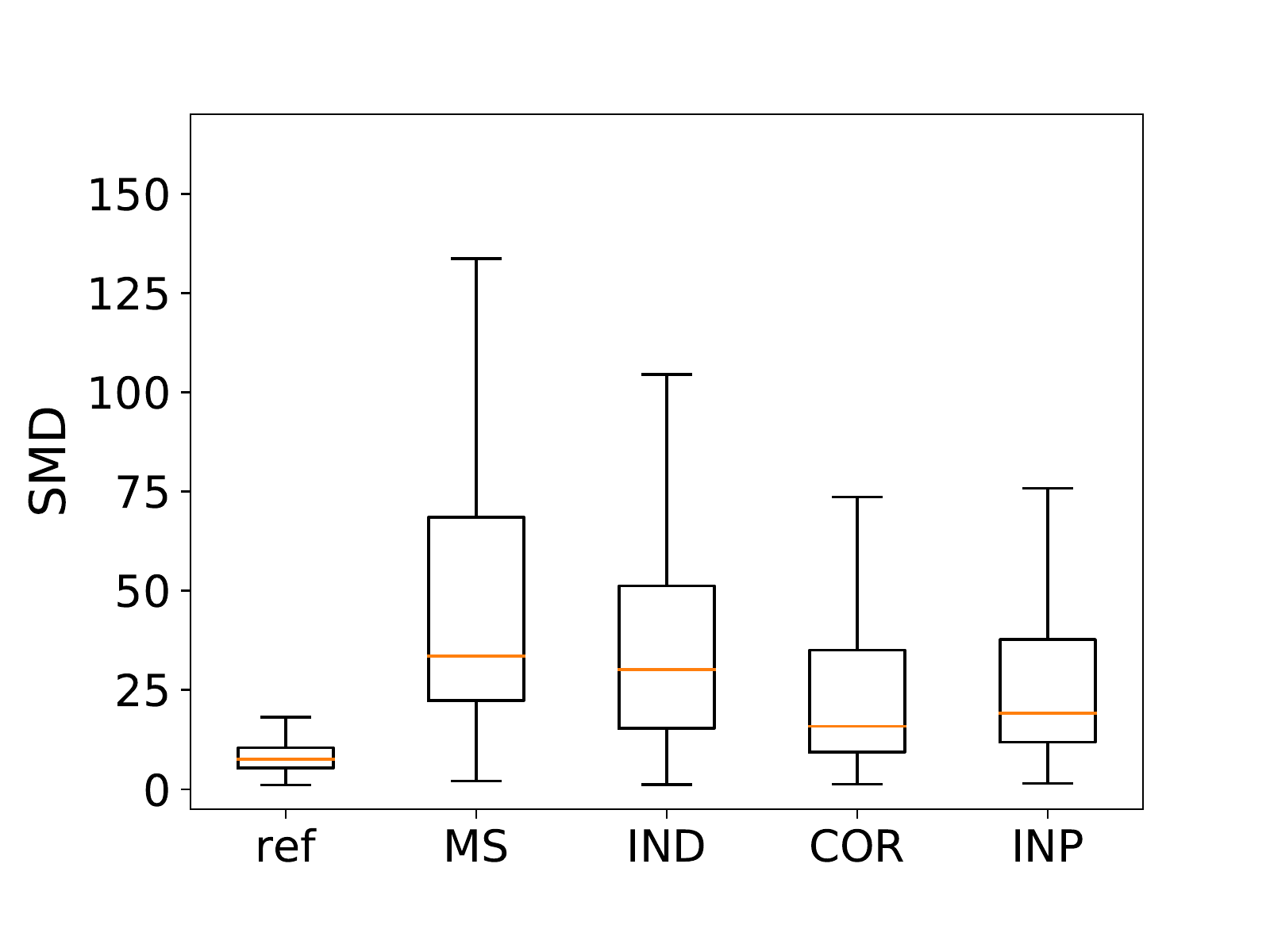}
	\vspace{-1.5\baselineskip}
        {\small{LHD 10}}
    \end{minipage}%
    \begin{minipage}[t]{0.304\textwidth}
        \centering
        \includegraphics[width=\textwidth, trim={7em 0 4em 0},clip]{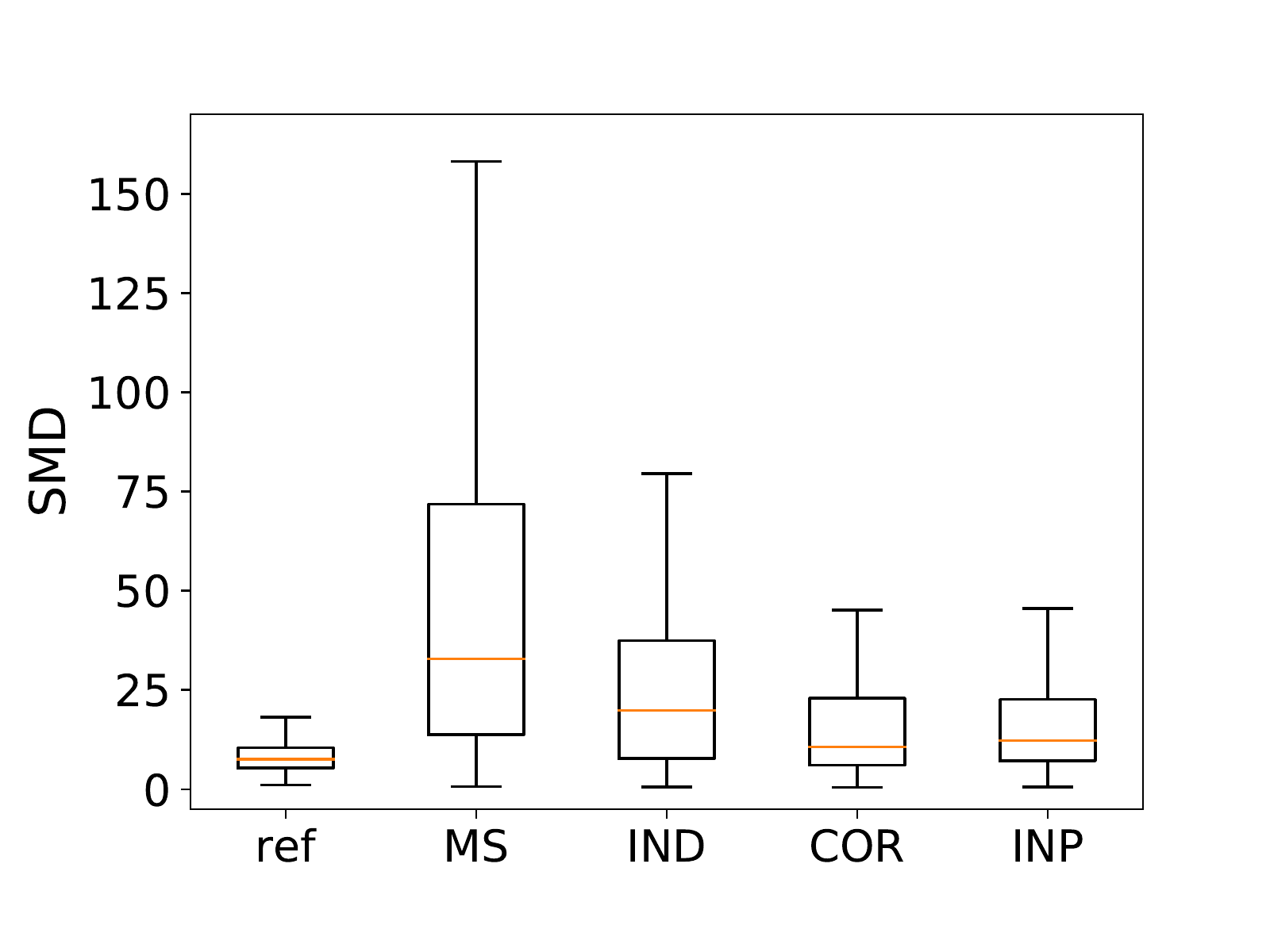}
	\vspace{-1.5\baselineskip}
        {\small{LHD 20}}
    \end{minipage}%
    \begin{minipage}[t]{0.304\textwidth}
        \centering
        \includegraphics[width=\textwidth, trim={7em 0 4em 0},clip]{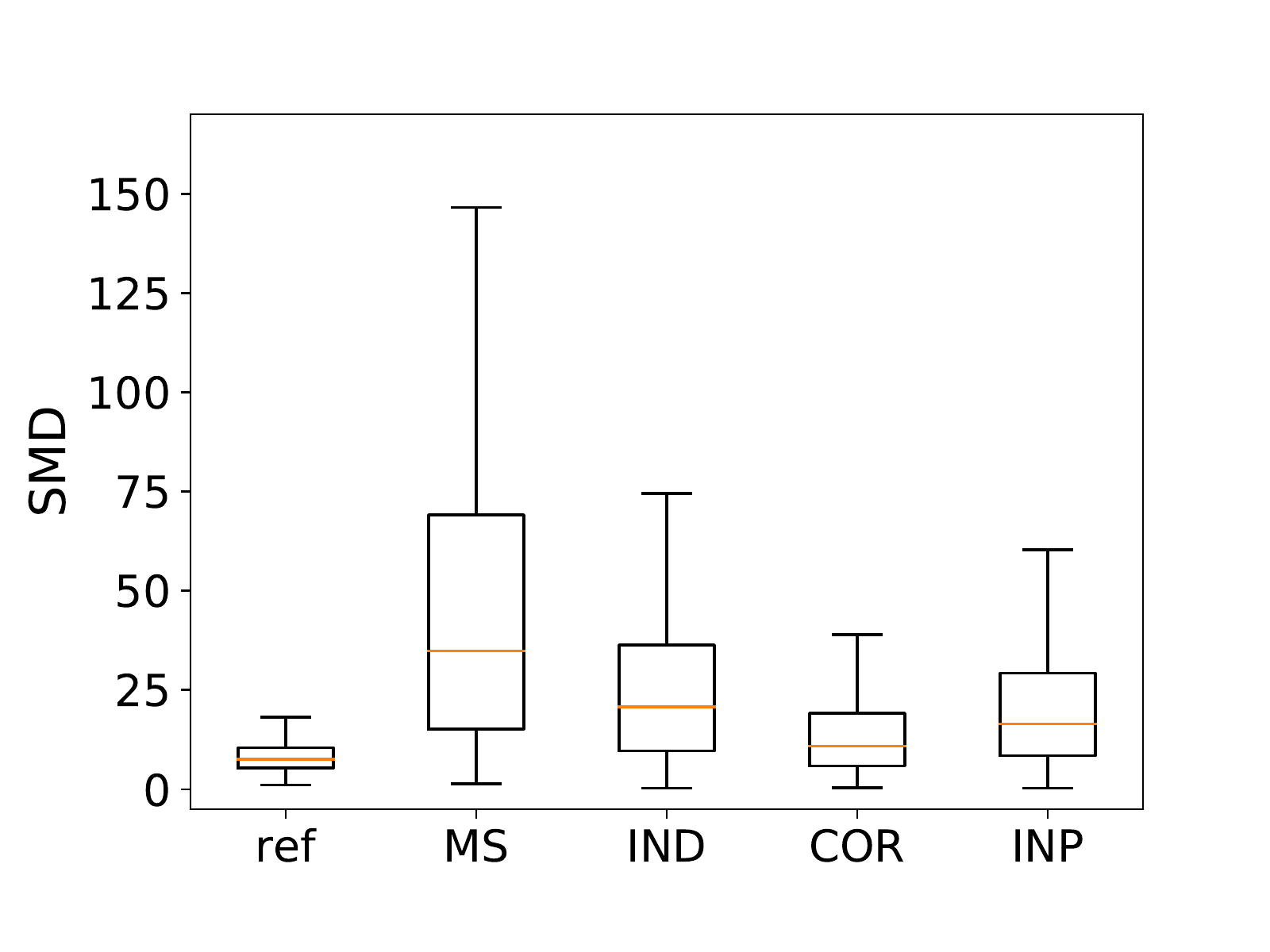}
	\vspace{-1.5\baselineskip}
        {\small{LHD 30}}
    \end{minipage}%
\vspace{1.\baselineskip}
\caption{Distribution of the squared Mahalanobis distances for the four multi-output emulators (MS, IND, COR, INP) for three different LHDs computed with 500 test points. The reference distribution (ref) is $\chi^2_8$. A single representative experiment is chosen for each of the three designs. The closer distributions are to the reference the better.}
\label{fig:smd_mult_out}
\end{figure*}

Figure \ref{fig:smd_mult_out} shows the box plots of the distributions of the squared Mahalanobis distances for the test points for the three designs. We only show the results of a single experiment for each design. The reference distribution $\chi^2_8$ has mean of $8$ and variance of $16$. We observe that although all of the emulators fail to correctly model the posterior covariance (a direct consequence of underestimating the predictive uncertainty), the IND, COR, and INP emulators all do better than MS with the last two coming closest to the reference. If we look at the means of the distributions, then for LHD 10, the average values over the 20 experiments are 54, 38, 27, and 29 for the MS, IND, COR, and INP emulators, respectively. The results shown represent the general trend that, on average, COR and INP emulators tend to  represent the predictive covariance better.   

In terms of the wall-clock time required for training and evaluating the four emulators, the most time-consuming emulator to train is INP for which the required training time grows exponentially with the  size of the training set. The next most time-consuming emulator to train is MS, however, it can be easily trained and evaluated in parallel since there is no overlap between the outputs. The time for training IND and COR emulators is approximately the same and is considerably less than for the other two since only a single GP needs to be trained. 


Based on the performed experiments, in the following we choose to work with IND and COR emulators due to their good performance on the three test statistics as well as good computational efficiency. The question of how to obtain the correlation matrix for the COR emulator still remains open. In our numerical experiments we did not observe significant differences in the emulator performance with the small changes to the correlation matrix $\BSigma_k$. In particular, the estimates of $\BSigma_k$ obtained with the INP emulator were similar for all  experiments. It appears that the variability in the results due to the training designs outweighs the variability from using approximate correlations. 

As mentioned previously, an alternative way of treating the separable form of the covariance is to assume a ``non-informative'' prior on $\BSigma_k$ as in \cite{SConti_AOHagan_2010a} which allows analytical integration of this matrix out of the predictive distribution. In addition, if the mean $\bmu(\cdot)$ is taken to be a generalized linear model with a flat prior, it can also be treated analytically. We deviated from the approach of \cite{SConti_AOHagan_2010a} for the sake of a simpler and more interpretable model.

\section{Numerical study of inference with adaptive GP emulators}\label{sec:inference_numerics}

In this section we evaluate the performance of the adaptive construction of the GP emulators introduced in Section \ref{sec:inference}. First, we solve  a synthetic inference problem for the same set of three parameters $\btheta=(F, T_0, \gamma)$. For this task, we generate synthetic measurement data with the post-processing model of Section \ref{sec:gimlet-model} and corrupt it with noise. Similarly to Section \ref{sec:multi_GP_numerics}, we use a tri-linear interpolation of the outputs of the post-processing model as the forward model for the GP construction and inference. We run Algorithm \ref{algo:adaptiveGP} and use the constructed GP emulator to obtain the posterior of the parameters $\btheta$ given the measurements. We compare the results obtained with the adaptive approach to those obtained using the GP emulators built using fixed design (see Section \ref{sec:LHD}). Having a relatively inexpensive forward model, we can obtain a reference posterior using the ``true'' likelihood $L(\btheta | \bd)$, which allows us to obtain quantitative measures of the quality of the GP-based posteriors. 

Following this detailed study, we use the adaptive algorithm to obtain parameter posteriors using observational data from \cite{MViel_et_al_2013a} and using the post-processing model \ref{sec:gimlet-model} as the forward model. These results are reported in Section \ref{sec:adaptive_gimlet_viel}.

\begin{figure*}[t]
    \centering
    \begin{minipage}[t]{0.45\textwidth}
        \centering
        \includegraphics[width=\textwidth, trim={0 0 2em 0},clip]{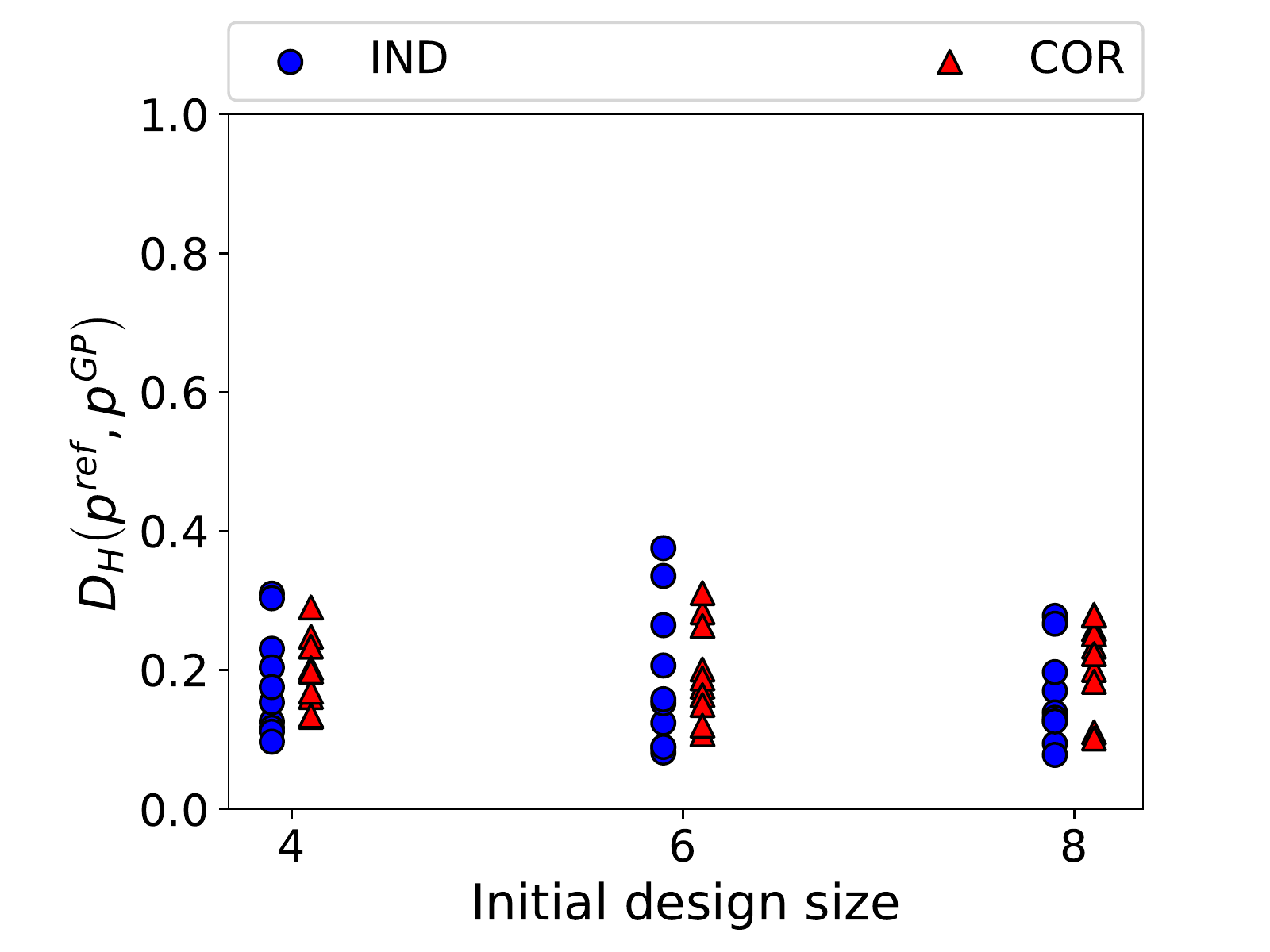}
	\vspace{-1.\baselineskip}
        {\small{Adaptive designs.}}
    \end{minipage}%
    \begin{minipage}[t]{0.44\textwidth}
        \centering
        \includegraphics[width=\textwidth, trim={1em 0 2em 0},clip]{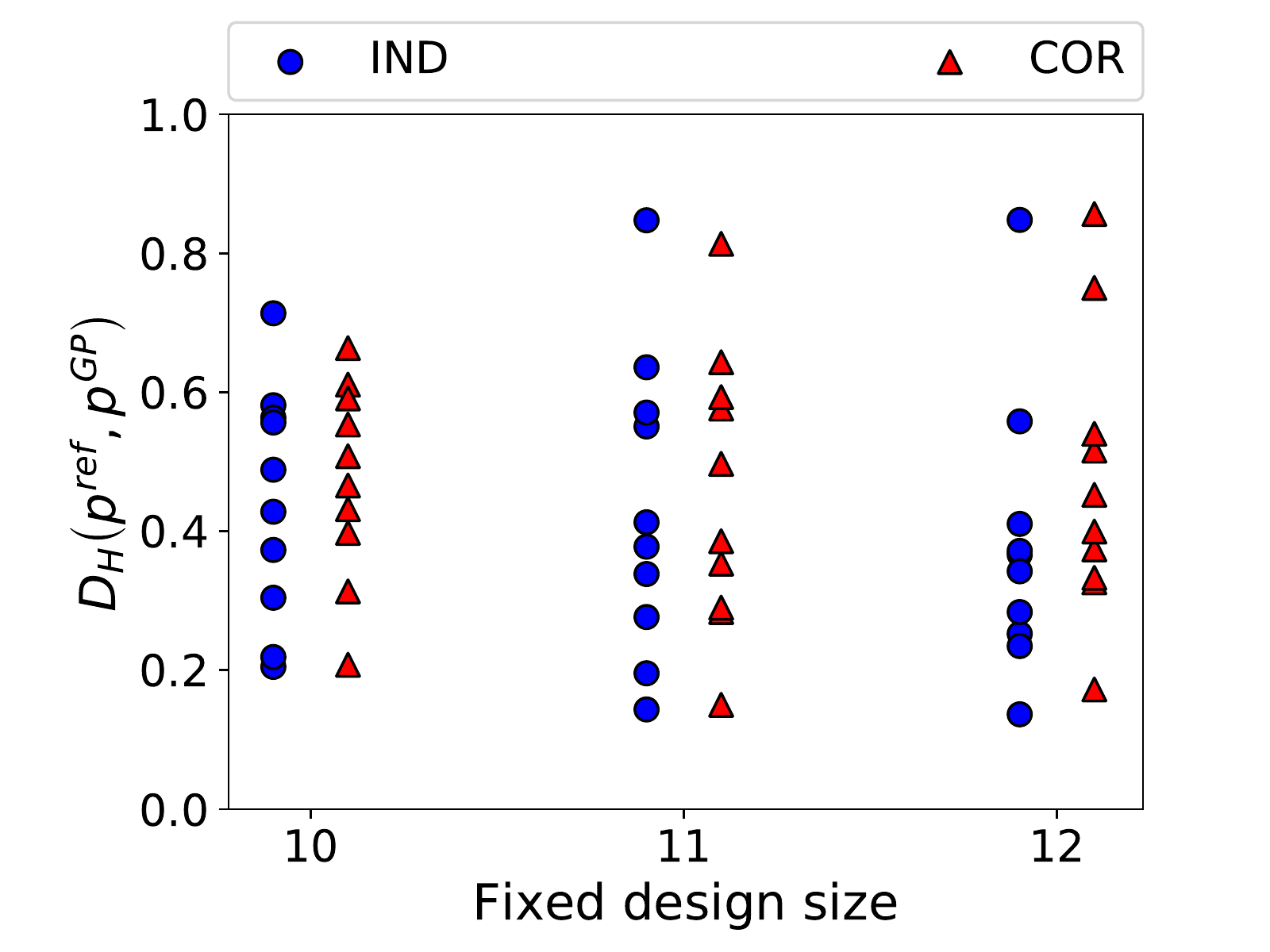}
	\vspace{-1.\baselineskip}
        {\small{Fixed designs.}}
    \end{minipage}%
    \vspace{1.\baselineskip}
    \caption{Hellinger distances $D_{H}\left(p^{ref}, p^{GP}\right)$ for the GP-based posteriors. While fixed-design-based emulators can occasionally produce good quality posteriors, the inconsistency of the results makes them a poor choice compared to adaptive designs.}
\label{fig:h_dists}
\end{figure*}

\begin{figure}[t]
    \centering
    \begin{minipage}[t]{0.45\textwidth}
        \centering
        \includegraphics[width=\textwidth, trim={1em 0 2em 0},clip]{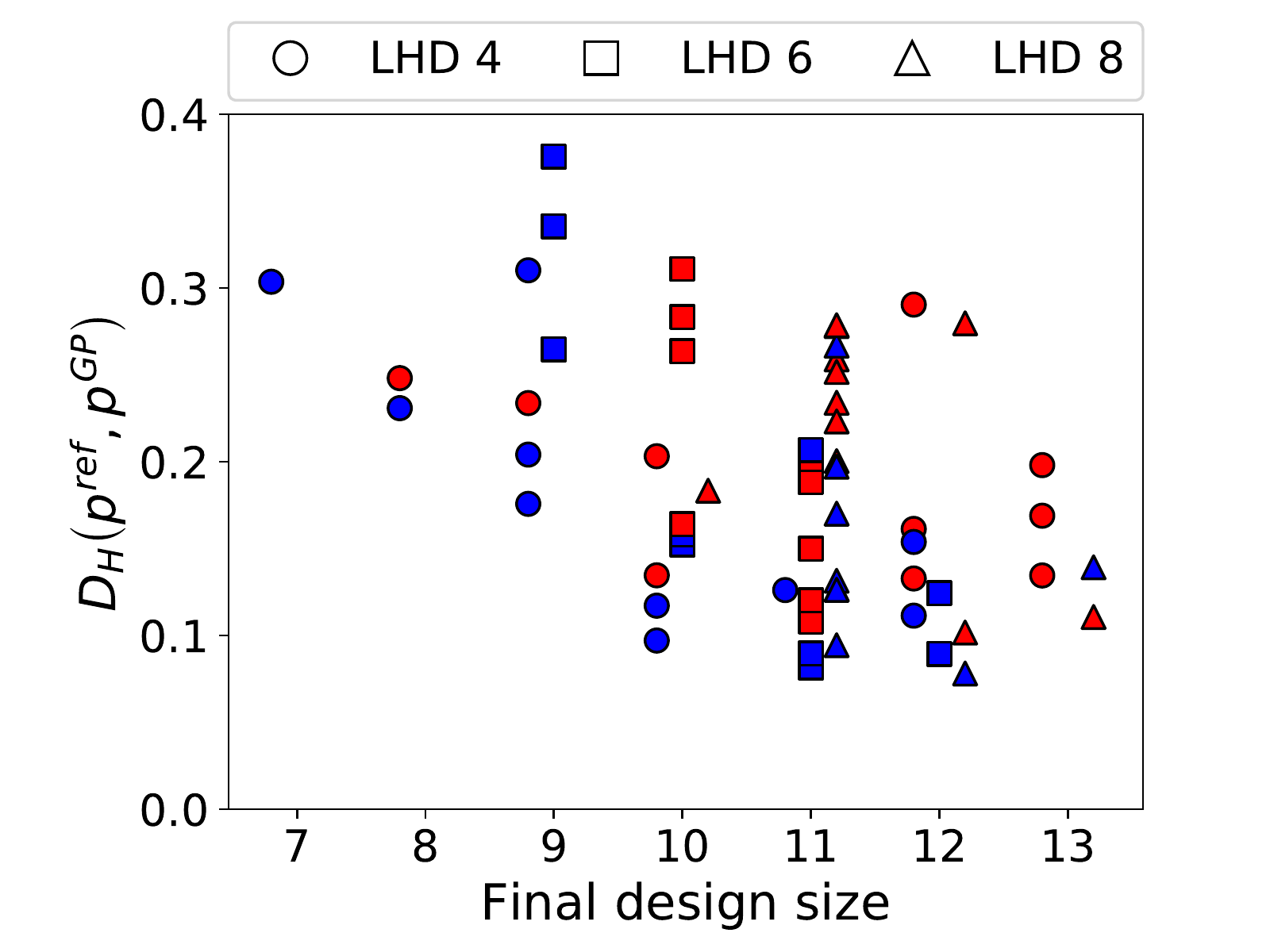}
	\vspace{-1.\baselineskip}
    \end{minipage}%
    \vspace{1.\baselineskip}
    \caption{Hellinger distances for the posteriors obtained with the adaptive designs ordered by the size of the final designs. In blue - IND, in red - COR approaches. Different markers correspond to different initial design sizes.}
\label{fig:divs_by_final}
\end{figure}

Finally, we use a simplistic version of our adaptive approach to construct posteriors for an extended set of parameters, namely $\btheta=(F, T_0, \gamma, \lambda_P)$, using the same Viel data and the THERMAL suite of Nyx simulations\footnote{http://thermal.joseonorbe.com/}. Here, we restrict the search space $\cX_\theta$ to the 75 points for the parameters $(T_0, \gamma, \lambda_P)$ in this dataset and $40$ values of $F$ giving us a total of $3{,}000$ possible values of $\btheta$. We use our adaptive algorithm to select a small subset of the points for constructing a GP emulator. We compare the results obtained with this restricted version of our algorithm to those in \cite{Walther2018} where all 75 points for $(T_0, \gamma, \lambda_P)$ and several values of $F$ were used for the GP construction. Note that we do \textit{not} expect to obtain identical results, as, besides differences in implementation (\cite{Walther2018} build a GP emulator using the PCA-based approach of \cite{SHabib_KHeitmann_DHigdon_CNakhleh_BWilliams_2007a}, see below), we also use only \cite{MViel_et_al_2013a} subset of measurement data for the given redshift. The reason for this approach is our focus on the inference method. However, even using a ``restricted'' version of our adaptive algorithm, we are able to effectively constrain the parameters using only a fraction of the available inputs and simulation results.

\subsection{State-of-the-art data-agnostic approach}
\label{sec:LHD}

Currently, space-filling LHDs  are widely used by the cosmological community  for selecting the training points for the construction of emulators. Typically, the number of training points is selected a priori and remains fixed. For example, in the first such work in cosmology \citep{SHabib_KHeitmann_DHigdon_CNakhleh_BWilliams_2007a}, the space-filling LHD with $128$ points is selected for the problem of inferring $5$ cosmological parameters from   matter power spectrum measurements. The authors employ a PCA-based approach to constructing multi-output emulators. Specifically, they use a singular value decomposition (SVD) of the simulations at the training points specified by the LHD. The weights of the SVD are then modeled as independent GP emulators. 

A similar approach is taken in \cite{Walther2018} for recovering thermal parameters of the IGM using the Ly$\alpha$ flux power spectrum. However, the training points do not exactly form an LHD, as one of the parameters, $\lambda_P$, is difficult to
independently vary in a way that is not correlated  with  the other thermal  state  parameters $T_0$ and $\gamma$.
This is because $\lambda_P$ probes the integrated thermal history
which is smooth for each individual physical model of heating and cooling of
the IGM during and after reionization process.
Of course, in principle one could generate models with abruptly changing instantaneous temperature such that the pressure smoothing does not have enough time to adjust, but we lack physical motivation for such models.
In addition, one of parameters---the mean flux of the Ly$\alpha$ forest---is not
part of the LHD as it can be easily rescaled in the post-processing, thus its sampling does not require running additional expensive simulations.

\subsection{Comparison of the adaptive method and the data-agnostic approach}

\begin{figure*}[htbp]
    \centering
    \begin{minipage}[t]{0.333\textwidth}
        \centering
        \includegraphics[width=\textwidth, trim={0 0 4em 0},clip]{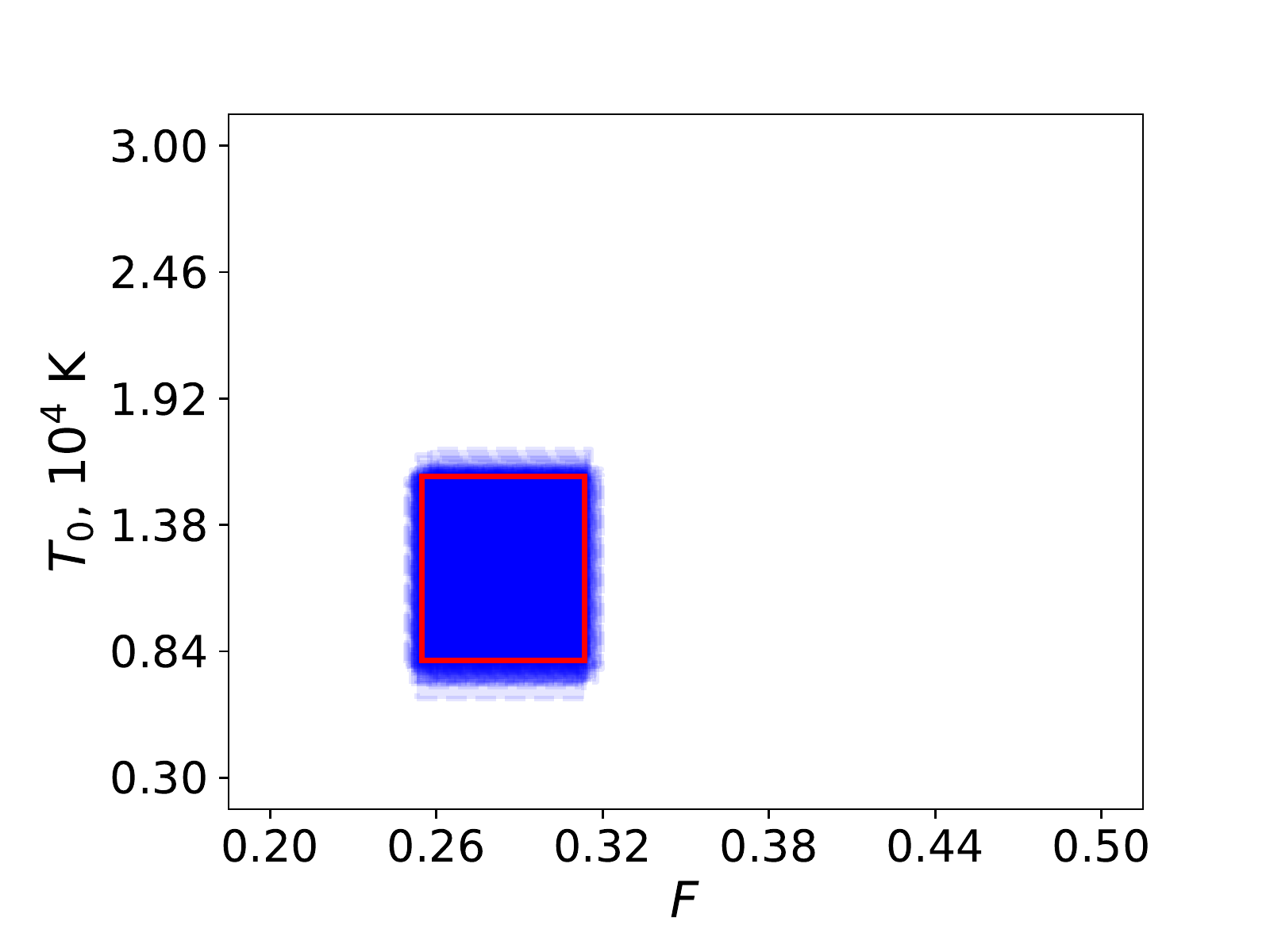}
	\vspace{-1.\baselineskip}
    \end{minipage}%
    \begin{minipage}[t]{0.325\textwidth}
        \centering
        \includegraphics[width=\textwidth, trim={1em 0 4em 0},clip]{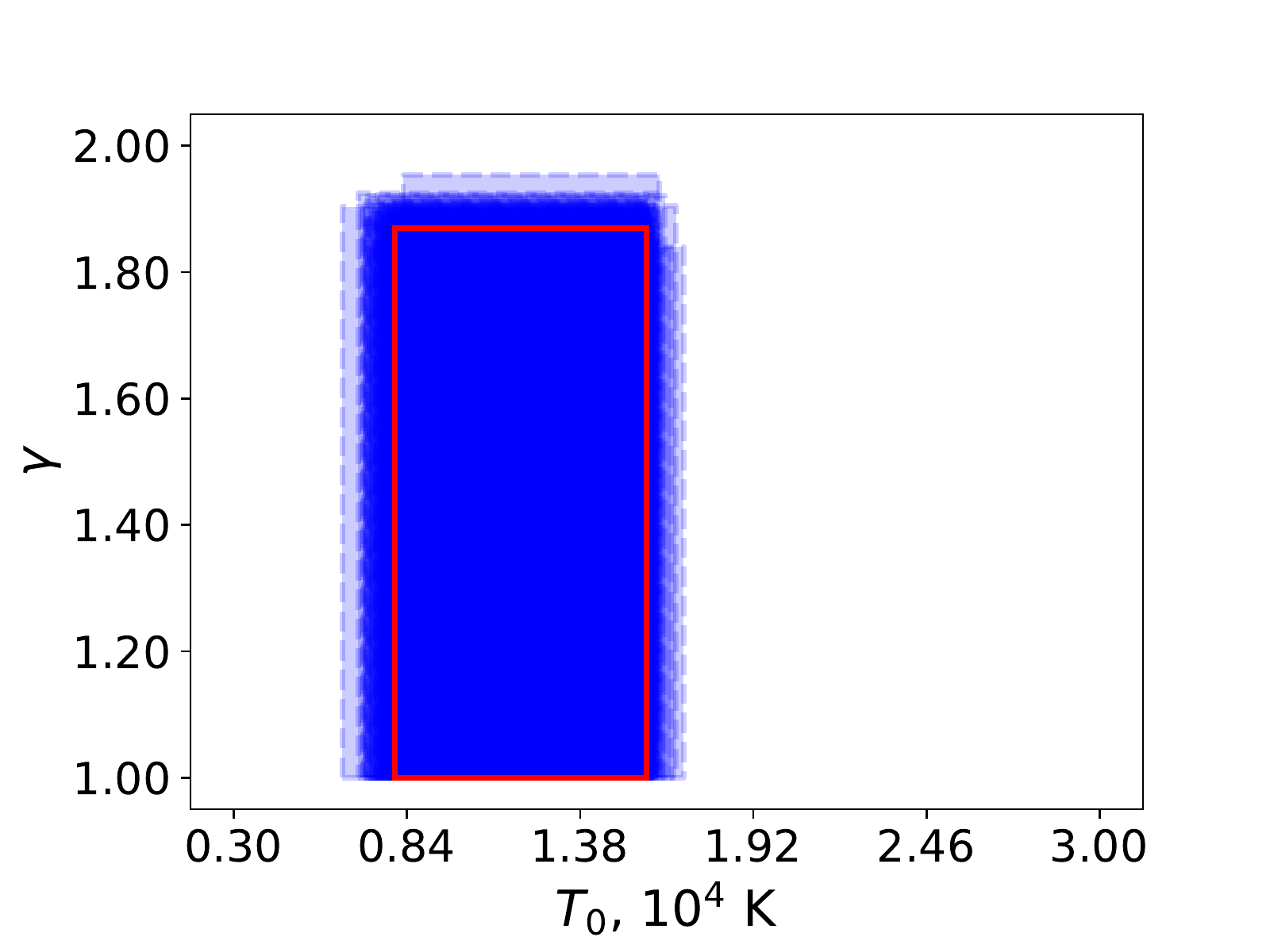}
	\vspace{-1.\baselineskip}
    \end{minipage}%
    \begin{minipage}[t]{0.325\textwidth}
        \centering
        \includegraphics[width=\textwidth, trim={1em 0 4em 0},clip]{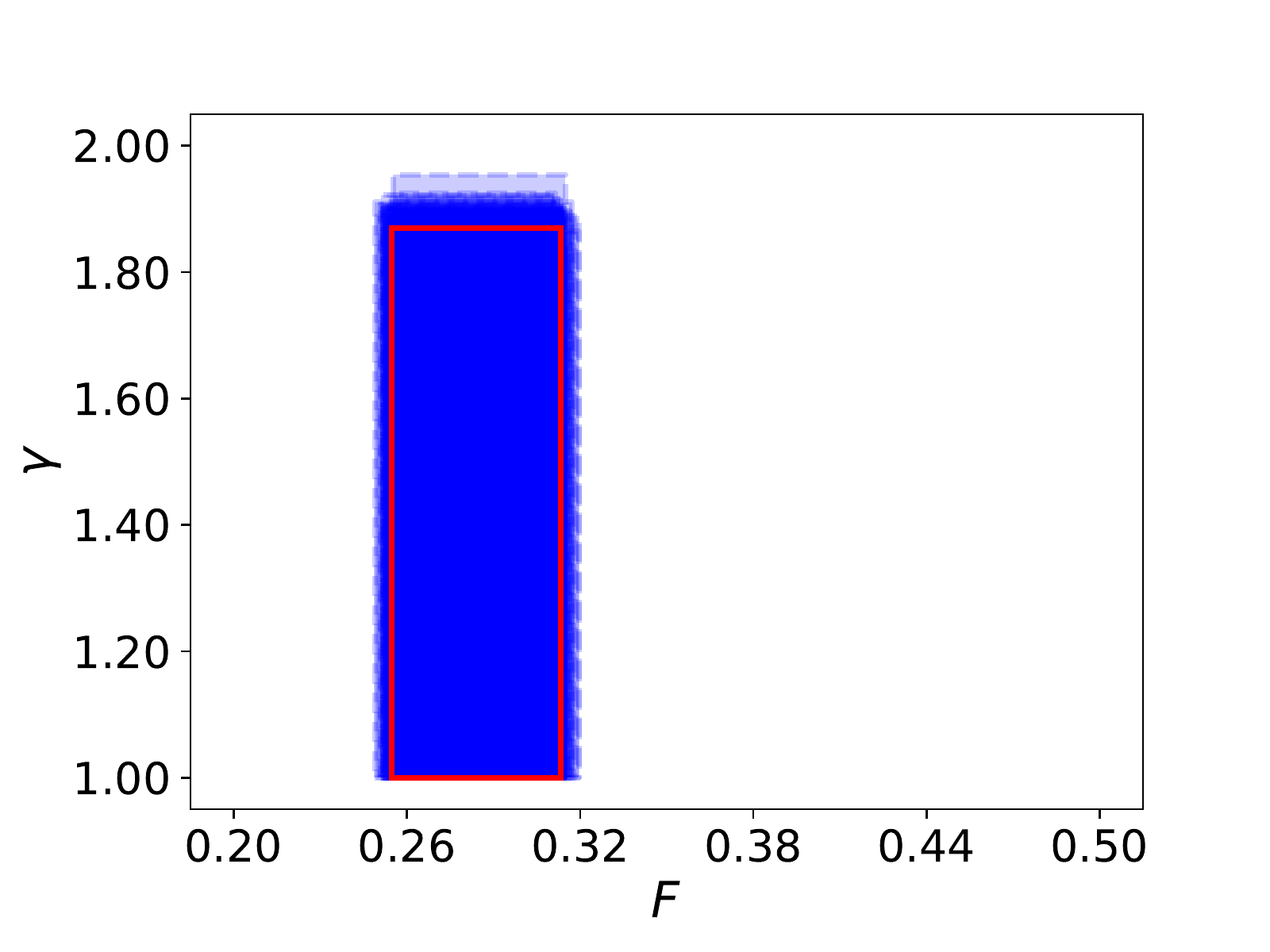}
	\vspace{-1.\baselineskip}
    \end{minipage}%
\caption{$95\%$ HPD intervals for the posteriors obtained with the adaptive designs (blue rectangles). The red rectangle represents the HPD intervals of the reference posterior.}
\label{fig:hpd_regions_adaptive}
\end{figure*}

\begin{figure*}[htbp]
    \centering
    \begin{minipage}[t]{0.333\textwidth}
        \centering
        \includegraphics[width=\textwidth, trim={0 0 4em 0},clip]{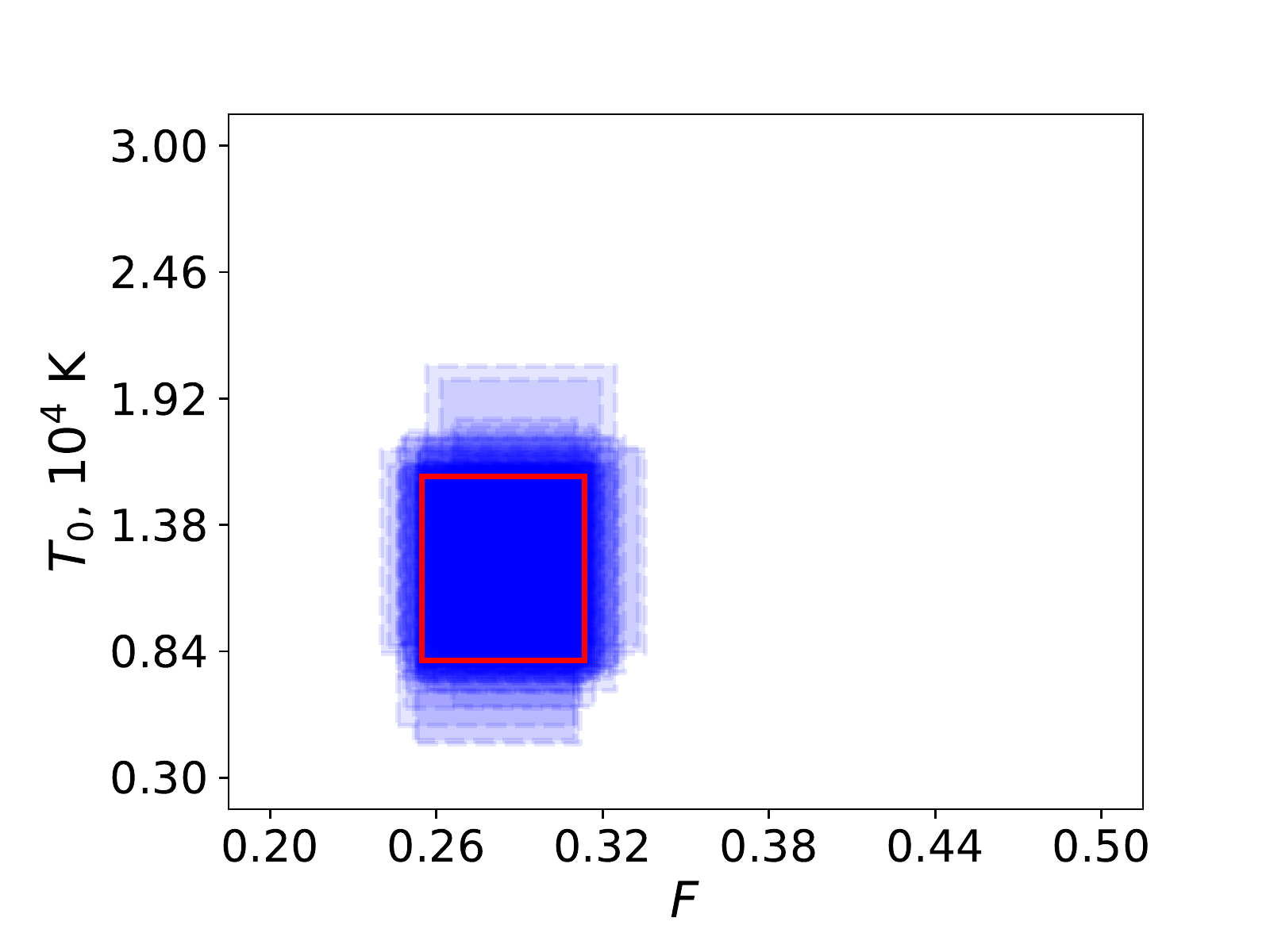}
	\vspace{-1.\baselineskip}
    \end{minipage}%
    \begin{minipage}[t]{0.325\textwidth}
        \centering
        \includegraphics[width=\textwidth, trim={1em 0 4em 0},clip]{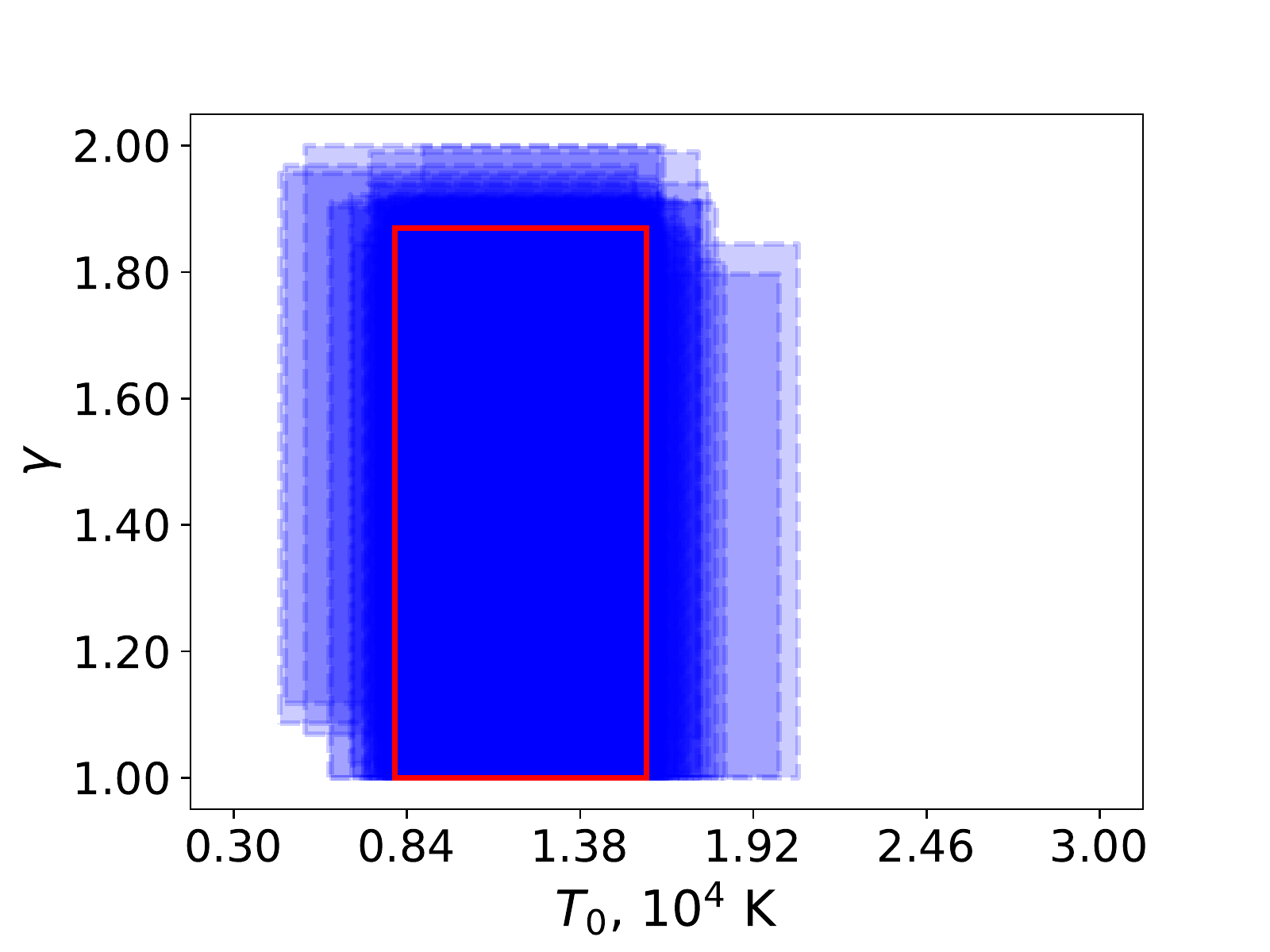}
	\vspace{-1.\baselineskip}
    \end{minipage}%
    \begin{minipage}[t]{0.325\textwidth}
        \centering
        \includegraphics[width=\textwidth, trim={1em 0 4em 0},clip]{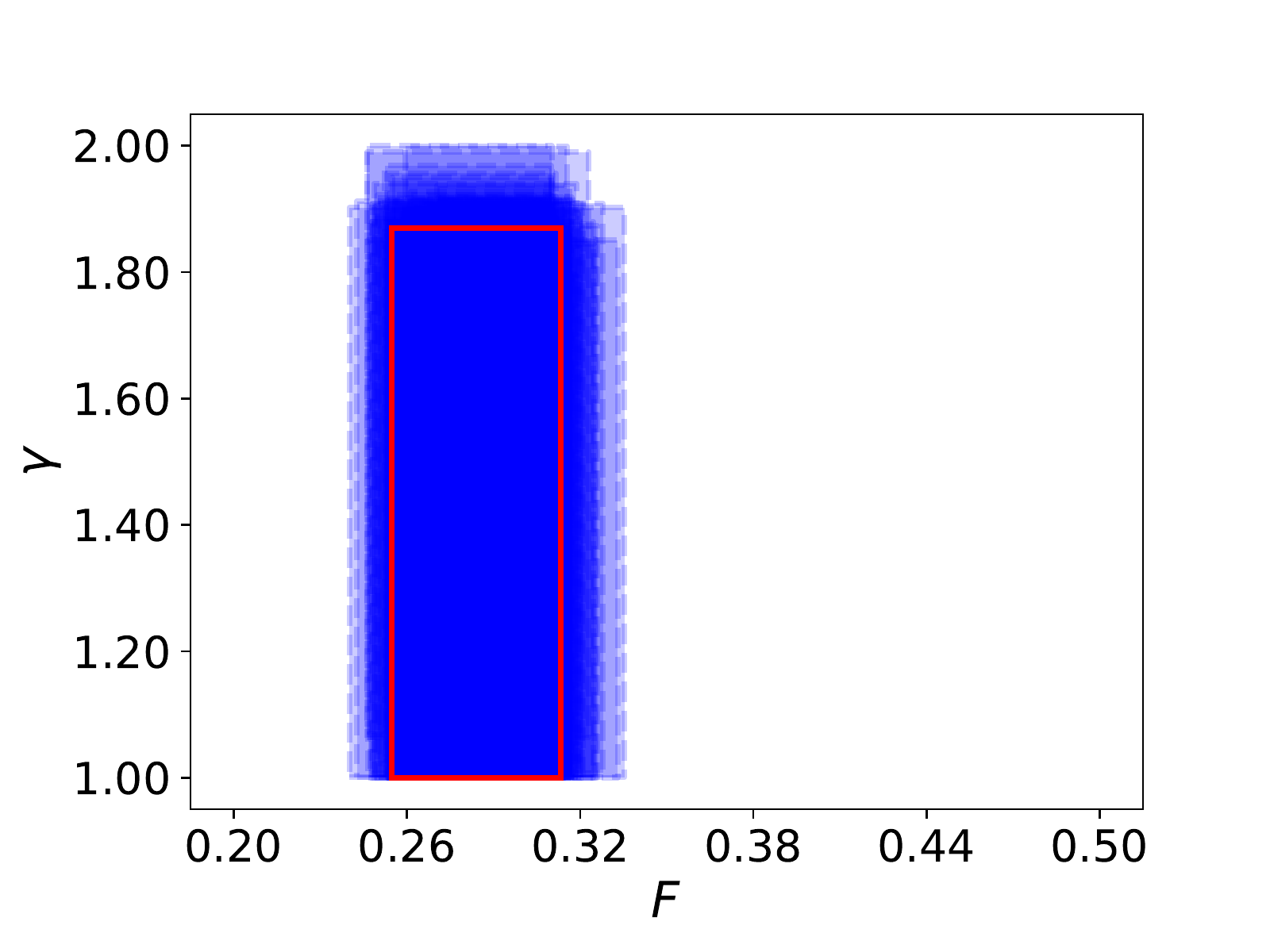}
	\vspace{-1.\baselineskip}
    \end{minipage}%
\caption{$95\%$ HPD intervals for the posteriors obtained with the fixed designs (blue rectangles). The red rectangle represents the HPD intervals of the reference posterior.}
\label{fig:hpd_regions_fixed}
\end{figure*}

We continue to use the post-processing model described in Section \ref{sec:gimlet-model} as the true ``forward model'' (with the same $q=8$ outputs).
We first generate the mock measurement by evaluating the model at a fixed $\btheta_{true}=(0.275, 1.245\times 10^4K, 1.6)^T$, and we corrupt the resulting measurement vector with noise from a multivariate Gaussian distribution $\cN_q(\bzero_q, \, \sigma_e^2\BI_q)$ with $\sigma_e = 0.01$. This level of measurement noise is consistent with the observational data \citep{MViel_et_al_2013a} that we use later.

In order to analyze how the size of the initial design for the adaptive algorithm influences the obtained solution, we perform experiments with 4, 6, or 8 design points in the three-dimensional parameter space $\btheta=(F, \, T_0, \, \gamma$). In   each iteration $s$ of  Algorithm \ref{algo:adaptiveGP}, we construct a GP surrogate using the current selection of design points, we solve the auxiliary optimization problem to maximize the expected improvement in fit function, and augment the design set with a single point that  provides the largest predicted improvement. We iterate until the relative expected improvement drops below a $1\%$ threshold.

For each selection of the size of the initial training design, we run our algorithm 10 times, each time selecting new initial design points using a maximin LHD.
For each of the initial designs, we run the algorithm with two emulator choices: IND and COR (see Section \ref{sec:multi_GPs}). For the COR emulator the inter-output correlation matrix $\BSigma_k$ is taken to be the same as the one used in Section \ref{sec:multi_GP_numerics} (see Figure \ref{fig:cor_matrix}). 
On average the final designs contain $10$--$11$ inputs regardless of the number of points in the initial design or the emulator choice. 

We perform MCMC sampling of the posterior using the integrated likelihood based on the constructed GP surrogates (see Appendix \ref{sec:math} for details). To obtain a quantitative measure of the quality of the obtained posteriors, we compute the Hellinger distance between the GP-based posteriors $p^{GP}(\btheta | \bd, \cD)$ and the reference posterior $p^{ref}(\btheta | \bd)$ obtained by a direct MCMC sampling with the ``true'' likelihood function, i.e., the likelihood of the measurement data that uses evaluations of our (post-processing) forward model. The Hellinger distance is a metric for evaluating differences between two probability distributions. It can be related to other commonly used distance measures, such as total variation distance and Kullback-Leibler divergence, see, e.g., \cite{MDashti_AMStuart_2016a}. It has also been recently studied in the context of posteriors obtained with Gaussian process emulators \citep{AMStuart_ALTeckentrup_2018a}. The Hellinger distance between $p^{ref}$ and $p^{GP}$ is defined as follows:
 \begin{align}\label{eq:H_distance}
 \begin{split}
        D_{H}(p^{ref}, p^{GP}) &= \\
        \bigg( \frac{1}{2}\int_{\cX_\theta} \big(&\sqrt{p^{ref}(\btheta | \bd)} - \sqrt{p^{GP}(\btheta | \bd, \cD)}\big)^2 d\btheta \bigg)^{1/2}.
\end{split}
\end{align}

To compute $D_H(p^{ref}, p^{GP})$ we approximate the densities $p^{ref}$ and $p^{GP}$ by fitting kernel-density estimates (KDEs) with Gaussian kernels to the generated samples from the respective posteriors and discretize the integral in equation  \eqref{eq:H_distance} using 3-dimensional Sobol' sequence with $10^4$ points. The results for the posteriors obtained with the adaptive GPs (10 runs for each initial design) are presented in  Figure \ref{fig:h_dists} on the left.

We compare the posteriors obtained with the adaptive approach to the posteriors obtained by training GP models wtih fixed maximin LHDs. Here, we fix the design sizes to be $10$, $11$, and $12$. As in the adaptive case, we train the GP emulators using both IND and COR approaches, and we repeat each experiment $10$ times. The results for the posteriors obtained with the fixed designs are shown in  Figure \ref{fig:h_dists} on the right.

The comparison of the results in Figure \ref{fig:h_dists} demonstrates the superiority of the adaptive approach. The adaptive approach is able to achieve results that are closer to the reference posterior in the Hellinger distance, and often with fewer design points. Furthermore, the results for the adaptive approach are less spread out, and thus making it more robust and consistent.

In Figure \ref{fig:divs_by_final}, we re-plot the data from Figure \ref{fig:h_dists} for the adaptive cases, and we show the final design sizes on the $x$-axis. As we can see, in most cases, the final design size is either 10, 11, or 12. There is no significant difference in the results for different initial design sizes. However, there appears to be more variability in the final design sizes when the initial design contains only 4 points (LHD 4)---the final designs  have between 7 to 13 points. We also observe a small trend of decreasing distance values as the final design size increases from 7 to 10. However, beyond 10 there is no significant difference in the results. The COR emulator appears less likely to terminate with a design consisting of less than 10 points, but in terms of the distance values, COR does not outperform the IND approach.

Results for the Hellinger distances confirm that there is little difference between the IND and the COR approaches for our current application. As far as the choice of the initial design size, starting with smaller designs (4 to 6 points for the current three-dimensional problem) leads to fewer forward model evaluations without compromising the quality of the result.

Additionally, we compare the highest posterior density (HPD) intervals of the posteriors obtained with the adaptive algorithm and with fixed LHDs. We estimate the $95\%$ HPD intervals from the posterior samples. The two-dimensional projections of the HPD intervals for all parameter pairs are shown in Figures \ref{fig:hpd_regions_adaptive} (adaptive cases) and \ref{fig:hpd_regions_fixed} (fixed cases). We overlay the HPD intervals from all  $60$ experiments (2 approaches $\times$ 3 initial designs $\times$ 10 repetitions) for each case. In red we show the HPD intervals of the reference posterior. We observe a good correspondence between the HPD intervals for the adaptive designs. For the fixed designs the posteriors are more diffused. This comparison reinforces the conclusion that the results obtained with fixed designs are inconsistent,  and, therefore,  less reliable.

\subsection{Results for the adaptive GP with post-processing model and Viel data}\label{sec:adaptive_gimlet_viel}

In this section we apply our adaptive GP approach to the problem of inferring the same three parameters $\btheta=(F, \, T_0, \, \gamma)$ using the post-processing model of Section \ref{sec:gimlet-model} as the forward model of the power spectrum, and  data from \cite{MViel_et_al_2013a} for the redshift of $z=4.2$.

The measurement data consists of seven values of the power spectrum for $k=\{5.01\times 10^{-3}, 7.95\times 10^{-3}, 1.26\times 10^{-2}, 1.99\times 10^{-2}, 3.16\times 10^{-2}, 5.01\times 10^{-2}, 7.95\times 10^{-2}\} {\rm km^{-1}s}$ with errorbars that we treat as $\pm1\sigma_k$. The measurement noise covariance, thus, has a diagonal form $\BSigma_E=\text{diag}[\sigma_1^2,\dots,\sigma_7^2]$. 
We take a uniform (flat) prior $p(\btheta)$ on all three parameters defined over the same box $\cX_\theta$ as in Section \ref{sec:multi_GP_numerics}:
\[
    \cX_\theta = [0.2, 0.5]\times[3\times 10^3K, 3\times 10^4K] \times[1.0, 2.0].
\]

We use the COR emulator with the same $\BSigma_k$ as in Section \ref{sec:multi_GP_numerics} and initialize Algorithm \ref{algo:adaptiveGP} with a maximin LHD with $4$ points. The stopping threshold for the algorithm is again set to $1\%$.  

\begin{figure}[t]
    \centering
    \begin{minipage}[t]{0.32\textwidth}
        \centering
        \strut\vspace*{-\baselineskip}\newline\includegraphics[width=\textwidth, trim={0.1em 0 2.8em 0},clip]{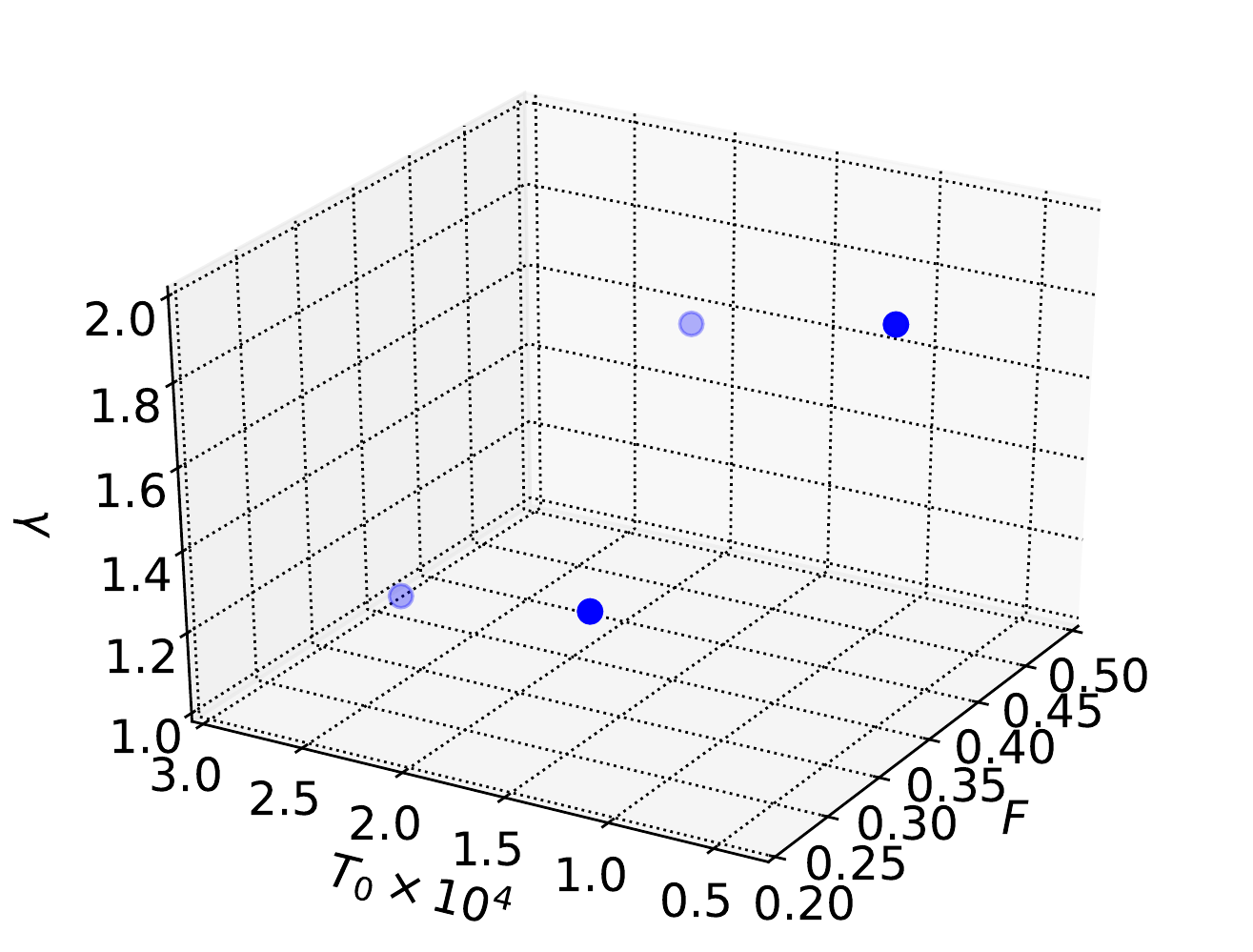}
	\vspace{0.\baselineskip}
	\vfill
        {\small{Initial design}}
    \end{minipage}%
    \begin{minipage}[t]{0.32\textwidth}
        \centering
        \strut\vspace*{-\baselineskip}\newline\includegraphics[width=\textwidth, trim={0.1em 0 2.8em 0},clip]{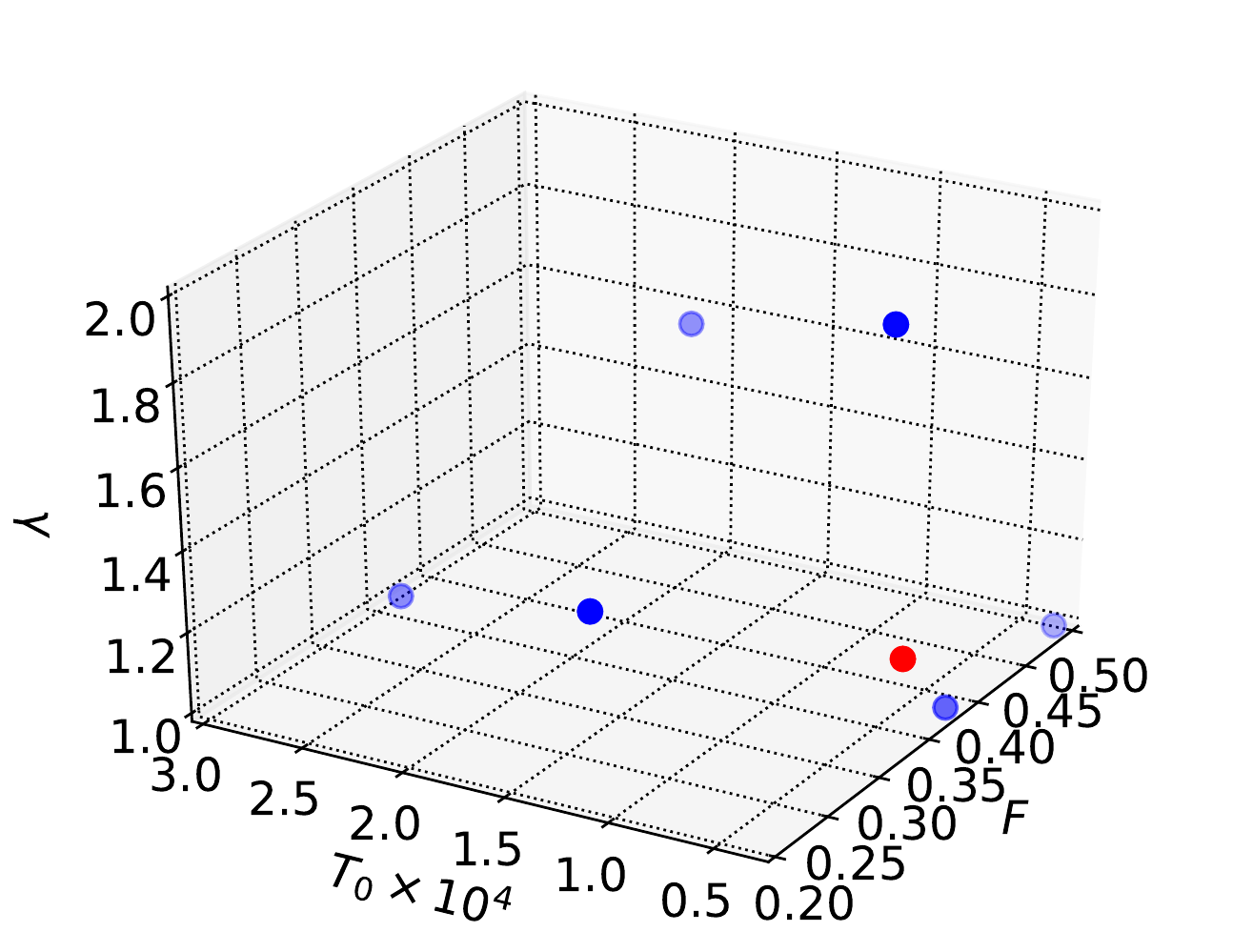}
	\vspace{0.\baselineskip}
	\vfill
        {\small{Design after iteration $s=3$}}
    \end{minipage}%
    \begin{minipage}[t]{0.32\textwidth}
        \centering
        \strut\vspace*{-\baselineskip}\newline\includegraphics[width=\textwidth, trim={0.1em 0 2.8em 0},clip]{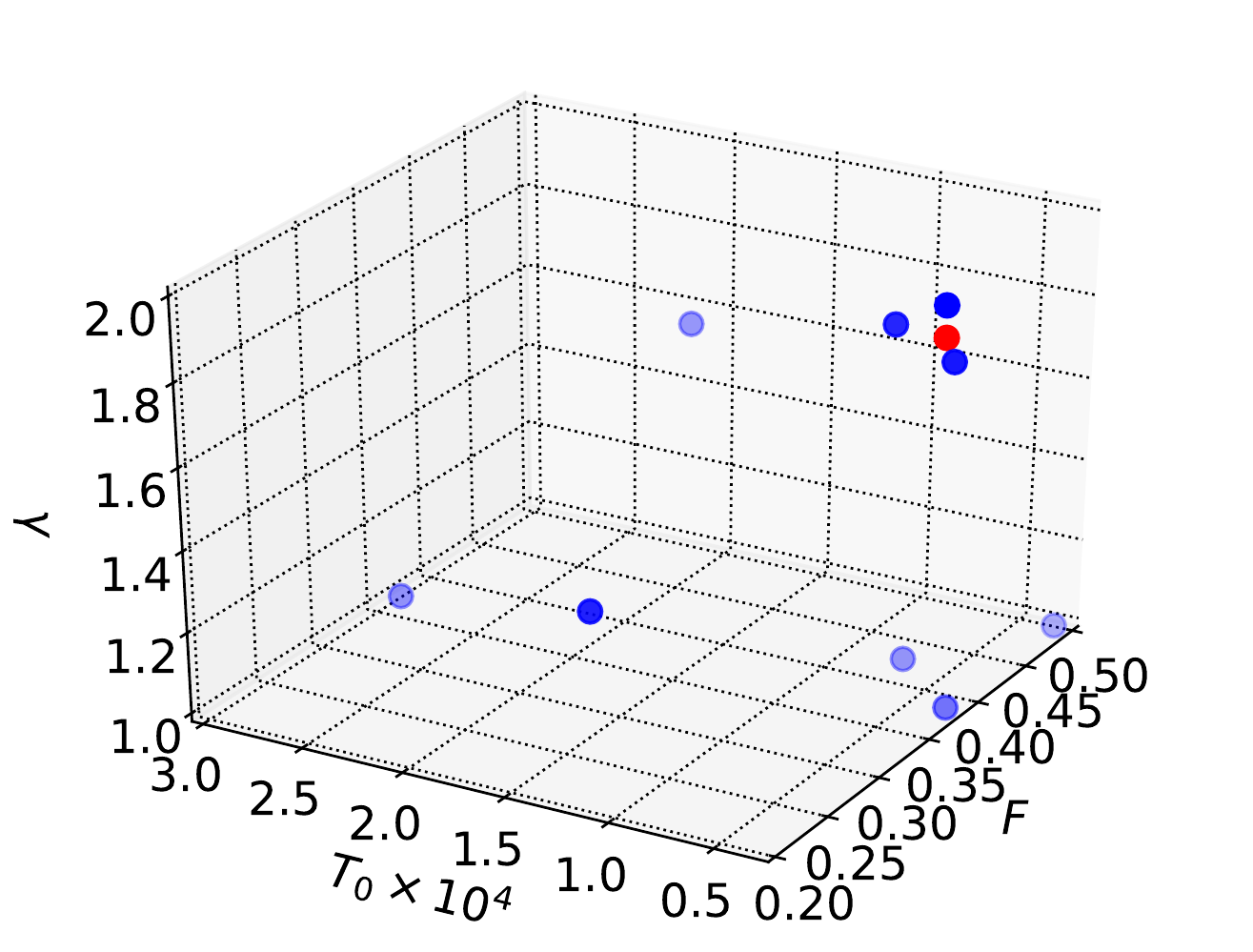}
	\vspace{0.\baselineskip}
	\vfill
        {\small{Final design}}
    \end{minipage}%
\caption{\small{Evolution of the designs for the adaptive GP using the post-processing model and Viel data.}}
    \label{fig:designs_gimlet_viel}
\end{figure}

Figure \ref{fig:designs_gimlet_viel} shows the design points at different iterations of the adaptive algorithm. The first figure shows the initial 4 points arranged in maximin LHD, the figure in the middle has three additional points  after three iterations of the adaptive algorithm, and the last figure shows the final design upon termination.

\begin{figure}[t]
    \centering
    \begin{minipage}[t]{0.45\textwidth}
        \centering
        \includegraphics[width=\textwidth, trim={0 0 1em 0},clip]{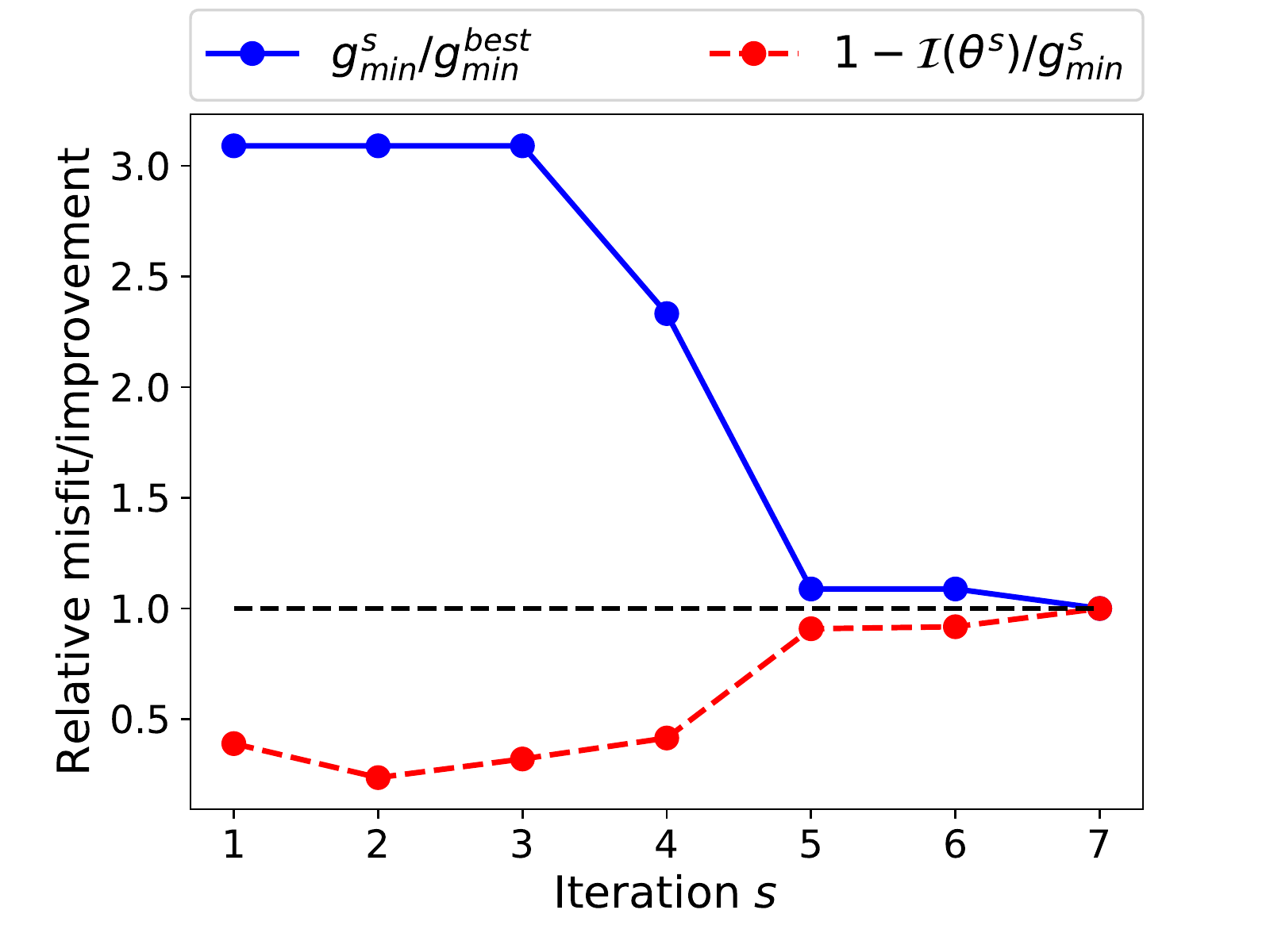}
    \end{minipage}%
    \caption{Iteration history of Algorithm \ref{algo:adaptiveGP} applied to post-processing model with Viel data.}
    \label{fig:iterations_adaptive_inversion_gimlet}
\end{figure}

Figure \ref{fig:iterations_adaptive_inversion_gimlet} shows iteration history of the algorithm with iteration $s=1$ corresponding to the initial design with four points. The blue line in this figure shows the value of the best misfit for the points in the training set in each iteration $s$, $g_{min}^s$, scaled by the best misfit value obtained upon convergence, $g_{min}^{best}$. A point with a better misfit value is not obtained in every iteration, e.g., in iterations 2 and 3 we have the same $g_{min}$ as initially. The points added to the design in these iterations serve the purpose of decreasing the overall uncertainty of the GP. The new point added to the training set $\cD$ after iteration $s=3$ provides a reduction in $g_{min}$ for the first time (see red dot in the middle figure in Figure \ref{fig:designs_gimlet_viel}). The red dotted line in  Figure \ref{fig:iterations_adaptive_inversion_gimlet} shows one minus the relative expected improvement in each iteration, i.e., $1 - \cI(\btheta^s)/g_{min}^s$. As the algorithm progresses, we expect both lines to approach 1. 

\begin{figure}[t]
    \begin{minipage}[t]{0.44\textwidth}
        \centering
        \includegraphics[width=\textwidth, trim={0.5em 0 0em 0},clip]{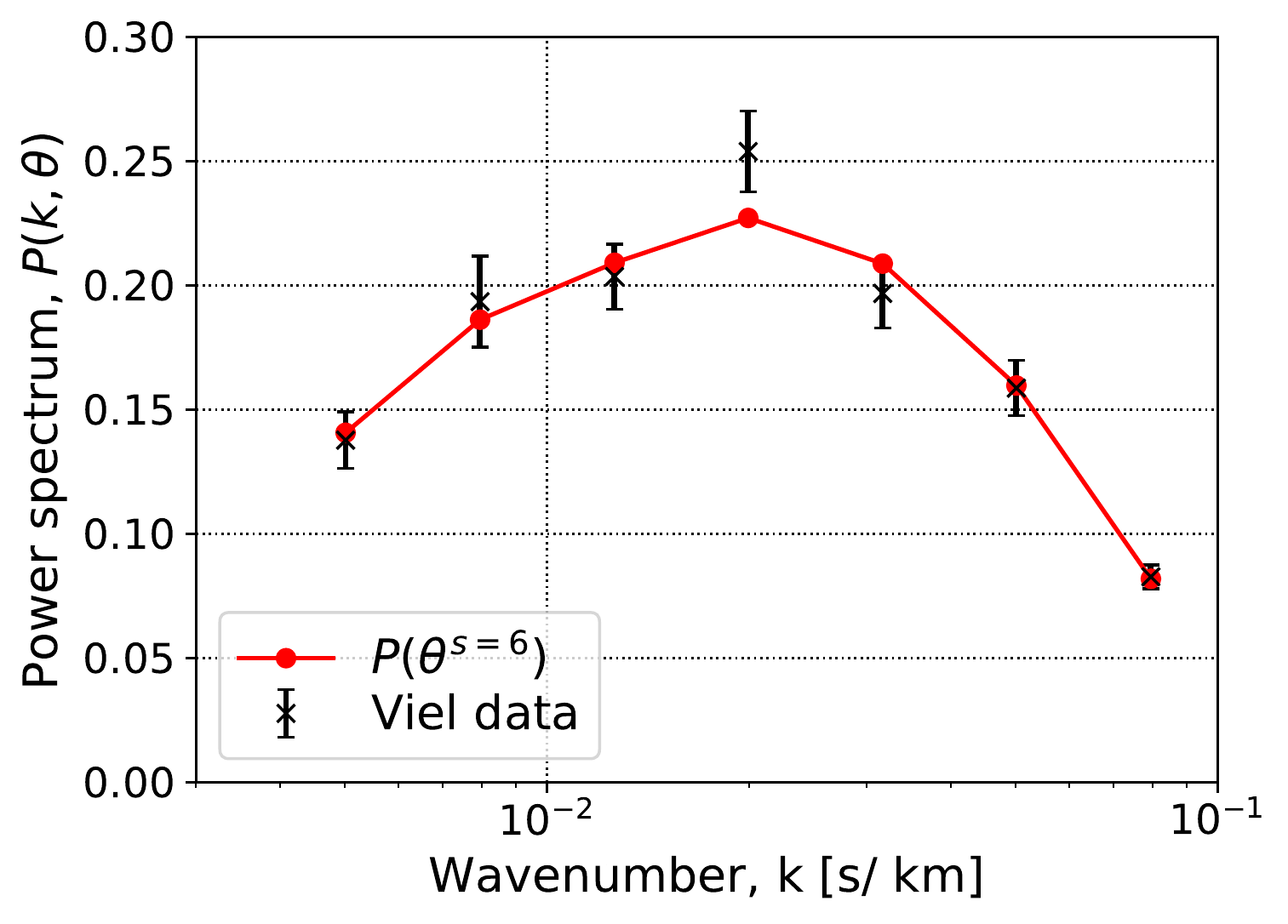}
	\vspace{-0.\baselineskip}
    \end{minipage}%
    \caption{$\BP(\btheta^{s=6})$ corresponding to the best found misfit for the adaptive GP with post-processing model and Viel data.}
\label{fig:fit_adaptive_inversion_gimlet}
\end{figure}

Figure \ref{fig:fit_adaptive_inversion_gimlet} shows the power spectrum $\BP(\btheta^{s=6})$ evaluated at the last $\btheta$ added by the algorithm in iteration $s=6$. This point corresponds to the smallest  misfit $g_{min}^{best}$ to the Viel data found by our algorithm. This point is shown as a red dot in the bottom panel of Figure \ref{fig:designs_gimlet_viel}.

\subsection{Results for the adaptive GP with Nyx simulations and Viel data}

\begin{figure}[htbp]
    \centering
    \begin{minipage}[t]{0.45\textwidth}
        \centering
        \strut\vspace*{-\baselineskip}\newline\includegraphics[width=\textwidth, trim={0 0 1em 0}, clip]{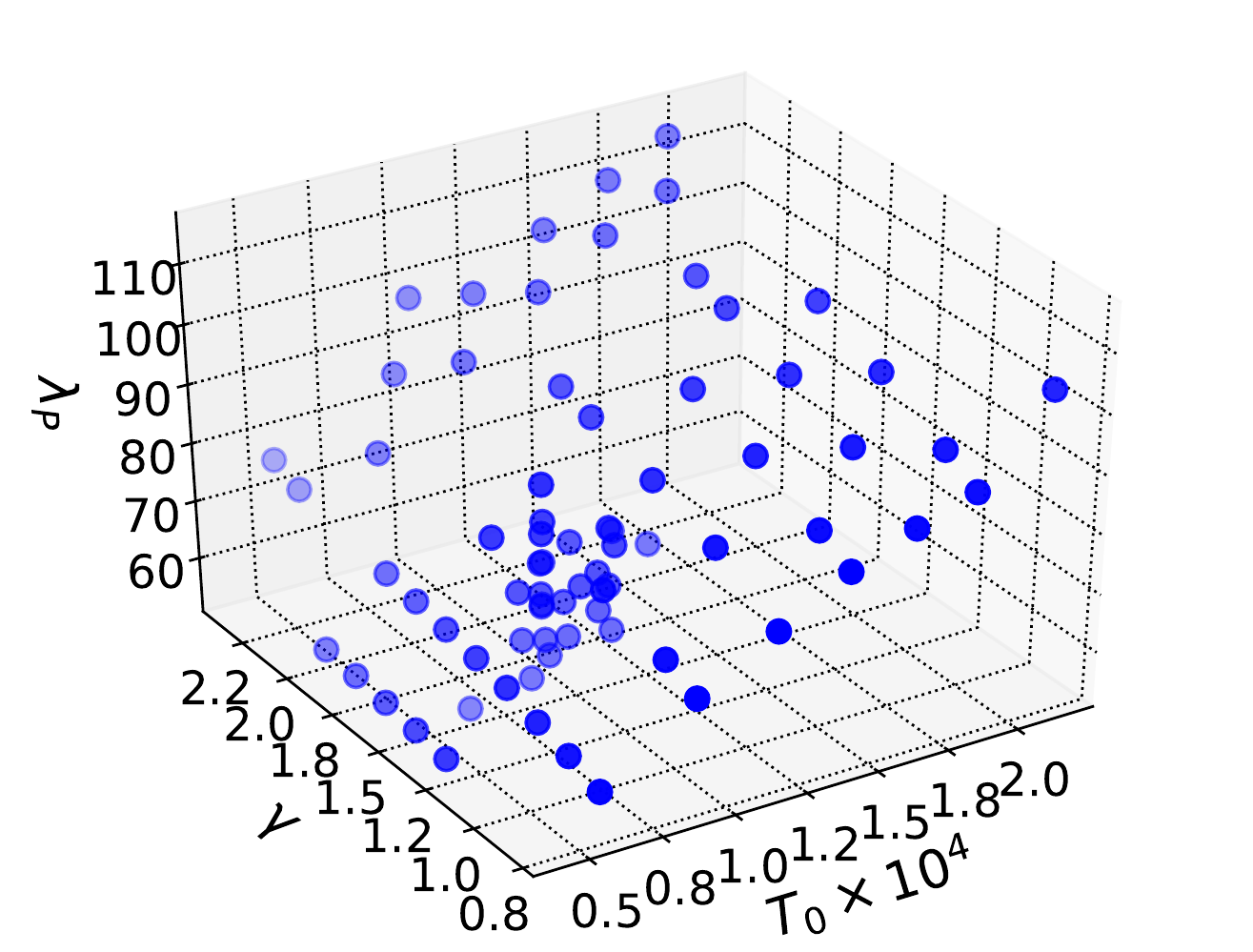}
    \end{minipage}%
    \caption{All 75 simulated models in the THERMAL suite.}
    \label{fig:thermal_suite_nyx}
\end{figure}

In this section we work with the THERMAL suite of Nyx simulations consisting of 75 models with given parameters $(T_0, \gamma, \lambda_P)$ (see Figure \ref{fig:thermal_suite_nyx}).
Mean flux we treat in post-processing as in virtually all Ly$\alpha$
power spectrum inference works.
To be specific, in this work we produce $40$ equidistant values for the parameter $F$ in the interval $[0.2, 0.5]$ for each of the 75 thermal models, thereby sampling the 4-dimensional parameter space $\btheta=(F, T_0, \gamma, \lambda_P)$.

\begin{figure}[t]
    \centering
    \begin{minipage}[t]{0.45\textwidth}
        \centering
        \strut\vspace*{-\baselineskip}\newline\includegraphics[width=\textwidth, trim={0 0 1em 0}, clip]{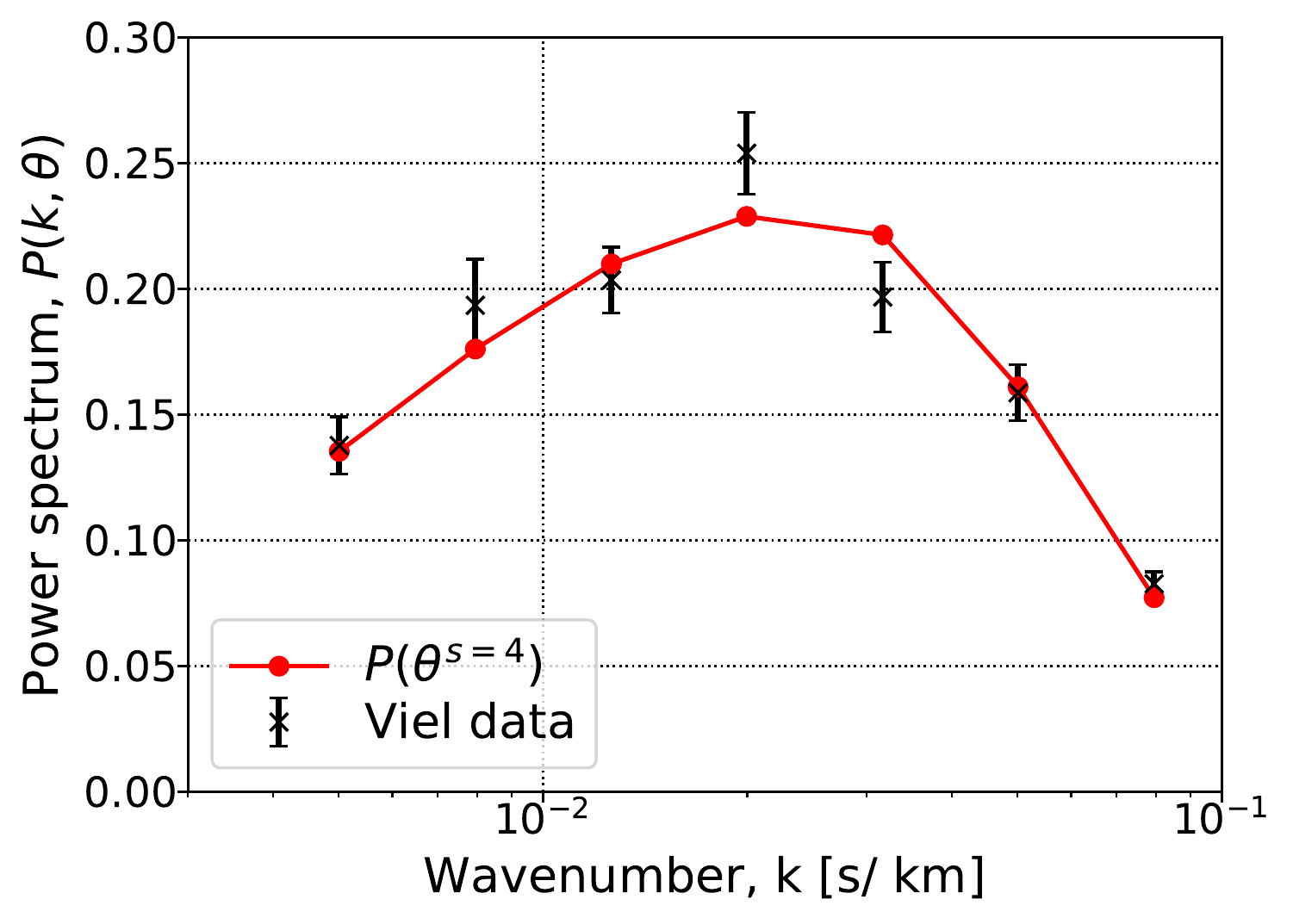}
    \end{minipage}%
    \caption{$\BP(\btheta^{s=4})$ corresponding to the best found misfit value for the adaptive GP with Nyx simulations and Viel data.}
    \label{fig:results_adaptive_inversion_nyx}
\end{figure}

We run a restricted version of Algorithm \ref{algo:adaptiveGP} with the possible selection of $\btheta$ limited to the existing thermal models. We start by selecting the first six points randomly, build a GP emulator using the IND approach, and evaluate the expected improvement in fit function $\cI(\btheta)$ for the remaining points (using Viel data). We select the  point that corresponds to the largest improvement in fit, and iterate until no further improvement can be made. In this regime we do not perform a direct optimization over the parameter space but select the inputs out of the available THERMAL data.
Our restricted algorithm terminates after iteration $s=4$ using a total of $10$ design points. 
Figure \ref{fig:results_adaptive_inversion_nyx} shows the plot demonstrating the fit of the $\BP(\btheta^{s=4})$ corresponding to the best found misfit value. 

\begin{figure*}[htbp]
    \centering
    \begin{minipage}[t]{0.55\textwidth}
        \centering
        \includegraphics[width=\textwidth, trim={0 0 0em 0},clip]{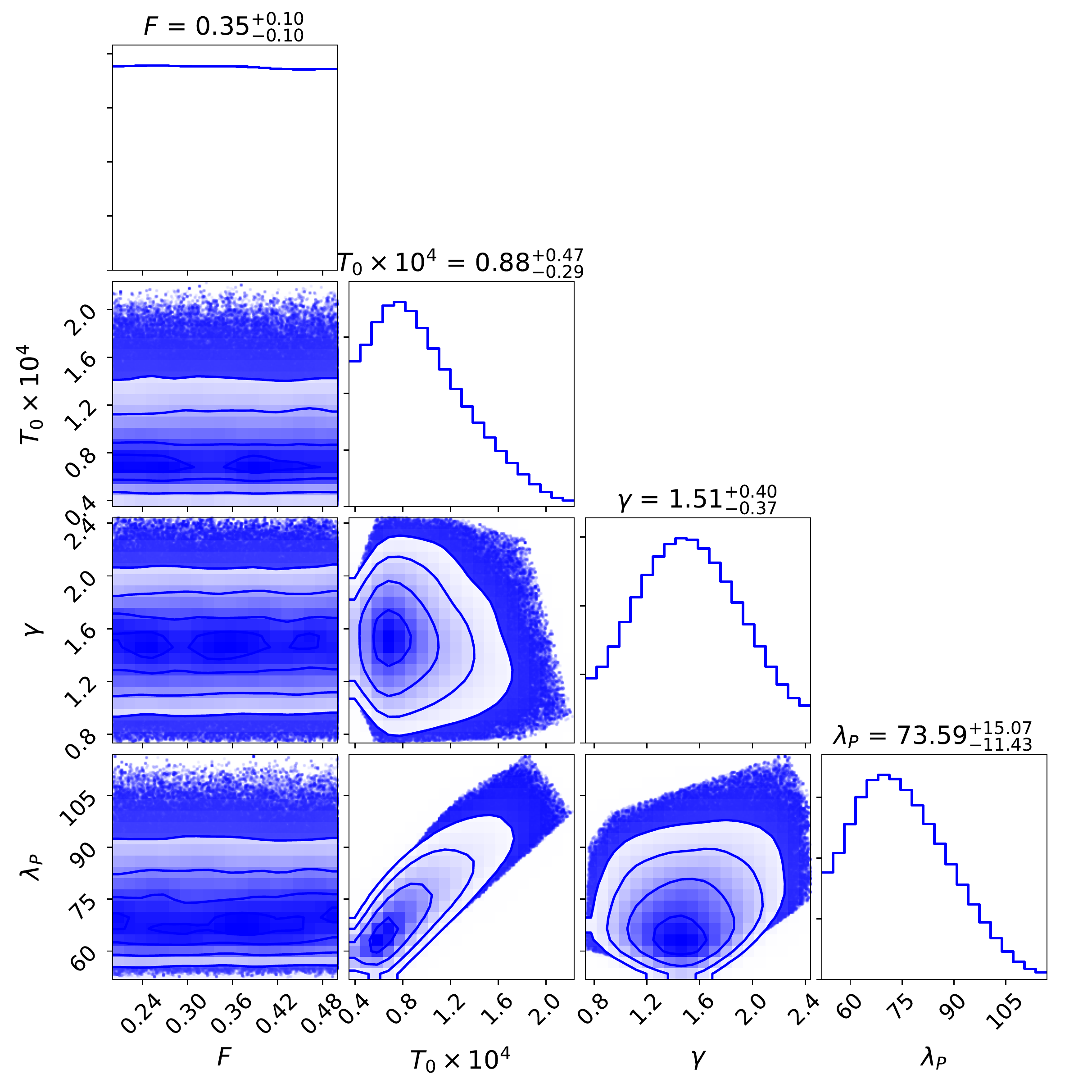}
    \end{minipage}%
    \caption{Prior $p(\btheta)$ for $\btheta=(F, T_0, \gamma, \lambda_P)$ \`a la \cite{Walther2018}}
    \label{fig:prior_truncated_walther}
\end{figure*}

The other important ingredient is the construction of the prior $p(\btheta)$.
We use a flat prior for $F$, $\log T_0$, $\gamma$, and $\log \lambda_P$ in a box constrained by the smallest and the largest values for each parameter. We then truncate this prior to the convex hull of the THERMAL grid points, as is done in \cite{Walther2018}.
The resulting truncated prior is shown in Figure \ref{fig:prior_truncated_walther}. 
This truncation is done to avoid GP extrapolation into a region of parameter space where this IGM model cannot produce an answer, for example, in case of very low $T_0$ but very high pressure smoothing scale (see also discussion in Section \ref{sec:LHD} about the $\lambda_P$ parameter).

\begin{figure*}[htbp]
\centering
    \begin{minipage}[t]{0.7\textwidth}
        \includegraphics[width=\textwidth, trim={0em 0 0em 0},clip]{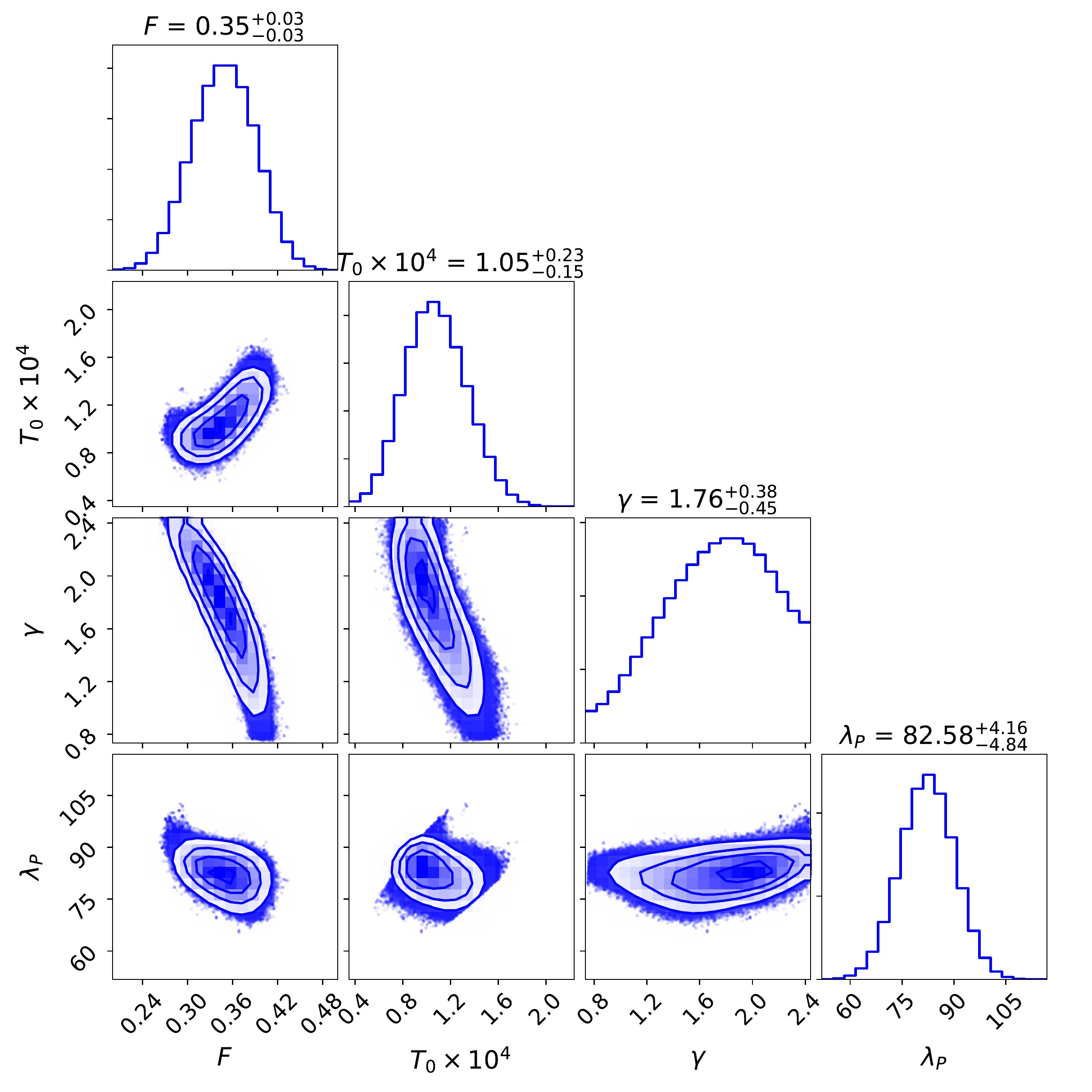}
    \end{minipage}%
    \caption{1D and 2D marginal posteriors for $\btheta=(F, T_0, \gamma, \lambda_P)$ obtained with a restricted version of Algorithm \ref{algo:adaptiveGP} using Nyx simulation and Viel data. Note that we apply smoothing to the plots of the marginal histograms which makes them look more Gaussian. The numbers above 1D histograms report $50\%$-quantiles of the marginal distributions plus/minus differences between $84\%$- and $50\%$-quantiles and $50\%$- and $16\%$-quantiles.}
\label{fig:posterior_adaptive_inversion_nyx_plus}
\end{figure*}

The posterior obtained with this prior using the likelihood based on the adaptively constructed GP is shown in Figure \ref{fig:posterior_adaptive_inversion_nyx_plus}. 
We observe that the resulting posterior is considerably more constrained than the prior, although we want to draw the reader's attention to the poor constraints on  parameter $\gamma$, which plays very little role at high redshifts (at most!)
when the density of Ly$\alpha$ absorbing gas is close to the cosmic mean.
Overall, the marginal ranges and central values for the parameters are in good agreement with the ones reported in \cite{Walther2018}.
Note, however, that we do not use BOSS \citep{NPD2013} measurements here, but only \cite{MViel_et_al_2013a} data,
as we want to avoid modeling of correlated Silicon III absorption which is present in the BOSS dataset.
We prefer maintaining the forward model as simple as possible as the goal of this work is
testing and improving inference schemes for the Ly$\alpha$ power spectrum.
In any case, the fact that our results presented here are completely consistent with
\cite{Walther2018} indicates that the BOSS dataset contributes negligible information about
the thermal state of the IGM at high redshifts.

\section{Conclusions}
\label{sec:conclusions}
In this work we described the use of an adaptive design of  GP emulators of the Lyman $\alpha$ flux power spectrum for solving inference problems for the thermal parameters of the IGM. To the best of our knowledge, while GP emulators are extensively used in the  cosmological community, the adaptive selection of training inputs has been considered only very recently in \cite{KKRogers_HBPeiris_APontzen_SBird_LVerde_AFontRibera_2019a} and in our work. Our motivation for this work is primarily the reduction of the number of computationally intensive simulations required to build a GP emulator. By prioritizing the regions of the parameter space that are consistent with the measurement data under the predictive model of the emulator, we obtain the desired reduction without sacrificing the quality of the parameter posteriors. A numerical study that we performed on a problem with an approximate model of the Lyman $\alpha$ forest power spectrum and with synthetic measurement data demonstrated that our adaptive approach obtains consistently good approximations of the parameter posterior and outperforms a similar-size fixed design approach based on maximin Latin hypercube designs.

We provided a complete framework for building multi-output GP emulators that  predict the power spectrum at the pre-selected modes $k$. 
Our numerical study demonstrates that the resulting multi-output emulators that either treat outputs as conditionally independent given the hyperparameters (IND) or explicitly model linear correlations between the outputs (COR)  are effective and computationally efficient. Furthermore, our approaches  allow us to train emulators using only highly limited number of training inputs, which in turn enables the adaptive selection of additional inputs.   

The initial results obtained with our adaptive approach are encouraging. Specifically, for the problem of inferring three thermal parameters of the IGM and mean flux using measurements of the power spectrum at seven values of $k$ our approach (constrained to the 75 available Nyx THERMAL simulations) required simulation outputs for only $10$ input values to constrain the parameters to the same level of accuracy as in \cite{Walther2018} that used substantially larger number of simulations.

Finally, we want to emphasize that we do not consider the ``classical'' parameterization of $\btheta=(F, T_0, \gamma, \lambda_P)$ to be the best for modeling the state of the IGM, but we nevertheless perform this type of analysis as it is straightforward to make comparisons of our results with previous works.
While these parameters have intuitive physical meaning in describing the thermodynamical state of the IGM, there are several practical problems with them.  First, they are \emph{output} rather than \emph{input} parameters which brings significant difficulties with implementations of sampling and iterative emulation procedure.
Second, these 4 parameters are parameterizing \emph{each time snapshot} instead of \emph{the physical model} itself. For that reason, we consider models which parameterize the time and duration of the reionization as well as associated heat input \citep{Onorbe2018} as better and we will be using those in future works.

\acknowledgments
Authors are grateful to Jose O\~norbe for making his Nyx simulations available to us, as well as for providing helpful comments and insights.  We thank Joe Hennawi and all members of the Enigma group\footnote{http://enigma.physics.ucsb.edu/} at UC Santa Barbara for insightful suggestions and discussions.
This research used resources of the National Energy Research Scientific Computing Center (NERSC), which is supported by the Office of Science of the U.S. Department of Energy under Contract no.~DE-AC02-05CH11231.
This work made extensive use of the NASA Astrophysics Data System and of the astro-ph preprint archive at arXiv.org.
\appendix
\section{Training GP emulators}
\label{sec:math}

Let $\btheta^{(j)}$, $j=1,\dots,n_{train}$, represent training inputs with $\BP\big(\btheta^{(j)}\big)$ being a $q$-vector of outputs corresponding to a particular input. Denote by $\by_i$ the $n_{train}$-vector of the normalized values of the $i$-th output, $i=1,\dots,q$, defined as follows
\[
    \by_i = \bigg(\widehat{P}_i\big(\btheta^{(1)}\big), \dots, \widehat{P}_i\big(\btheta^{(n_{train})}\big) \bigg)^T,
\]
where
\begin{equation}\label{eq:training_normalization}
    \widehat{P}_i(\btheta) = \frac{P_i(\btheta) - m_i}{\V_i^{1/2}}
\end{equation}
with
\[
    m_i = \frac{1}{n_{train}} \sum_{j=1}^{n_{train}} P_i\big(\btheta^{(j)}\big), \quad \V\nolimits_i = \frac{1}{n_{train}} \sum_{j=1}^{n_{train}} \bigg(P_i\big(\btheta^{(j)}\big)  - m_i \bigg)^2.
\]
The normalized training outputs together form an output matrix $\BY = [\by_1, \dots, \by_q] \in \real^{n_{train}\times q}$. Finally, the vectorized form of $\BY$ is obtained by stacking the normalized training outputs into a $(n_{train}\cdot q)$-vector $\overline{\by} = \text{vec}(\BY)$.

The set of all training inputs we denote by $\btheta_{train} = \{\btheta^{(j)}, j=1,\dots,n_{train}\}$, and the combined set of training inputs and outputs, or training data, by $\cD = \{\btheta_{train}, \BY\}$.

Training a GP emulator requires specifying the hyperparameters $\bpsi$ of its kernel. These are characterised by the posterior distribution (note that conditioning on $\BSigma_k$ is implicit but not necessary since it's fixed)
\begin{equation}\label{eq:hyperparameter_posterior}
    p(\bpsi \,|\, \cD) = \frac{p(\cD | \bpsi)p(\bpsi)}{p(\cD)},
\end{equation}
where 
\[
    p(\cD | \bpsi) = p(\BY | \btheta_{train}, \bpsi) = \cN_{n_{train}\times q}(\BY | \bmu^{norm}(\btheta_{train}), c(\btheta_{train}, \btheta_{train} ; \bpsi), \BSigma_k^{norm})
\]
is the likelihood of the training data $\cD$ under the matrix-normal distribution defining the Gaussian process (see Section \ref{sec:GPs}) and $p(\cD) = \int p(\cD|\bpsi)p(\bpsi) d\bpsi$ is referred to as \textit{evidence}. Note that $\bmu^{norm}$ is a normalized version of the mean function obtained by applying the linear transformation \eqref{eq:training_normalization}, and $\BSigma_k^{norm}$ is the inter-output correlation matrix. In order to somewhat simplify this notation let us denote the covariance matrix for the training inputs by $\BC_{\psi} = c(\btheta_{train}, \btheta_{train} ; \bpsi)$. Also, recall that we take $\bmu(\cdot)\equiv 0$. Thus, we have
\[
    p(\BY | \btheta_{train}, \bpsi) = \cN_{n_{train}\times q} (\BY | \bzero_{n_{train}\times q}, \BC_{\psi}, \BSigma_k^{norm}).
\]
In a vectorized form we can express the likelihood above as a regular multivariate normal density
\begin{equation}\label{eq:likelihood_data_vectorized}
    p(\overline{\by} | \btheta_{train}, \bpsi) = \cN_{n_{train}\cdot q}(\overline{\by} | \bzero_{n_{train}\cdot q}, \BSigma_k^{norm}\otimes\BC_{\psi}).
\end{equation}
How do we obtain the hyper-posterior \eqref{eq:hyperparameter_posterior}? Since no analytical form for this posterior exist, we describe it via a \textit{particle approximation} \cite[Section~2.6]{IBilionis_NZabaras_2016a}. That is we approximate the hyper-posterior with a weighted sum of Dirac delta functions centered at samples $\bpsi^{(j)}$:
\begin{equation}\label{eq:hyper_particle_approx}
    p(\bpsi | \cD) \approx \sum_{j=1}^{n_{\psi}} w^{(j)} \delta(\bpsi - \bpsi^{(j)})
\end{equation}
with weights $w^{(j)}\geq 0$ and $\sum_{j=1}^{n_{\psi}} w^{(j)} = 1$.

One way to obtain such a particle approximation is by maximizing the likelihood of the data given by \eqref{eq:likelihood_data_vectorized}. This leads to a single-particle approximation
\[
    p(\bpsi | \cD) \approx \delta(\bpsi - \bpsi^*_{MLE}),
\]
where 
\[
    \bpsi^*_{MLE} = \argmax_{\bpsi\in\cX_{\psi}} p(\cD | \bpsi)
\]
is the maximum likelihood estimator (MLE) of the hyperparameter vector. In the case of a flat prior on the hyperparameters this estimator coincides with a maximum a posteriori (MAP) estimator. MLE approach is convenient to work with, since the covariance matrix $\BC_{\psi}$ needs to be only formed once, however, it might lead to somewhat over-confident estimates of predictive uncertainties of the GP emulator \cite{}. In the case of a sharply peaked likelihood $p(\cD | \bpsi)$ the MLE estimator can be sufficient. 
Another way of obtaining the particle approximation of the hyper-posterior is by sampling it using MCMC techniques. This way provides a more complete picture of the hyper-posterior, albeit at an additional computational cost.

Whether using MLE or MCMC approach to obtaining hyper-posterior $p(\bpsi | \cD)$, we need to be able to evaluate the logarithm of the likelihood function \eqref{eq:likelihood_data_vectorized} (note that by applying logarithm we preserve the order relation and obtain a better-behaved function). In the following we derive the expression for the log-likelihood of the data and explain how it can be efficiently computed.

Let $\BSigma_{tot}=\BSigma_{k}^{norm}\otimes \BC_{\psi}$ and let $a_{i,j}$ denote the entries of $(\BSigma_{k}^{norm})^{-1}$, then
\begin{align*}
	\log p(\overline{\by} \,|\, \btheta_{train}, \bpsi) &= -\frac{1}{2} \overline{\by}^T \BSigma_{tot}^{-1} \overline{\by} - \frac{1}{2} \log | \BSigma_{tot} | -\frac{n_{train}\cdot q}{2} \log (2\pi) \\
	&= -\frac{1}{2}\sum_{i=1}^q \sum_{j=1}^q a_{i,k} \by_i^T \BC_{\psi}^{-1} \by_j - \frac{1}{2} \big( n_{train} \log |\BSigma_{k}^{norm}| + q \log |\BC_{\psi}| \big) \\
	&\hphantom{\,=\,} - \frac{n_{train}\cdot q}{2} \log(2\pi) \\
	&= -\frac{1}{2} \text{tr}((\BSigma_{k}^{norm})^{-1}\BY^T\BC_{\psi}^{-1}\BY) - \frac{1}{2} \big( n_{train} \log |\BSigma_{k}^{norm}| + q \log |\BC_{\psi}| \big) \\
	&\hphantom{\,=\,} - \frac{n_{train}\cdot q}{2} \log(2\pi).
\end{align*}
In our implementation we first compute
\[
	\BA = \BL^T\backslash(\BL\backslash\BY),
\]
where $\BC_\psi=\BL\BL^T$ is the Cholesky decomposition of the input covariance, then compute 
\[
	\BB = \BY^T\BA,
\]
and set
\[
	\BD = \BS^{T}\backslash (\BS \backslash \BB),
\]
where $\BSigma_{k}^{norm} = \BS\BS^T$ is the Cholesky decomposition of the output correlation matrix. Then
\[
	\log p(\overline{\by} \,|\, \btheta_{train}, \bpsi) = -\frac{1}{2} \bigg(\text{tr}(\BD) + 2n_{train}\sum_{i=1}^q \log(\BS_{i,i}) + 2q\sum_{i=1}^{n_{train}}\log(\BL_{i,i}) + qn_{train}\log(2\pi)\bigg).
\]
The expression above can be further simplified in certain cases. For example, for the IND emulator that treats outputs as independent given the $\BC_{\psi}$ matrix, the output correlation matrix $\BSigma_k^{norm} = \BI_q$, and the computation of $\text{tr}(\BD)$ does not require cross-terms $\by_i\BC_{\psi}^{-1}\by_j$ for $i\neq j$.

\section{Obtaining predictions}

In order to obtain a prediction for an un-tried input $\btheta$, we apply the standard GP formulas obtained by conditioning on the data $\cD$.
Furthermore, by exploiting the Kronecker product structure of the covariance, we can apply standard GP formulas to each output separately. Indeed, as shown in \cite{EVBonilla_MKChai_CWilliams_2008a},
\begin{align*}
	\bmm^{norm}(\btheta; \cD, \bpsi) &= (\BSigma_{k}^{norm}\otimes \bc_\psi)^T (\BSigma_{k}^{norm}\otimes \BC_{\psi})^{-1}\overline{\by} \\
	&= ((\BSigma_{k}^{norm})^T\otimes\bc_\psi^T)((\BSigma_{k}^{norm})^{-1}\otimes \BC_{\psi}^{-1})\overline{\by}\\
	&= ((\BSigma_{k}^{norm}(\BSigma_{k}^{norm})^{-1})\otimes(\bc_\psi^T\BC_\psi^{-1}))\overline{\by} \\
	&= (\bc_\psi^T\BC_\psi^{-1}\by_1,\dots, \bc_\psi^T\BC_\psi^{-1}\by_q)^T \\
	&= (m(\btheta; \cD_1, \bpsi), \dots, m(\btheta; \cD_q, \bpsi))^T \in\real^q,
\end{align*}
where superscript $\textit{norm}$ indicates that this is the predictive mean of the GP fitted to the normalized outputs, and $\bc_{\psi} = c(\btheta, \btheta_{train}; \bpsi)\in\real^{n_{train}}$. For the predictive covariance we get
\begin{align*}
	\BSigma_{GP}^{norm}(\btheta; \cD, \bpsi) &= c(\btheta, \btheta; \bpsi)\BSigma_{k}^{norm} - (\BSigma_{k}^{norm}\otimes\bc_{\psi})^T((\BSigma_{k}^{norm})^{-1}\otimes\BC_{\psi}^{-1})(\BSigma_{k}^{norm}\otimes\bc_{\psi}) \\
	&= (c(\btheta, \btheta; \bpsi) - \bc_\psi^T\BC_\psi^{-1}\bc_\psi)\BSigma_{k}^{norm} = \V(\btheta; \cD, \bpsi)\BSigma_{k}^{norm}.
\end{align*}
Upon re-scaling we obtain:
\[
    \BP^{GP}(\btheta) | \cD, \bpsi, \BSigma_k \sim \cN_{q}(\BP^{GP}(\btheta) | \bmm(\btheta; \cD, \bpsi), \BSigma_{GP}(\btheta; \cD, \bpsi))
\]
with
\[
	\bmm(\btheta; \cD, \bpsi) = (\V\nolimits_1^{1/2}m(\btheta; \cD_1, \bpsi) + m_1, \dots, \V\nolimits_q^{1/2}m(\btheta; \cD_q, \bpsi) + m_q)^T,
\]
and
\begin{align*}
    \BSigma_{GP}(\btheta; \cD, \bpsi) &= \BV^{1/2}\BSigma_{GP}^{norm}(\btheta; \cD, \bpsi)\BV^{1/2} \\
    &= \V(\btheta; \cD, \bpsi)(\BV^{1/2}\BSigma_{k}^{norm}\BV^{1/2}) \\
    &= \V(\btheta; \cD, \bpsi)\BSigma_{k},
\end{align*}
where $\BV = \text{diag}[\V_1, \dots, \V_q]\in\real^{q\times q}$. Finally, integrating out the hyperparameters $\bpsi$ (recall the particle approximation \eqref{eq:hyper_particle_approx}) we obtain
\begin{equation}\label{eq:GP_predictive_distribution_sample_approximation}
    \BP^{GP}(\btheta) | \cD, \BSigma_k \sim \sum_{j=1}^{n_{\psi}} w^{(j)} \cN_q\big(\BP^{GP}(\btheta) | \bmm(\btheta; \cD, \bpsi^{(j)}), \BSigma_{GP}(\btheta; \cD, \bpsi^{(j)})\big)
\end{equation}

\section{Inference using GP emulators}

Suppose now that we are given a vector of observations $\bd\in\real^q$ and a distribution of the measurement noise $\cN_q(\bzero_{q}, \BSigma_E)$ with a known covariance $\BSigma_E$. Upon substituting the true response $\BP(\cdot)$ with the GP emulator $\BP^{GP}(\cdot)$ in the likelihood of the measurement data, and integrating with respect to the GP distribution \eqref{eq:GP_predictive_distribution_sample_approximation}, we obtain (see \cite{TTakhtaganov_JMueller_2018a} for details) the so-called $\cD$-restricted likelihood
\[
    L(\btheta | \bd, \cD) = \sum_{j=1}^{n_{\psi}} \frac{s^{(j)}}{n_{\psi}} \exp\bigg[ - \frac{g\big(\btheta; \cD, \bpsi^{(j)}\big)}{2}\bigg],
\]
where $s^{(j)} = (2\pi)^{-q/2}|\BSigma_E + \BSigma_{GP}(\btheta; \cD, \bpsi^{(j)})|^{-1/2}$, and $g(\btheta; \cD, \bpsi)$ is a data misfit function defined as
\[
    g(\btheta; \cD, \bpsi) = (\bd - \bmm(\btheta; \cD, \bpsi))^T(\BSigma_E + \BSigma_{GP}(\btheta; \cD, \bpsi))^{-1}(\bd - \bmm(\btheta; \cD, \psi)).
\]
When performing inference the likelihood $L(\btheta| \bd, \cD)$ needs to be repeatedly evaluated for different values of $\btheta$.
Instead of using the Cholesky factorization of the matrix appearing in the definition of $g(\btheta; \cD, \bpsi)$, we compute the misfit efficiently as follows. 

First, we cover the case of homoscedastic measurement noise, i.e., when $\BSigma_E = \sigma_E^2\BI_q$. Denote the matrix appearing in $g(\btheta; \cD, \bpsi)$ as
\[
    \BSigma_{lik}(\btheta; \cD, \bpsi) = \BSigma_E + \BSigma_{GP}(\btheta; \cD, \bpsi).
\]
Plugging-in $\BSigma_E$ and $\BSigma_{GP}$ we get
\[
    \BSigma_{lik}(\btheta; \cD, \bpsi) = \V(\btheta; \cD, \bpsi)\bigg(\BSigma_k + \frac{\sigma_E^2}{\V(\btheta; \cD, \bpsi)} \BI_q \bigg).
\]
We have the sum of a symmetric matrix and a constant times the identity matrix. In this case, the inverse of $\BSigma_{lik}$ can be efficiently computed using the eigendecomposition of the $\BSigma_k$ matrix. Let
\[
	\BSigma_{k} = \BQ\BLambda\BQ^T, \quad \text{with } \BQ^{-1}=\BQ^T, \, \BLambda = \text{diag}[\lambda_1, \dots, \lambda_q].
\]
Then
\[
	\BSigma_{lik}^{-1}(\btheta; \cD, \bpsi) = \frac{1}{\V(\btheta; \cD, \bpsi)} \BQ \bigg[ \BLambda + \frac{\sigma_E^2}{\V(\btheta; \cD, \bpsi)}\BI_q \bigg]^{-1} \BQ^T.
\]
In order to compute the misfit function $g(\btheta; \cD, \bpsi)$ we first compute
\[
	\bvv = \BQ^T(\bd - \bmm(\btheta;\cD, \bpsi)) \in \real^q,
\]
then 
\begin{align*}
	g(\btheta; \cD, \bpsi) &= \frac{1}{\V(\btheta; \cD, \bpsi)} \bvv^T\BD^{-1}\bvv \\
	&= \sum_{i=1}^q \frac{1}{\V(\btheta; \cD, \bpsi)\lambda_i + \sigma_E^2} v_i^2,
\end{align*}
where $\BD =  \BLambda + (\sigma_E^2 / \V(\btheta; \cD, \bpsi) \BI_q$ is a diagonal matrix. Thus, for each $\btheta$ and $\bpsi$ the computation of the data misfit function requires $\cO(q^2)$ operations.

For a general $\BSigma_E$, we use the generalized eigendecomposition
\[
	\BSigma_E\BU = \BSigma_{k}\BU\BLambda,
\]
which leads to the following form for the inverse of $\BSigma_{lik}$:
\[
	\BSigma_{lik}^{-1}(\btheta; \cD, \bpsi) = \frac{1}{\V(\btheta; \cD, \bpsi)} \BU \left[ \BI_q + \frac{1}{\V(\btheta; \cD, \bpsi)} \BLambda \right]^{-1} \BU^T.
\]
Then
\[
	g(\btheta; \cD, \bpsi) = \sum_{i=1}^q \frac{1}{\V(\btheta; \cD, \bpsi) + \lambda_i} v_i^2,
\]
with $\bvv = \BU^T(\bd - \bmm(\btheta; \cD, \bpsi))$.

\bibliography{main}

\end{document}